\documentclass[a4paper,fleqn,usenatbib,useAMS]{mnras}
\usepackage{ae,aecompl}
\usepackage{float}
\usepackage[final]{graphicx}
\usepackage{subcaption}
\usepackage{booktabs}
\usepackage{graphicx}	% Including figure files
\usepackage{amsmath}	% Advanced maths commands
\usepackage{amssymb}	% Extra maths symbols
\usepackage{txfonts}
\usepackage[T1]{fontenc}
\usepackage{upgreek}
\usepackage{pdflscape}
\usepackage{afterpage}
\newcommand{\molh}{$\mathrm{H_{2}}$}
\newcommand{\Spitzer}{\textit{Spitzer}}
\newcommand{\WISE}{\textit{WISE}}
\newcommand{\W}{\textit{W}} %all WISE colorbands need to be italic.
\newcommand{\micronm}{$\upmu\mathrm m$}
\newcommand{\eqwpah}{{EQW[PAH~6.2~\micronm]}}
\title{Mid-Infrared Spectroscopic Evidence for AGN Heating Warm Molecular Gas}

\author[E. L. Lambrides et al.]{Erini L. Lambrides,$^{1,2}$\thanks{E-mail: \href{mailto:erini.lambrides@jhu.edu}{erini.lambrides@jhu.edu}}
Andreea O. Petric,$^{2}$
Kirill Tchernyshyov,$^{1}$
Nadia L. Zakamska$^{1}$
\newauthor Duncan J. Watts$^{1}$
\\
$^{1}$ Department of Physics \& Astronomy, Johns Hopkins University, Bloomberg Center, 3400 N. Charles St., Baltimore, MD 21218, USA\\\
$^{2}$ Institute for Astronomy, University of Hawaii, 65-1238 Mamalahoa Hwy, Kamuela, Hawaii 96743, USA
}

\date{Accepted XXX. Received YYY; in original form ZZZ}

\pubyear{2018}

\begin{document}
\label{firstpage}
\pagerange{\pageref{firstpage}--\pageref{lastpage}}
\maketitle

\begin{abstract}

We analyse 2,015 mid-infrared (MIR) spectra of galaxies observed with \textit{Spitzer}'s Infrared Spectrograph, including objects with growing super-massive black holes and objects where most of the infrared emission originates from newly formed stars. We determine if and how accreting super-massive black holes at the centre of galaxies -- known as active galactic nuclei (AGN) -- heat and ionize their host galaxies' dust and molecular gas. We use four MIR diagnostics to estimate the contribution of the AGN to the total MIR emission.  We refer to galaxies whose AGN contribute more than 50 per cent of the total MIR emission as AGN-dominated. We compare the relative strengths of PAH emission features and find that PAH grains in AGN-dominated sources have a wider range of sizes and fractional ionizations than PAH grains in non-AGN dominated sources. We measure rotational transitions of \molh\ and estimate \molh\ excitation temperatures and masses for individual targets, \molh\ excitation temperatures for spectra stacked by their AGN contribution to the MIR, and the \molh\ excitation temperature distributions via a hierarchical Bayesian model. Using the hierarchical Bayesian model, we find an average 200K difference between the excitation temperatures of the \molh S(5) and \molh S(7) pure rotational molecular hydrogen transition pair in AGN-dominated versus non-AGN dominated galaxies. Our findings suggest that AGN impact the interstellar medium of their host galaxies. 
%We analyse 2015 mid-infrared spectra of active galaxies to determine if and how accreting super-massive black holes at the centre of galaxies (active galactic nuclei) impact the interstellar medium of their hosts. We assess the AGN's impact on the gas and dust of their host galaxies by compiling the largest sample of extragalactic objects with mid-infrared spectroscopic data, and performing a suite of diagnostics and statistical tests. We use several MIR diagnostics to estimate the AGN contribution to the total IR emission and refer to sources whose AGN contribution is more than 50\% of the total IR emission as AGN dominated. We conduct a census of the warm dust emission as modelled by polycyclic aromatic hydrocarbons. We find that dust in AGN dominated galaxies appears to have a more diverse dust grain distribution than non-AGN dominated galaxies. We interpret our findings as evidence of distinct differences in the star-forming molecular gas in AGN host galaxies.
%We  provide the largest census of rotational molecular hydrogen emission. We find an average 200 K difference between the excitation temperatures of the higher pure rotational molecular hydrogen transitions in AGN dominated versus non-AGN dominated spectra. We interpret our molecular hydrogen temperature differences as evidence of AGN host galaxies having a warmer or more dense warm molecular hydrogen component. Both results 

\end{abstract}

\begin{keywords}
galaxies: active - galaxies: ISM - 
galaxies: starburst - infrared: galaxies - techniques: spectroscopic - surveys
\end{keywords}

\section{Introduction} \label{sec:intro} 

The evolution of central supermassive black holes (SMBHs) appears connected to the histories of the host galaxies that harbour them. Observations suggest that there are SMBHs in all galaxy bulges and their masses are proportional to the masses of the host bulges \citep[see][for reviews]{fabian12,kormendyaa,heckmanaa}. Furthermore, star-formation and SMBH growth have similar evolutions \citep[see][for a review]{madau}. Theory suggests that feedback from growing SMBHs/active galactic nuclei (AGN) is able to successfully reproduce the properties of local massive galaxies \citep[see][for review]{silk12}, and explain the observed galaxy scaling relations and the quenching of star-formation in massive galaxies \citep{silk98,fabian99,king03,hopkins06,illustris2017}.

There is mounting observational evidence for AGN interacting with the gas and dust of their host galaxies. Some AGN appear to ionize the interstellar medium (ISM) up to several kiloparsecs away from the central black hole \citep{greene2011,greene2012,liu2013,cresci,villar,karouzos,dominika}. Strong radio galaxies have been observed injecting energy into the molecular gas of their host galaxies \citep[e.g.][]{appleton,ogle,nesvadba,guillard}. Molecular outflows have been observed in powerful quasars \citep{feruglio10,cicone12,stone16}. Evidence for feedback effects in host galaxies that harbour lower luminosity AGN has been mixed, but these surveys were on relatively small numbers of AGN \citep[e.g.][]{petric,Hill,stierwalt,petric18}. In this paper, we use mid-infrared (5.2--38.0~\micronm) spectra of a sample of 2,015 galaxies, 942 of which are galaxies whose IR emission comes predominantly from the AGN, to investigate the impact of the AGN on the warm molecular gas and dust components of the ISM in their host galaxies. 
\newpage
The ISM fuels star-formation and AGN activity. The primary sources for heating the ISM in AGN host galaxies are newly formed stars and supernovae \citep[e.g.][]{weedman}, AGN \citep{sanders,elvis,elitzur12}, and old stars \citep{buat,rowan,sauvage92,sauvage94}. To estimate the impact AGN have on the ISM, we first estimate how much the AGN contributes to the total mid-infrared (MIR) emission. We use a range of diagnostics developed from studies of normal galaxies, luminous AGN, and luminous infrared galaxies using data from the \textit{Infrared Space Observatory} \citep[for a review]{genzel} and the \textit{Spitzer Space Telescope}'s Infrared Spectrograph \citep{armus07,spoon,petric}.

Optical diagnostics \citep[e.g.][]{baldwin,kauffmann} can provide distinctions between star-formation (SF) and accretion processes, but are not ideal for objects with significant dust obscuration or for composite objects with both significant AGN and SF activity \citep{trump15}. MIR diagnostics are less sensitive to dust obscuration. MIR empirical methods that can be used to disentangle an AGN-dominated from an SF-dominated galaxy include the ratio of the continuum to dust emission features, the relative fluxes of high- to low-ionization emission, and the slope of the MIR continuum. These diagnostics were derived using observations of pure star-formation and pure-AGN samples \citep{genzel,laurent,armus06,smith07,spoon}. In this paper we use the 6.2~\micronm\ polycyclic aromatic hydrocarbon (PAH) equivalent width, hereafter \eqwpah, to quantify AGN activity.

%Here we use measurements of PAH emission features to investigate the impact AGN have on the ISM of their host galaxies by comparing them to theoretical models of PAH emission in different environments, and observations of PAH features other samples of galaxies. 

PAHs are organic compounds whose emission features in physics laboratories are similar to MIR features in astronomical spectra \citep{leger,allamandola}. PAH emission features are ubiquitous in MIR spectra of regions with recent star-formation \citep{tielens05}.
%PAHs are commonly attributed as the main carrier of the unidentified infrared bands \citep{leger,allamandola}. 
PAHs radiate through IR fluorescence after being excited by a single ultraviolet photon and may play an important role in the energy balance of the ISM. Several models predict the impact of radiation on the ionization and grain sizes of PAHs \citep{li2001,draine07}. %Utilizing these models, the PAHs' dust properties, e.g. their ionization state and grain size, can be used to assess the impact of the ionizing source on the ISM. 
%Although there is not an exact known correspondence between MIR features and PAH properties \citep{sadjadi,zhang}, examination of PAH emission in nearby galaxies suggests certain empirical trends \citep{smith07,sales}. 
Although the relations between the PAH features and their environments are not completely understood \citep{sadjadi,zhang}, empirically we measure low \eqwpah\ in galaxies with AGN \citep{smith07,sales}. This property is a powerful diagnostic of the AGN's contribution to the MIR emission.  %, making their emission suitable for estimating the UV radiation field strength. 

%The Infrared Spectrograph \citep[IRS]{irs} aboard the \textit{Spitzer  Space Telescope} has the MIR wavelength coverage (5.2--38.0 \micronm) necessary to estimate the AGN's contribution to the IR emission, the properties of the warm dust from PAH emission, the strength of the 9.7 \micronm\ silicate feature, and the emission from rotational transitions of warm molecular hydrogen (\molh).

%The warm molecular gas - ISM component , and in the MIR, \molh\ is an important coolant. 
% The rotational transitions of \molh\ observed in MIR spectra trace the mass and temperature distribution of the molecular gas heated to 100--1000 K. \molh\ can be excited through three primary mechanisms: (1) Far ultraviolet heating, in which photons radiatively pump the \molh\ into its electronically excited states; (2) inelastic collisions, in which collisions maintain the lowest pure rotational levels in thermal equilibrium in regions where the gas density and temperature is high enough, (3) X-ray heating in which hard X-rays penetrate into zones opaque to UV photons and radiatively excite \molh. 

In star-forming galaxies, \molh\ and PAH emission are tightly correlated \citep{roussel}. \molh\ is the dominant component of the warm, dense, star-forming molecular gas of galaxies. \molh\ can be excited through three primary mechanisms: (1) far ultraviolet heating, in which photons radiatively pump the \molh\ into its electronically excited states; (2) inelastic collisions, in which collisions maintain the lowest pure rotational levels in thermal equilibrium in regions where the gas density and temperature is high enough; and (3) X-ray heating, in which hard X-ray photons penetrate into UV-opaque zones and radiatively excite \molh. 

In normal galaxies, \molh\ is predominantly heated by far-ultraviolet photons in photon-dominated regions (PDRs) \citep{hollenbach}. For PDRs with $n_\mathrm{H} \gtrsim 10^{4} \,\mathrm{cm}^{-3}$, collisions maintain the lowest rotational levels ($J\lesssim5$), keeping the PDRs in thermal equilibrium \citep{burton}. This makes their populations consistent with Boltzmann distributions, which makes the \molh\ emission a robust thermal probe. Other sources of \molh\ excitation include small-scale shocks \citep{neufeld}, extra-nuclear large-scale shocks from galactic gravitational interactions \citep{appleton,cluver,ogle12}, and X-ray heating \citep{roussel}. % However, in normal galaxies the bulk of \molh\ emission is due to the radiative pumping via FUV photons in PDRs \citep{rigopoulou,higdon,roussel}. 

Some AGN host galaxies appear to have more \molh\ emission relative to that of other coolants such as PAHs or [\ion{Si}{ii}] emission, suggesting that at least some of the \molh\ does not originate in PDRs. This may indicate that AGN impact the molecular component of their host's ISM \citep{rigopoulou,higdon,zakamska,petric,shipley,Hill}. While observational studies have provided evidence of some AGN injecting the additional energy required to heat the molecular gas, the small sample size of these studies makes it difficult to assess whether this scenario is representative. Our large catalogue of AGN resolves this.%The ISM's dust properties, e.g. its ionization state and grain size,  can be used to assess the impact an AGN has on the ISM. Broad MIR dust-bands are commonly fit with PAH models. While PAH emission features are still somewhat mysterious, it appears that PAHs are made up of tens to hundreds of carbon atoms in planar lattices, and are associated with strong emission features in the MIR spectrum of star-forming or AGN-starburst composite galaxies \citep{leger, allamandola}. PAHs radiate through IR fluorescence after being excited vibrationally by a single ultraviolet photon \citep{tielens05}, making their emission suitable for estimating the UV radiation field strength. 

In galaxies where the AGN contributes most of the IR emission, there is an excess of warm \molh\ emission relative to PAH emission \citep{rigopoulou}. Subsequent studies using \Spitzer's Infrared Spectrograph confirmed the trend of excess  \molh\ emission in Ultra Luminous InfraRed Galaxies (ULIRGs) with IR luminosities above $10^{11}\,\mathrm L_{\sun}$, and a subset of slightly less luminous LIRGs \citep{zakamska,Hill,stierwalt, petric18}. \citet{ogle12} find excess \molh\ emission in over 30 per cent of the their sample of radio galaxies. However, \citet{higdon} analyse a similar sample of ULIRGs, and do not find a relationship between the warm \molh\ mass and the \textit{IRAS} 25 to 60~\micronm\ flux density ratio (an empirical AGN contribution diagnostic), despite finding an excess of warm \molh\ relative to the PAH emission. 

In this paper we present \molh\ and PAH emission measurements in active galaxies observed with the \Spitzer\ IRS low resolution ($R= {{\lambda}/{\Delta \lambda}} \sim 60$) modules. Our sample consists of a wide range of infrared luminosities ($\nu L_\nu[24\ \micron]\sim10^{8}\textrm{--}10^{12}\, \mathrm L_{\sun}$), which allows us to test if the \molh\ to PAH ratio increases as a function of the AGN's contribution to the total IR emission of the galaxy, and if the temperatures of the warm \molh\ are different in AGN host galaxies versus SF dominated galaxies. We use the pure rotational transitions of \molh\ observed in the MIR to estimate the masses and temperatures of 100--1000~K molecular gas. We then look for differences between \molh\ in AGN-dominated galaxies and \molh\ in SF-dominated systems. 

In \autoref{sec:sample} we describe the data acquisition, reduction, and analysis algorithms. In \autoref{sec:results} we present our AGN selection methods, PAH properties of our sample, and molecular hydrogen properties of our sample. We show a significant difference between the temperatures of the higher \molh\ transitions in AGN and SF-dominated systems via three independent analysis methods. In \autoref{sec:discussion} we discuss the implications of AGN host galaxies containing higher \molh\ temperature distributions than galaxies dominated by SF processes, and we summarize our findings in \autoref{sec:conclusion}. We use an $h = 0.7$, $\Omega_{m} = 0.3$, $\Omega_{\Lambda} = 0.7$ cosmology throughout this paper. To evaluate the statistical significance of correlations, we use the Spearman rank test ($r_{s}$), and report the probability of a null hypothesis as $p_{s}$, the probability of two sets of data being uncorrelated. We use the two-sample Kolmogorov--Smirnov test ($D_\mathrm{KS}$) to evaluate if two underlying distributions come from the same distribution, and report the probability of the two distributions being the same as $p_\mathrm{KS}$.

\section{Sample, Data, and Measurements} \label{sec:sample}

\subsection{Data Mining}

The Infrared Spectrograph (IRS) aboard the \textit{Spitzer  Space Telescope} has four separate modules that cover 5.2--3.8~\micronm: Short-Low (SL), Short-High (SH), Long-Low (LL), and Long-High (LH) \citep{irs}. Here we amass spectra obtained with the low resolution modules, SL ($60  < R < 128$) and LL ($57  < R < 126$). Each low-resolution module is divided into two in-line sub-slits (i.e. two spectroscopic orders per module): SL1 ($7.46\ \micron < \lambda < 14.29\ \micron$), SL2 ($5.13\ \micron < \lambda < 14.29\ \micron$), LL1 ($19.91\ \micron<\lambda < 39.90\ \micron$), and LL2 ($13.90\ \micron< \lambda < 21.27\ \micron$). Some data contain bonus segments in the first order of each module (SL1 Bonus Segment - $7.3\ \micron < \lambda < 8.7\ \micron$ and LL2 bonus segment - $19.4\ \micron < \lambda < 21.7\ \micron$).

Each observation has an associated unique identifier, an AORkey, which we used to find the observation within the \Spitzer\  mission, including coordinates, observation type, and all other relevant information \Spitzer\  releases associated with the object. 
The \textit{Spitzer  Space Telescope} team stores this information from all \Spitzer\ observations on the NASA/IPAC Infrared Science Archive (SHA). We begin by mining the abstracts from the accepted cold mission \Spitzer\  proposals. We use a technique known as `web scraping' to extract data from websites by parsing the html source of the website. We extract all observing programs that contain the following keywords in their abstract text: AGN, Radio Galaxy, QSO, Quasar, Starburst Galaxy, and ULIRG/LIRG. We use the SHA to retrieve IPAC tables with relevant object and observation information (i.e. coordinates, instrument mode, AORkey, etc.) for every program identification number. For the 439 programs, we find a total of 3,793 AORkeys. This paper focuses only on the low-resolution IRS mode, which includes  2,807 AORkeys. Finally, after acquiring redshifts (which we describe in more detail in \autoref{sec:fluxcal}) and only using spectra with detection levels $\ge 3\sigma$, we obtain our final sample of 2,015 targets. 

\begin{figure}
\centering
\includegraphics[width=\columnwidth]{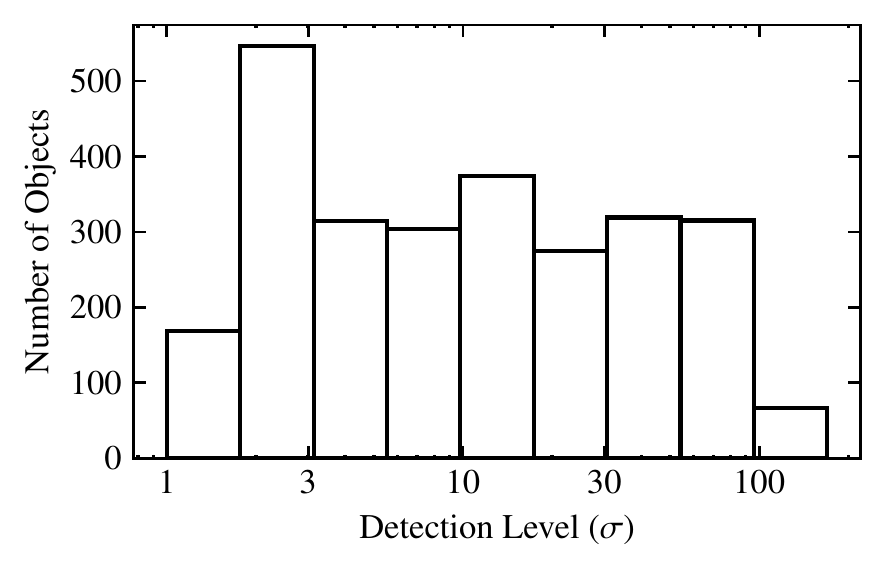}
	\caption{Quality of the sample via CASSIS pipeline:
    The detection level is the maximal average signal-to-noise ratio among the module/order/nod spectra, and is a data product provided with the reduced spectra. We only use spectra with detection levels $\ge 3\sigma$ and accurate redshift determination, which is discussed in \autoref{sec:sampprop}. This leaves us with 2,015 objects.
    }
\label{fig:detection}
\end{figure}

We use the \Spitzer\ low-resolution reduced spectra provided by the Combined Atlas of Sources with \Spitzer\ IRS Spectra \citep[CASSIS]{cassis}. The majority of our sample does not have reduced spectra via the enhanced products of \Spitzer\ in the SHA, so we use only the CASSIS reduced spectra for consistency. The CASSIS pipeline handles a variety of different observations via an automatic extraction algorithm that accounts for each signal's detection quality, as well as its spatial extent. The spectral extraction pipeline performs optimal extraction for point-like sources, and a tapered column extraction for extended sources (defined as being greater than $2$ arcmin in spatial extent). The optimal extraction method uses the point spread function profile to weigh the pixels in the spatial profile, while the tapered column extraction integrates the flux in a spectral window that expands with wavelength. The algorithm employed in the CASSIS pipeline % quantifies the detection level of an object: it 
approximates an uncertainty $\sigma$ for each spectrum by finding the maximal average signal-to-noise ratio among the module/order/nod spectra. We show the quality of the spectra in our sample in \autoref{fig:detection}.    

\subsection{Stitching}
In 25 per cent of our spectra, we find a difference between the flux in the spectral region of 13.9 to 14.2~\micronm\ as measured in the SL and LL data respectively. This is partially due to the different widths of the SL and LL slits: SL1 has a width of 3.7 arcsec and LL2 a width of 10.5 arcsec. This causes different parts of a source to be observed by the two slits. For example, at $z = 0.3$ the two slit widths correspond to physical distances of 16.5 kpc and 46.7 kpc respectively.  

We use the overlap region to scale the SL spectra to the LL measurements. The range of redshifts ($0.002<z<3.0$) in our sample causes the potential break to occur at different rest-frame wavelengths. We develop automated methods to calculate the necessary scalings and account for possible emission features near the overlap region. We use a 1~\micronm\  window size, centred on the wavelength location of the slit boundaries, to ensure we include enough flux points from each order. We assume the continuum is linear in this small spectral window, then look for and eliminate any emission lines. We then fit a line to the SL and LL overlap separately, estimate the flux from these fits, and estimate a scaling factor to bring the SL overlap emission up to the LL overlap value. To mask out any potential lines in our overlap windows, we proceed as follows. We calculate the forward finite difference for each pair of flux points, i.e. $(f_{\nu,i+1} - f_{\nu,i}) / \Delta \lambda$, where $f_{\nu}$ is the flux density and $\lambda$ the corresponding wavelength array. We exclude any points whose difference is greater than a standard deviation of the finite difference array. After this step, we perform an additional check by fitting a linear continuum using least squares minimization on each of the spectral segments. If the slopes of the spectral segments are not consistent to within a standard deviation of each segment's fit, we iteratively remove points until the slope of the line fits this criterion. We provide the resultant scale factors in \autoref{tab:egsampleprops}.

%\afterpage{
%    \clearpage
\begin{table*}
    %\centering
    %\caption{Full Catalogue Properties: We list the properties of ten
        %representative objects below. The full version of this table is
    %available in on-line format.}

%\begin{subtable}{\textwidth}
    %\centering
\caption{Example Sample Properties: We list the AORkey (\Spitzer\ IRS identification number), RA and Dec, the level of the detection as provided by the CASSIS reduction pipeline, and the cross-matches \WISE\ and 2MASS magnitudes. The full version of this table is available in on-line format.}
\label{tab:egsampleprops}
\begin{tabular}{|r|r|r|r|l|l|r|r|r|r|r|r|r|}
AORkey   & RA            & Dec            & Detection & $z$            & SL1--LL2 Scale & \W1 & \W2 & \W3 & \W4 & $J$ & $H$ & $K_{s}$ 
\\
& (deg) & (deg) & ($\sigma$) & & &(mag)&(mag)&(mag)&(mag)&(mag)&(mag)&(mag)\\
\hline
4935168  & 186.727 & $-0.878$  & 109      & 0.0073 & 1.126  & 10.75  & 9.49  & 3.89  & 0.32  & 13.18        & 12.45        & 11.86        \\
6650880  &  69.961 & $-48.721$ & 48       & 0.2035 & 1      & 14.15 & 12.92  & 8.63  & 5.64  & 16.38        & 15.72        & 15.00        \\
22115072 & 139.977 & 32.933  & 15       & 0.0499 & 1.727  & 11.29 & 10.85 & 6.51   & 3.84  & 14.02        & 13.28        & 12.79        \\
4671744  & 186.265 & 12.887  & 13       & 0.0034 & 1      & 8.02  & 8.02  & 7.14  & 5.91  & 10.66        & 10.05        & 9.81         \\
4985600  & 253.245 & 2.401   & 109      & 0.0245 & 1.097  & 9.34  & 8.62  & 4.53  & 1.27  & 11.94        & 11.20        & 10.57        \\
22079488 & 133.908 & 78.223  & 29       & 0.0047 & 1.987  & 8.54   & 8.46  & 6.38  & 4.30    & 10.38        & 9.55         & 9.40         \\
18526208 & 184.740 & 47.304  & 42       & 0.0015 & 1.250  & 8.54   & 8.18  & 5.48  & 3.31  & 11.07        & 10.54        & 10.07        \\
25408512 & 171.848 & $-29.258$ & 39       & 0.0239 & 1.125  & 10.88 & 10.54 & 6.50  & 3.86   & 13.67        & 12.86        & 12.37        \\
20316160 & 86.796 & 17.563   & 80       & 0.0186 & 1.192  & 10.26 & 9.79  & 5.11  & 2.04  & 13.16        & 12.23        & 11.46        \\
22087680 & 187.509 & 13.637  & 7        & 0.0045 & 1.606  & 8.85  & 8.91  & 7.99  & 6.58  & 10.69        & 9.90         & 10.02       
\end{tabular}
%\end{subtable}
%\end{table*}

\vspace{1cm}

%\begin{table*}
%\begin{subtable}{\textwidth}
    \centering
\caption{Example Molecular Hydrogen Results: We list the AORkey and \molh S(0), \molh S(1), \molh S(2), \molh  S(3), \molh S(5), \molh S(6), \molh S(7) line luminosities in units of $10^{39}\ \mathrm{erg\,s^{-1}}$ with their respective errors for $ \ge 2\sigma$ detections of 10 example objects. For $<2\sigma$, we only report the upper limit. The full version of this table is available in on-line format.}
\label{tab:egmolh}
\begin{tabular}{r|rr|rr|rr|rr|rr|rr|rr}
AORkey   & $L[$\molh S(0)$]$  & $L[$\molh S(1)$]$    & $L[$\molh S(2)$]$    & $L[$\molh S(3)$]$    & $L[$\molh S(5)$]$    & $L[$\molh S(6)$]$    & $L[$\molh S(7)$]$  &  \\
 & ($10^{39}\ \mathrm{erg\ s^{-1}}$)
 & ($10^{39}\ \mathrm{erg\ s^{-1}}$)
 & ($10^{39}\ \mathrm{erg\ s^{-1}}$)
 & ($10^{39}\ \mathrm{erg\ s^{-1}}$)
 & ($10^{39}\ \mathrm{erg\ s^{-1}}$)
 & ($10^{39}\ \mathrm{erg\ s^{-1}}$)
 & ($10^{39}\ \mathrm{erg\ s^{-1}}$)
\\
\hline
4935168  & $28.9     \pm 11.89        $&       $14.59   \pm 2.51        $ & $5.16    \pm 2.57         $&  $5.34    \pm 0.47        $ &   $<8.9     $ & $< 7.88  $ & $<10.08 $ \\
6650880  &  $<2419.43 $& $544.47  \pm 222.07      $ &  $<677.1  $& $231.52  \pm 84.55       $ &  $304.17   \pm 347.44      $ &  $< 1083.2$ & $<885.41$ \\
22115072 &    $<290.4      $& $295.51  \pm 131.33      $ & $43.48   \pm 25.74        $&  $91.32   \pm 22.97       $ &  $93.95    \pm 67.61       $  & $< 224.95$ & $<178.68$ \\
4671744  &  $<0.09    $& $0.28    \pm 0.04        $ & $0.09    \pm 0.03         $& $0.38    \pm 0.03        $  & $1.03     \pm 0.07        $    &$< 0.28  $ &$ 0.19    \pm 0.09       $   \\
4985600  & $102.43   \pm 41.22        $&  $908.96  \pm 26.26       $ & $300.52  \pm 21.8         $& $513.38  \pm 13.3        $  & $1192.94  \pm 41.3        $ &  $456.98   \pm 36.12       $ & $203.6   \pm 29.01      $   \\
22079488 &  $<1.55    $& $4.1     \pm 0.63        $ & $1.57    \pm 0.38         $&  $3.27    \pm 0.47        $  & $6.69     \pm 1.24        $  & $< 5.44  $ &   $<4.75  $ \\
18526208 & $0.18     \pm 0.03         $& $1.91    \pm 0.07        $ & $0.18    \pm 0.04         $& $0.3     \pm 0.05        $  & $0.11     \pm 0.09        $ &  $0.12     \pm 0.08        $ & $<0.33  $ \\
25408512 &  $<28.63   $& $72.57   \pm 11.67       $ & $17.14   \pm 6.72         $&  $25.09   \pm 6.83        $  &  $<53.85   $ & $< 42.81 $ & $<40.28 $ \\
20316160 & $49.4     \pm 11.62        $& $107.34  \pm 10.61       $ &  $<14.19  $& $25.29   \pm 3.54        $  & $33.01    \pm 13.36       $  & $< 54.89 $ &$<34.9  $ \\
22087680 &  $<0.32    $& $0.33    \pm 0.12        $ &  $0.21    \pm 0.08         $&  $0.76    \pm 0.1         $ &   $<0.88    $ & $< 0.76  $ & $<0.9   $
\end{tabular}
%\end{subtable}
%\end{table*}

\vspace{1cm}
%\begin{table*}
%\begin{subtable}{\textwidth}
    \centering
\caption{Example PAH Results: We list the AORkey, the \eqwpah\ (and upper limits in the case of $<2\sigma$ $L[\mathrm{PAH\ 6.2\ \micron }]$ detection), the PAH 6.2~\micronm\, PAH 7.7~\micronm\, and PAH 11.3~\micronm\ line luminosities in units of $10^{41}\ \mathrm{erg\,s^{-1}}$ with their respective errors for $\ge 2\sigma$ detections of 10 example objects or upper limits for $<2\sigma$, and silicate feature strength $\tau_\textrm{9.7 \micron}$. The full version of this table is available in on-line format.}
\label{tab:egpah}
\begin{tabular}{rrrrrrrrrr}
AORkey   & \eqwpah &  $L[\mathrm{PAH\ 6.2\ \micron }]$ & $L[\mathrm{PAH\ 7.7\ \micron }]$ &  $L[\mathrm{PAH\ 11.3\ \micron }]$  & $\tau_{9.7\,\mathrm{\umu m}}$\\
& (\micronm)  & ($10^{41}\ \mathrm{erg\ s^{-1}}$) &  ($10^{41}\ \mathrm{erg\ s^{-1}}$) & ($10^{41}\ \mathrm{erg\ s^{-1}}$) \\
\hline
4935168  & $<0.24$      & $                      < 0.83                  $     & $48.50               \pm 5.40                    $ & $1.18                  \pm 0.026 $         & 3.10            \\
6650880  & 0.61                 & $189.64              \pm 7.94                       $    & $841.37              \pm 54.84                   $ & $114.34               \pm 5.31 $         & 1.88            \\
22115072 & 1.5                  & $72.21               \pm 1.89                  $  & $317.58              \pm 3.70                    $ & $48.32               \pm 1.01 $         & 0.65            \\
4671744  & 0.05                 & $0.014                \pm 0.002                   $  & $0.018                \pm 0.007                    $ & $0.034                 \pm 0.001  $         & 0.25            \\
4985600  & 0.51                 & $57.52               \pm 0.91                  $  & $273.98              \pm 15.61                    $ & $57.18                \pm 0.58 $           & 1.78          \\
22079488 & $<0.11$      & $                      < 0.43                  $            & $0.50                \pm 0.08                    $ & $0.35                 \pm 0.01 $           & 0.30          \\
18526208 & 0.03                 & $0.0092                \pm 0.0021                  $  & $0.0049              \pm 0.0039                    $ & $0.0175                 \pm 0.0012 $         & $-0.10$            \\
25408512 & 0.05                 & $0.65                \pm 0.34                  $  & $3.14              \pm 0.58                    $ & $1.39                 \pm 0.21 $               &  0.57     \\
20316160 & 1.99                 & $51.26               \pm 0.40                  $& $125.66               \pm 4.29                    $ & $27.51                \pm 0.21  $            & 0.91         \\
22087680 & 0.05                 & $0.026                \pm 0.006                   $  & $0.084              \pm 0.011                    $ & $0.084                 \pm 0.002 $          &  0.20          
\end{tabular}
%\end{subtable}
\end{table*}

%\clearpage
%}

\subsection{Flux Calibration} \label{sec:fluxcal}
We test our order stitching algorithm and flux calibration of the CASSIS spectra via a flux comparison to the \textit{Wide-field Infrared Survey Explorer} (\WISE) fluxes. \WISE\ imaged the sky at four wavelengths: 3.4 (\W1), 4.6  (\W2), 12  (\W3), and 22~\micronm\ (\W4) with angular resolutions $6.1$, $6.4$, $6.5$ and $12$  arcsec, respectively \citep{wright}. The IRS SL and LL slits provide complete spectral coverage of the \W3 and \W4 bands respectively. We cross-match our \Spitzer\  sample with the \WISE\  All-Sky catalogue using the NASA/IPAC Infrared Science Archive (IRSA). We employ a cone search with a tolerance of $6$ arcsec to maximize sample overlap while minimizing false matches. We verify that our objects are correctly cross-matched by comparing the coordinates of the associated Two Micron All-Sky Survey \citep[2MASS,][]{2MASS} observations where possible (also given in IRSA) and the IRS spectrum coordinates. The 2MASS photometric bands have aperture sizes smaller than that of the \WISE\ bands, corresponding to smaller uncertainties in the position of the object. We find complete coverage of \WISE\ 22~\micronm\ photometry for our sample, and 82 per cent of our sample with all \W1, \W2, \W3 and \W4 measurements with $\mathrm{S/N > 5}$. 

\begin{figure}
\includegraphics[width=\columnwidth]{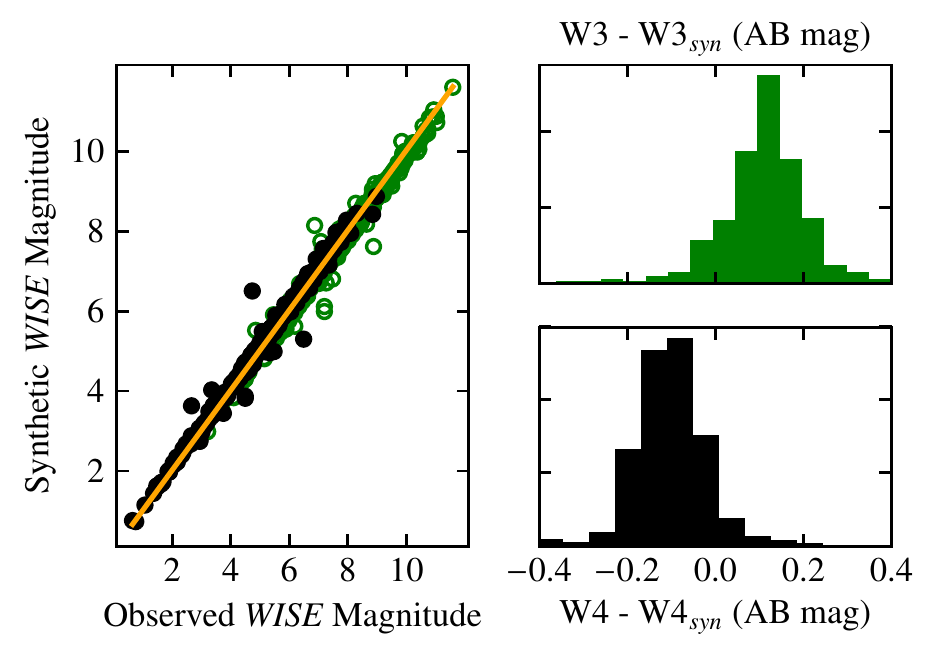}
	\caption{Spectrophotometry Test: In the leftmost plot, the filled black circles and the empty green circles are the 22 and 12~\micronm\ \WISE\ bandpasses respectively. We calculate the synthetic \WISE\ photometry by convolving the observed IRS spectrum with the relevant \WISE\ bandpass transmission curve. We plot an orange line with a slope of one for visual reference. In the rightmost plots, we show the distribution of the difference of the observed to synthetic photometry for each band. The median  differences between the \WISE\ synthetic and observed 12 and 22~\micronm\ bands are $0.11$ and $-0.10$ mag respectively.
    }
\label{fig:specphot}    
\end{figure}

We calculate the synthetic \W3 and \W4 magnitudes from our IRS spectra to test the flux calibration of the reduced IRS spectra and to test our spectral order scaling factors. We expect the offset between the synthetic and observed magnitudes to be within random error of the magnitude measurements if the spectra are correctly calibrated and stitched. We calculate the synthetic flux using
\begin{equation}
f_\mathrm{\nu, synth}= \frac{\int f_{\nu}(\nu) S(\nu)\, \mathrm d\nu}{\int S(\nu)\, \mathrm d\nu},
\end{equation}
where $f_\mathrm{\nu, synth}$ is the measured flux density averaged over the filter profile, $f_{\nu}(\nu)$ is the calibrated flux density, and  $S(\nu)$ is the filter's sensitivity response. We convert synthetic fluxes to Vega magnitudes using the zero points given in \citet{jarrett}. The median differences between the \WISE\ synthetic and observed 12 and 22~\micronm\ bands are $0.11$ and $-0.10$ mag respectively. We find  these offsets do not significantly affect our analyses, and in the following paragraph we describe this as well as the fraction of objects in our sample that are most susceptible to the aperture differences between the \WISE\ and IRS passbands. We show the offset between the observed and synthetic magnitude for the \W3 and \W4 bands in \autoref{fig:specphot}. %We then apply a multiplicative scale factor to bring the IRS measurements into agreement with the \WISE\  photometry. 

We use the ratio of observed to synthetic \WISE\ photometry to test for potential aperture biases. If an object is extended outside the IRS slit area, then the gas and dust measurements would be artificially smaller for that object. The angular resolution of the 22~\micronm\ \WISE\ photometric data is 12 arcsec, implying that the ratio of observed to synthetic will increase if the object is extended in the SL module which has a width of 4.5 arcsec. Less than 10 per cent of our sample has 22~\micronm\ (\W4) observed to synthetic ratios greater than 1.0, and our gas and dust relationships do not significantly change as a function of the ratio. We use the \W4 bandpass to calculate the synthetic magnitude at 24~\micronm\ via linear interpolation as follows:
\begin{equation}
f_{\nu}(24\ \micron) = f_{\nu}(22\ \micron)\left(\frac{24\ \micron}{22\ \micron}\right)^{\alpha}
\end{equation}
where $f_{\nu}(22\ \micron)$ is the \W4 band rest-frame synthetic flux and $\alpha$ is the spectral index calculated from the IRS spectroscopy between 15 and 30 microns.   

We use the 24~\micronm\ photometry estimate to derive the 24~\micronm\ luminosities used throughout our analysis. We provide these luminosities in \autoref{tab:egsampleprops}.

\subsection{Redshifts} \label{sec:sampprop}

\begin{figure}
\centering
\includegraphics[width=\columnwidth]{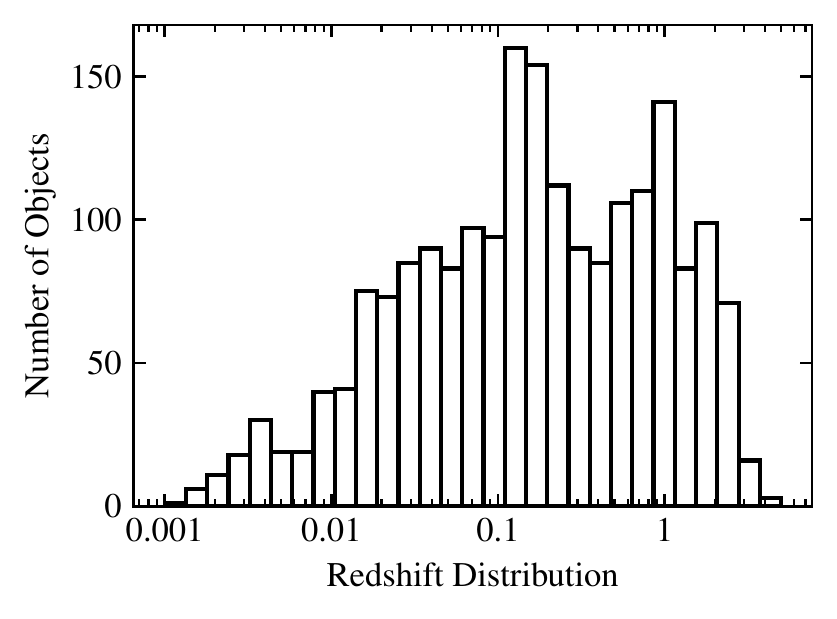}
	\caption{Redshift distribution from the cross-matched IDEOS objects. The median and mean redshifts for our sample are 0.15 and 0.4 respectively.}
\label{fig:zdist}
\end{figure}

In \autoref{tab:egsampleprops} we provide the AORkeys, coordinates, redshift, and other general sample properties. The Infrared Database of Extragalactic Observables (IDEOS) has a redshift catalogue for all the spectra in CASSIS \citep{ideos}. The IDEOS redshift catalogue was compiled by comparing with the NASA/IPAC Extra-galactic Database redshifts and optical counterparts, providing IRS redshifts with accuracy $\sigma_z\sim0.0011$. Over 85 per cent of our initial sample of 2,807 objects have reliable redshift measurements, and we show the distribution of redshifts in \autoref{fig:zdist}. The remaining objects have poor redshift determinations, so we exclude them from our sample. The median and mean redshifts for the objects in our sample with secure redshifts are 0.15 and 0.4 respectively.

\subsection{\textit{K} Luminosities}
\begin{figure}
\centering
\includegraphics[width=\columnwidth]{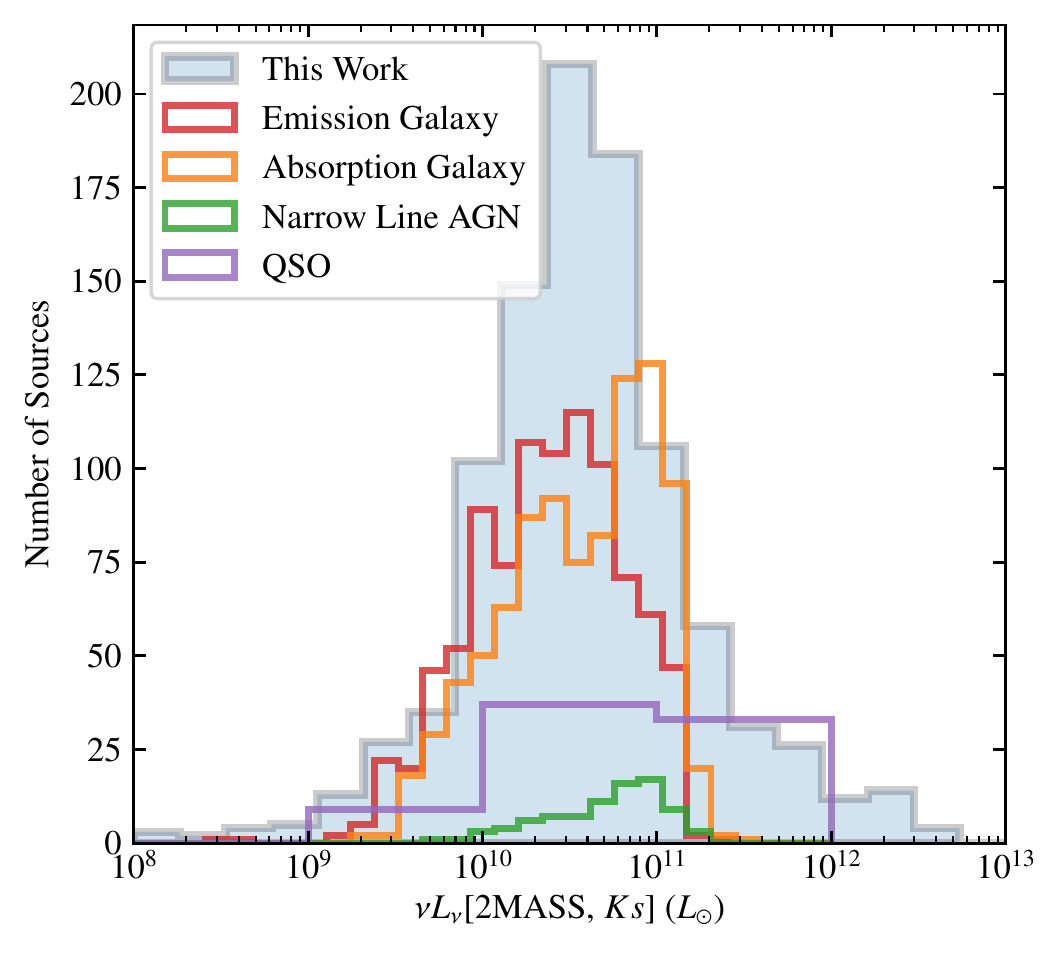}
	\caption{$K_s$-band luminosity distribution for the objects in our sample with $z < 0.05$ derived using 2MASS $K_s$-band. We show the luminosity distributions of selected galaxy sub-samples with \citet{maddox} for reference.}
\label{fig:kdist}
\end{figure}

In addition to the cross-matched \WISE\ photometry of our spectra, we use $J$, $H$ and $K_s$ bands photometry from the Two Micron All-Sky Survey (2MASS) survey (see \citealt{jarrett} for details on the \WISE--2MASS cross-matching collection).  For objects with $z< 0.5$ we calculate the absolute magnitudes in the $K$ rest-frame by employing $K$-corrections from \citet{kcor}. We provide the K-band  luminosities to estimate how our sample compares with larger, more complete samples of galaxies. 
We compare our distribution of K-band luminosities to that of complete samples of nearby narrow line AGN, QSO, emission line, and absorption line galaxies from \citet{maddox} \autoref{fig:kdist}.    

\citet{maddox} identify Type 2 Seyfert galaxies by the presence of narrow high-ionization emission lines, quasars by the presence of one emission line of full width at half maximum of at least $1500\ \mathrm{km\,s^{-1}}$ and $M_{i}< 22.4$ mag, star-forming galaxies by having at least one narrow emission line, and absorption line galaxies by having no emission lines and visible stellar absorption features. We calculate the absolute magnitudes from the published apparent magnitudes in \citet{maddox}, and compare their distributions with ours in \autoref{fig:kdist}. We calculate $K$-corrections using the methods of \citet{kcor}. \citet{maddox} exclude sources with $K < 11.5$ to prevent false UKIDSS detections and $K > 17$ because at $K \ge 17$ UKIDSS photometric errors increase significantly. We perform KS two sample test between the $K_s$-band distribution of our entire sample and those of galaxies in \citet{maddox}: Emission Line Galaxy ($D_\mathrm{KS} = 0.19$, $p_\mathrm{KS} \ll .001$), Absorption Line Galaxy ($D_\mathrm{KS} = 0.12$, $p_\mathrm{KS}\ll .001$), Narrow Line AGN ($D_\mathrm{KS} = 0.24$, $p_\mathrm{KS} = 0.0003$), QSO ($D_\mathrm{KS} = 0.5$, $p_\mathrm{KS} \ll 0.001$). In \autoref{sec:3_1}, we test if the $K_s$-band distribution of our AGN dominated objects differs from the $K_s$-band distribution of our SF dominated objects.

\subsection{Emission Line Measurements} \label{sec:emission}

We measure the emission lines listed in \autoref{tab:median_errors}.  We denote the \molh\ emission lines as \molh S($J$) for a transition from rotational level $J+2$ to $J$. All of the \molh\ features are unresolved, so the linewidths are set by the IRS spectral resolution and are listed in \citet{smith07}. 
The line resolution changes after we apply a rest-frame correction. To account for this, we determine a fitting window by
choosing only the points that are three Gaussian widths away relative to the linewidth of the feature.  We allow the line centre of the feature in the rest-frame to vary 0.03~\micronm\ to take into account wavelength calibration uncertainty \citep{smith07}. We perform a linear least squares regression to find the best-fit parameters for our model, parametrized as
\begin{equation}
f_{\nu}(\lambda) = B + C(\lambda - \lambda _{c}) + De^{-(\lambda - \lambda _{c})^{2}/2\sigma^{2}},
\end{equation} 
where $B$, $C$, $D$ are the fitted constants, $\lambda$ the wavelength array, $\lambda _{c}$ the line centre, and $\sigma$ the line resolution according to its wavelength location on the IRS spectrograph. We list the number of detections of each fitted line and their median signal-to-noise ratio in \autoref{tab:median_errors}, and a subset of the values themselves in \autoref{tab:egmolh}.  We compare our molecular hydrogen measurements with \citet{higdon} and \citet{Hill}, and find agreement within 0.2 dex. 

\begin{table}
   \caption{Number of $2\sigma$ or greater detections and the median signal-to-noise ratio of the detections. Although we do not use the fine-structure lines in this paper, we provide our fluxes for ease of comparison to other published samples and analyses.}
	\begin{centering}
  	\begin{tabular}{ | l | r |  r |}
    \hline
    Line & Detection & Median SNR \\ \hline
    [Ar\textsc{ ii}]6.985 \micronm\ & 668 & 4.5 \\ \hline
    [Ar\textsc{ iii}]8.991 \micronm\ & 220 & 3.4 \\ \hline
    [S \textsc{iv}]10.511 \micronm\ & 585 & 4.4 \\ \hline
    [Ne\textsc{ ii}]12.81 \micronm\ & 1135 & 8.9 \\ \hline
    [Ne\textsc{ iii}]15.56 \micronm\ & 889 & 6.2 \\ \hline
    [S \textsc{iii}]18.71 \micronm\ & 609 & 5.6 \\ \hline
    [O \textsc{iv}]25.910 \micronm\ & 520 & 7.3 \\ \hline
	[Fe \textsc{ii}]25.989 \micronm\ & 494 & 6.7 \\ \hline
	[S  \textsc{iii}]33.48 \micronm\ & 395 & 5.7 \\ \hline
	H$_{2}$S(0)28.212 \micronm\ & 73 & 2.7 \\ \hline
    H$_{2}$S(1)17.03 \micronm\ & 585 & 7.0 \\ \hline
    H$_{2}$S(2)12.279 \micronm\ & 159 & 4.0 \\ \hline
    H$_{2}$S(3)9.665 \micronm\ & 512 & 5.8 \\ \hline
    H$_{2}$S(5)6.909 \micronm\ & 244 & 2.7 \\ \hline
    H$_{2}$S(6)6.109 \micronm\ & 70 & 7.5 \\ \hline
	H$_{2}$S(7)5.511 \micronm\ & 82 & 4.8 \\ \hline
    \end{tabular}
    \label{tab:median_errors}
    \end{centering} 
\end{table}

\subsection{Continuum and Dust Features} \label{sec:drude}

PAH molecules consist of planar lattices of aromatic rings containing tens to hundreds of carbon atoms. The absorption of UV photons excites their vibrational modes, which can contribute dramatically to the MIR emission. In stochastic dust grain heating models, the relative strengths of the PAH bands are dependent on the distribution of grain sizes and ionization states \citep{li2001,draine07}.
The \eqwpah\ feature probes the contribution of the AGN to the MIR spectrum. The PAH 6.2~\micronm\ feature appears to originate from SF hot dust \citep{peeters}, and the 6~\micronm\ continuum is in a wavelength regime where the reprocessed light from the hot torus dominates. Therefore, the \eqwpah\ should be some possibly non-linear function of the ratio of SF-sourced energy to AGN torus-sourced energy \citep{spoon}.  PAHs generate the broad emission features at 6.2, 7.7, and 11.3~\micronm\ \citep{allamandola}, and these features contribute up to 30 per cent of the total MIR flux in galaxies whose star-formation processes dominate \citep{smith07}. 

% only here to stop pdfendlink error.
%\newpage

We model the PAH features using individual and blended Drude profiles \citep{smith07,Hill}
\begin{equation}
f_{\nu}^{(r)} = \frac{b_{r} \gamma_{r}^{2}}{(\lambda / \lambda_{r} - \lambda_{r} / \lambda)^{2} + \gamma_{r}^{2}},
\end{equation}
where $b_{r}$ is the fractional intensity, $\gamma_{r}$ is the fractional FWHM, and $\lambda_{r}$ the central wavelength. The integrated intensity of the Drude profile is
\begin{equation}
f^{(r)} = \int f_{\nu}^{(r)}\,\mathrm d\nu = \frac{\pi c b_{r} \gamma_{r}}{2 \lambda_{r}}.
\end{equation}
The rest-frame equivalent width of the Drude profile is
\begin{equation}
\mathrm{EQW} = \frac{\pi}{2} \frac{b_{r}}{f_{\nu}^\mathrm{cont}} \gamma_{r},
\end{equation}
where $f_{\nu}^\mathrm{cont}$ is the continuum flux density. We use the tabulated values for $\gamma_{r}$ as presented in \citet{smith07}.  For the most AGN-dominated spectra ({\eqwpah\ $< 0.01$~\micronm}), we find a non-negligible contribution from the [\ion{Ne}{vi}] line which is blended with the 7.7~\micronm\ feature. We fit an additional Gaussian to account for this potential line. For the $6.2$, $7.7$, and $11.3$~\micronm\ we have $2\sigma$ detections for 51, 58, and 56 per cent respectively for our sample. In \autoref{tab:egmolh} and \autoref{tab:egpah}, we show example \molh\ and PAH fluxes for 10 objects. We used the results of \citet{reyes} and \citet{zakamska08} extensively in training and refining our fitting procedures for both the emission line measurements and dust features.

PAHs trace the contribution of young B stars in PDRs \citep{peeters}. The PAH 11.3~\micronm\ feature's continuum is easier to constrain than that of the 7.7~\micronm\ feature. As shown in \citet{peeters17}, and tested on a large sample of extragalactic IRS low-resolution observations in \citet{stock17}, the full decomposition of the $\textrm{7--9}$~\micronm\ PAH emission includes two components that are more similar to a dust continuum rather than to the 7.7~\micronm\ complex emission  described in \citet{li2001}. The emission of this dust continuum, referred to as a plateau, occurs in spatially distinct regions from the PAH emission, and overall behaves independently. Although there is also a 10--15~\micronm\ plateau, the emission in this region is less pronounced so that the 11.3~\micronm\ feature is only marginally affected.  The 6.2~\micronm\ feature is in the wavelength regime where the AGN processes contribute to the continuum amplitude. Thus, we use the 11.3~\micronm\ feature to trace star-formation in our objects.

Other PAH measurement techniques widely used in the literature include: (1) direct integration of the feature super-imposed on a polynomial pseudo-continuum excluding other potentially contaminating lines or features (used in \citealt{brandl}), and  (2) simultaneous estimation of the contributions of PAHs, ions, molecules and old stellar populations to the observed spectra, e.g. \textsc{pahfit} (\citealt{pahfit}, used by \citealt{smith07,odowd,shipley}) and \textsc{cafe} (\citealt{cafe},  used by \citealt{stierwalt}).  We calculate the systematic offset between methods (1), (2), and our Drude measurements for our high signal-to-noise  stacked spectra presented in \autoref{sec:stacks}, and summarize the results in \autoref{tab:pah_offset}.

\begin{table}
   \caption{\eqwpah\ mean  per cent difference between the direct method/\textsc{pahfit} and the Drude profile method to estimate the fluxes and EQW of PAH emission features: For our stacked sample, the direct method yields slightly smaller equivalent widths than \textsc{pahfit}.}
   \centering
  	\begin{tabular}{ | l | l |  l |}
    \hline
    Method & EQW < 0.27\ \micronm & EQW > 0.27\ \micronm \\
     & (AGN Dominated) & (SF Dominated)
    \\ \hline
    $\mathrm{Drude - Direct}$ & 14\% & 52\% \\ \hline
    $\mathrm{Drude - \textsc{pahfit}}$ & $-66\%$ & 20\% \\ \hline
    \end{tabular}
    \label{tab:pah_offset}
\end{table}

\subsection{\Spitzer\ Stacks} \label{sec:stacks}

\begin{figure*}
\begin{center}
\includegraphics[width=\textwidth]{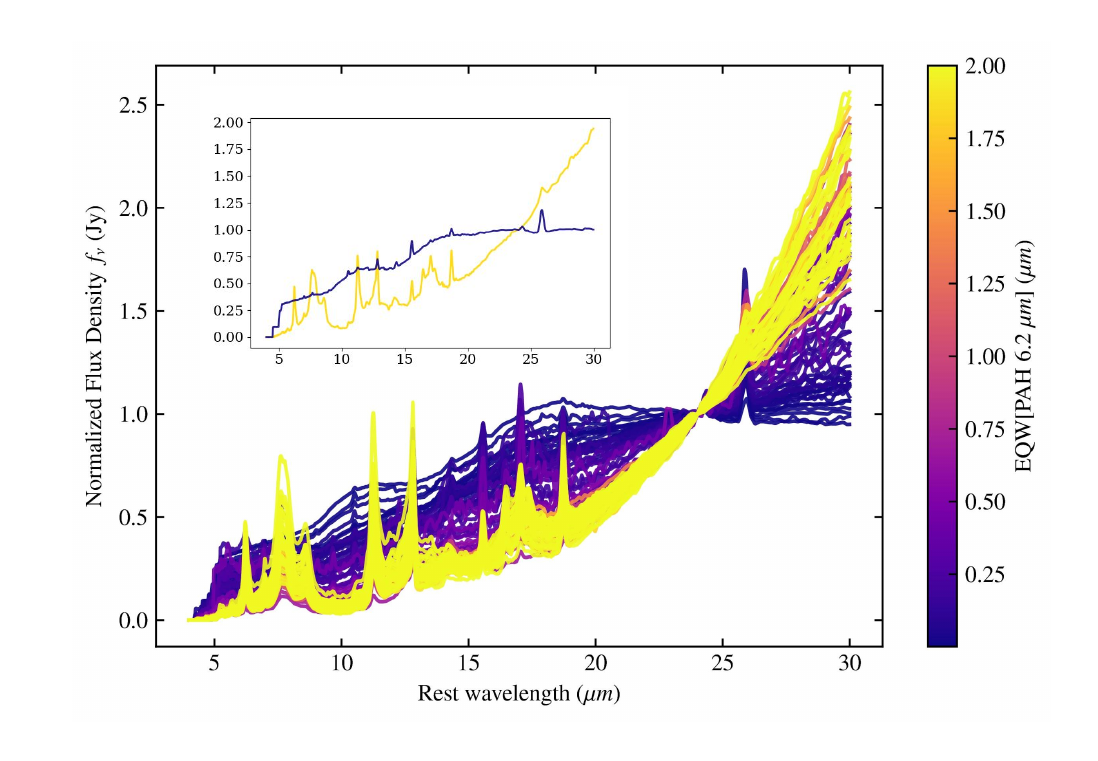}
	\caption{Stacked \Spitzer\  spectra binned by \eqwpah: we split our sample into 100 evenly populated bins of \eqwpah. We normalize each pre-stacked individual spectrum by its IRS $f_{\nu}$[24~\micronm]. We use the blue-to-yellow gradient colormap throughout this work to indicate the \eqwpah, with blue corresponding to AGN-dominated and yellow SF-dominated. The inset shows only two spectra from the stacks, a low EQW (blue) and high EQW (yellow) stack, and is meant to provide an easy comparison between the stacks. We provide the entirety of the stacked spectra in ASCII format in the on-line version of this publication.}
\label{fig:stacks_100}
\end{center}
\end{figure*}

We stack a subset of our 2,015 \Spitzer\ spectra in 100 equally populated bins of \eqwpah. We only include objects with  $z \leq 0.3$ to ensure the relevant features are not redshifted out of our wavelength range. After applying our $z$ constraints, each bin contains 12 objects. After binning our sample by \eqwpah, we determine a weight for each individual spectrum given by its average signal-to-noise ratio in the region around the \eqwpah\ feature. We assume the weight must be greater than or equal to 0.2, then normalize each spectrum by its rest-frame $L_{\nu}$[24~\micronm] and perform a weighted average.
%and add each spectrum multiplied by its respective weight.
We check that the width of the unresolved lines (the emission lines listed in \autoref{tab:median_errors}) are equal to the \Spitzer\ IRS minimum widths allowed by the instrument's spectral resolution,  and find that the widths vary negligibly from bin to bin. This is a check on the accuracy of our redshifts. 
The median absolute deviation of the spectra in each wavelength bin is less than 10 per cent of each bin's flux.
%We calculate the median absolute deviation for each bin, and find the dispersion of the spectra per wavelength bin is within 10 per cent. 
We display these spectra, colour-coded by \eqwpah, in \autoref{fig:stacks_100}.

We use the stacked spectra to identify and quantify differences between three methods to estimate the PAH emission. We use full spectral decomposition via \textsc{pahfit}, direct integration, and Drude model fitting. For the direct integration method we measure the associated continuum of the 6.2, 7.7, and 11.3~\micronm\ features by performing a linear interpolation while excluding ice features and other emission lines that fall in the immediate vicinity of the PAH \citep{spoon}. For the 6.2~\micronm\ feature we interpolate between 6.0 and 6.5 \micron, for the 7.7~\micronm\ feature we interpolate between 7.3 and 8.3~\micronm, and for the 11.3~\micronm\ feature we interpolate between 11.0 and 11.8 \micron. For \textsc{pahfit}, we input rest-frame calibrated (SL1--LL2 scale corrected, bonus order combined) spectra. We describe the Drude method in \autoref{sec:drude}.

We show the median and mean differences between the two methods and the Drude method in \autoref{tab:pah_offset}. We split the stacks into \eqwpah\ $< 0.27$~\micronm\ and \eqwpah\ $> 0.27$~\micronm.  Since the ISO mission in the 1990s, the equivalent widths of PAHs have been used to separate AGN from SF dominated objects \citep{genzel}. More recently, \citep{diamond,petric,zakamska2016} verified the efficiency of this technique by comparing multiple MIR diagnostics including PAH EQW, MIR colours, and relative high to low ionization emission line fluxes. Here we continue with this approach, but we test the robustness of our measurement as a function of measurement method: assuming a Drude model \citep{draine03,smith07}, direct integration \citep{brandl}, and simultaneous estimation using \textsc{pahfit} \citep{odowd}.

In \autoref{fig:stacks_100}, the difference in the relative continuum emission in the 6.2 \micronm\ region is clear. In \autoref{sec:3_1}, we show this region is a good differentiator between AGN versus SF dominated spectra via comparison to other MIR diagnostics. Thus it is important to test the consistency of the different PAH fitting algorithms on high signal to noise spectra with varying amounts of PAH emission. We test whether the algorithms agree for different AGN contributions to the MIR. We subtract the direct and \textsc{pahfit} measured EQW values from the Drude profile values and find the median and mean of the differences. The treatment of the continuum around the PAH emission feature accounts for most of the differences between PAH EQW estimates obtained from the three different methods. Direct methods tend to underestimate the continuum for the most SF-dominated spectra, unless one fits separately in the 7.7 and 11.3~\micronm\ regions the 5--10 and 10--15 \micron\ plateaus \citep{peeters2017}. We choose the Drude method because it is less sensitive to potential poor quality pixel values (unlike the direct method) and estimates the continuum more consistently than \textsc{pahfit}.   

%\textsc{pahfit} is a tool that is specifically for low resolution IRS spectroscopy. Since \textsc{pahfit} fits the entire spectrum simultaneously, one must ensure they have accounted for dominant features that can affect continuum levels, i.e.~silicate and ice absorption features must be fit. {\bf{Not sure what you mean in this last sentence, as is it implies that to use PAHFIT you have to remove some features. Maybe just say PAHFIT does not include silicate and ice absorption.. I agree with the ice part, but are you sure about silicates?}} Poor spectral quality {\bf{do you mean SNR or something else?, if yes just say low snr tranlates to poor fits, or not even say that.. or just say PAHFIT is not robust for low SNR spectra... or something like that}} can contribute to poor fits to the local continuum.  {\bf{I'd keep only this last sentence}}

\section{Results} \label{sec:results}

\subsection{The AGN contribution to the MIR emission}
\label{sec:3_1}

A significant fraction of MIR emission in AGN host galaxies comes from dust heated by $\lambda<10\ \micron$ photons \citep[e.g.][]{nenkova}. We adopt the empirical thresholds of AGN contribution to the MIR presented in \citet{laurent}, \citet{peeters}, \citet{brandl}, and \citet{armus07}. If the \eqwpah\ is less than 0.27~\micronm, the AGN contributes more than 50 per cent of the MIR emission and we refer to those sources as AGN-dominated. If the \eqwpah\ is larger than 0.27~\micronm\ but less than 0.54~\micronm, we classify the spectrum as a composite object with signatures of both AGN and SF. If the \eqwpah\ is greater than $0.54\ \micron$, then we classify the object as SF dominated. Subsection \ref{sec:stacks} visually demonstrates that the PAH 6.2~\micronm\ feature effectively differentiates between AGN and SF dominated MIR spectra: when we select AGN dominated targets on the basis of their \eqwpah\ we also find them to be AGN dominated on the basis of their continuum slopes between 15 to 30 \micronm. Using the \eqwpah\ selection method we find: 40 per cent AGN dominated, 12 per cent composite, 48 per cent SF dominated.

\begin{figure}
\begin{center}
\includegraphics[width=\columnwidth]{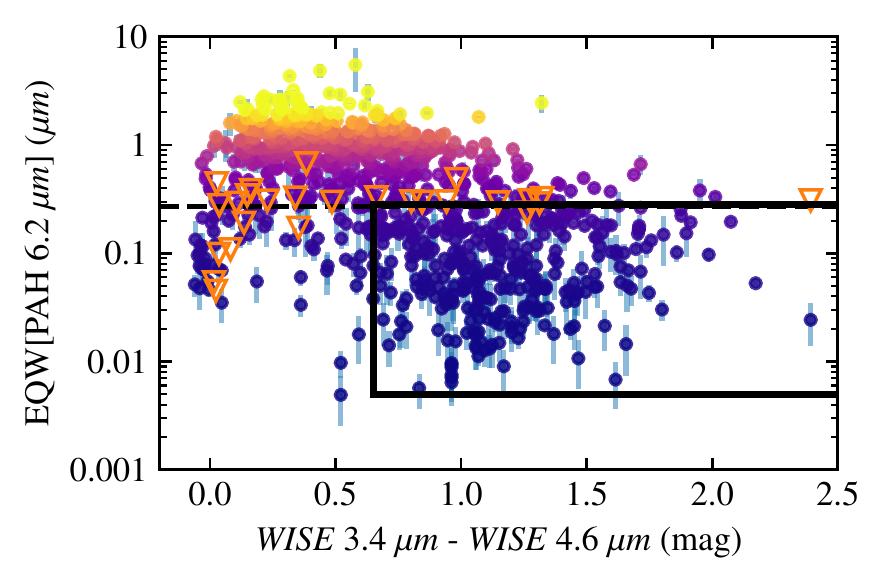}
	\caption{AGN selection comparison: Using a \WISE\ colour cut that is dependent on the \W2 magnitude as outlined in \citet{assef}, we capture 80 per cent of our objects that satisfy the AGN criterion \eqwpah$< 0.27$ \micronm. The solid black box encapsulates roughly all of the objects that satisfy both the EQW threshold and \WISE\ colour cut. The dashed black line marks the EQW threshold of AGN MIR dominance at 0.27~\micronm. The orange triangles are \eqwpah\ upper limits for objects with $2\sigma$ detections of \molh S(3), PAH 7.7~\micronm, and PAH 11.3~\micronm. The colours of the points are the same as in previous figures, with blue  denoting AGN-dominated objects and yellow denoting SF-dominated objects, defined by having small and large values of \eqwpah\ respectively.}
\label{fig:wisesel} 
\end{center}
\end{figure}

Photometric observations in the MIR have been used to find AGN with \Spitzer\ \citep{lacy04,stern05,martinez05,lacy07,stern05,donley12,eisenhardt,lacy15} and \WISE\ \citep{stern,assef}. As with most selection methods, there is a trade-off between completeness and reliability \citep[e.g.][]{petric,assef}. We use \citet{assef2017}'s \WISE\ AGN selection criterion, which is 90 per cent reliable and 17 per cent complete. We compare this criterion to the \eqwpah\ selection in \autoref{fig:wisesel}. Of the 2,105 objects in our overall Spitzer sample, 52 per satisfy the colour criteria by \citet{assef2017}. Of the \citet{assef2017} selected objects, 65 per cent are classified as AGN using \eqwpah. Conversely, of the 2,105 objects in the overall sample that satisfy the \eqwpah\ criterion, 80 per cent are selected. The \eqwpah\ criterion is calibrated to rule out SF--AGN composites. Selecting with a less stringent thrshold, \eqwpah\ $<$ 0.54 \micronm\ (i.e. AGN-dominated and SF--AGN composites), we get 52.4 per cent of our total sample classified as AGN, and are in good agreement with the \citet{assef2017} selection in the MIR colour selected sub-sample. Although this fraction is not impressively high, it is in qualitative agreement with other studies that demonstrate that spectroscopically-selected AGN are recovered by color selection methods at roughly the same rate \citep{yuan}.

The completeness of a selection method can depend on the AGN type. Using the \WISE\ colour wedge as defined in \citet{mateos12} on a sample of Type 2 quasars, \citet{yuan} find that only 34 per cent of these fit the \citet{mateos12} AGN selection criterion, which is 90 per cent reliable and 17 per cent complete. In \autoref{fig:wisesel}, there is a grouping of 26 objects with small equivalent widths but with \WISE\ colours that suggest they are star-forming (\eqwpah\ $< 0.27$  \micronm\ and $W1-W2 < 0.1$). We perform a literature search with the coordinates of these 26 objects, and find that 10 are FRI radio galaxies from the 3C sample \citep{ogle}. \citet{gurkan} found that \WISE\ colour wedges tend to miss these low-luminosity radio galaxies. Furthermore, \citet{blecha} find that \WISE\ colour-cuts that are too stringent (i.e. $W1 - W2 > 0.8$) can miss AGN in late stage mergers. As seen in \autoref{fig:wisesel}, 10 per cent of low \eqwpah\ objects would be missed with the above colour-cut. Due to its consistency with different AGN host-galaxy classes, this justifies our use of \eqwpah\ as AGN-dominated spectra selection criterion.   

\begin{figure}
    \includegraphics[width=\columnwidth]{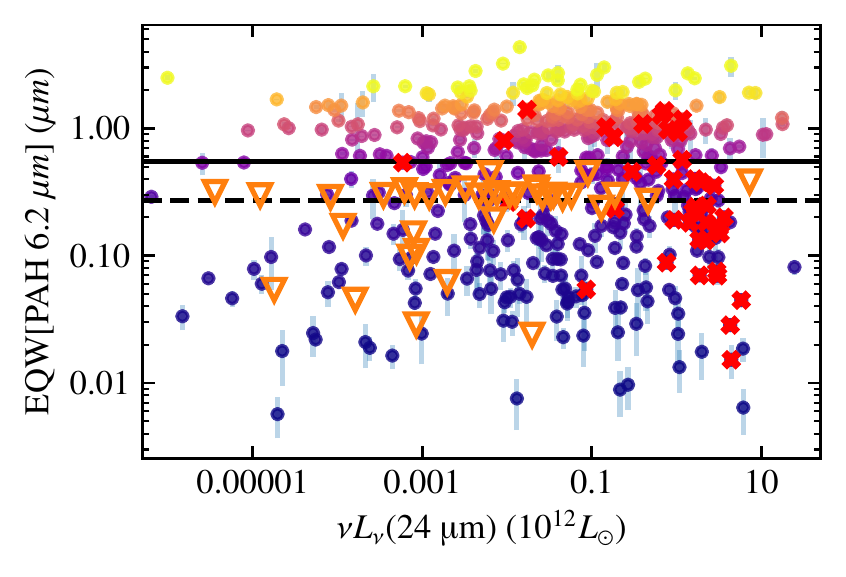}
\caption{24~\micronm\ luminosity selection: The solid black horizontal line is
    the mean EQW of normal star-forming galaxies as described in \citet{brandl}. 
    The dashed black horizontal line is the empirical AGN dominance EQW
    classifier. We find that although the majority of objects with large 24~\micronm\ luminosities have small 6.2~\micronm\ equivalent widths, most do
    not follow this trend; 80 per cent of our targets have low \eqwpah\ and 24~\micronm\ luminosities $< 10^{11}\,\mathrm L_{\sun}$. Bold red crosses show the 70
    objects that follow the trend found by \citet{desai} for ULIRGs.
    The orange triangles are \eqwpah\ upper limits for objects with $<2\sigma$ detections of \molh S(3), PAH 7.7~\micronm\ and PAH 11.3~\micronm.
The colours of the points are the same as in previous figures, with blue  denoting AGN-dominated objects and yellow denoting SF-dominated objects, defined by having small and large values of \eqwpah\ respectively.}
    \label{fig:desai}
\end{figure}

\begin{figure*}
\includegraphics[]{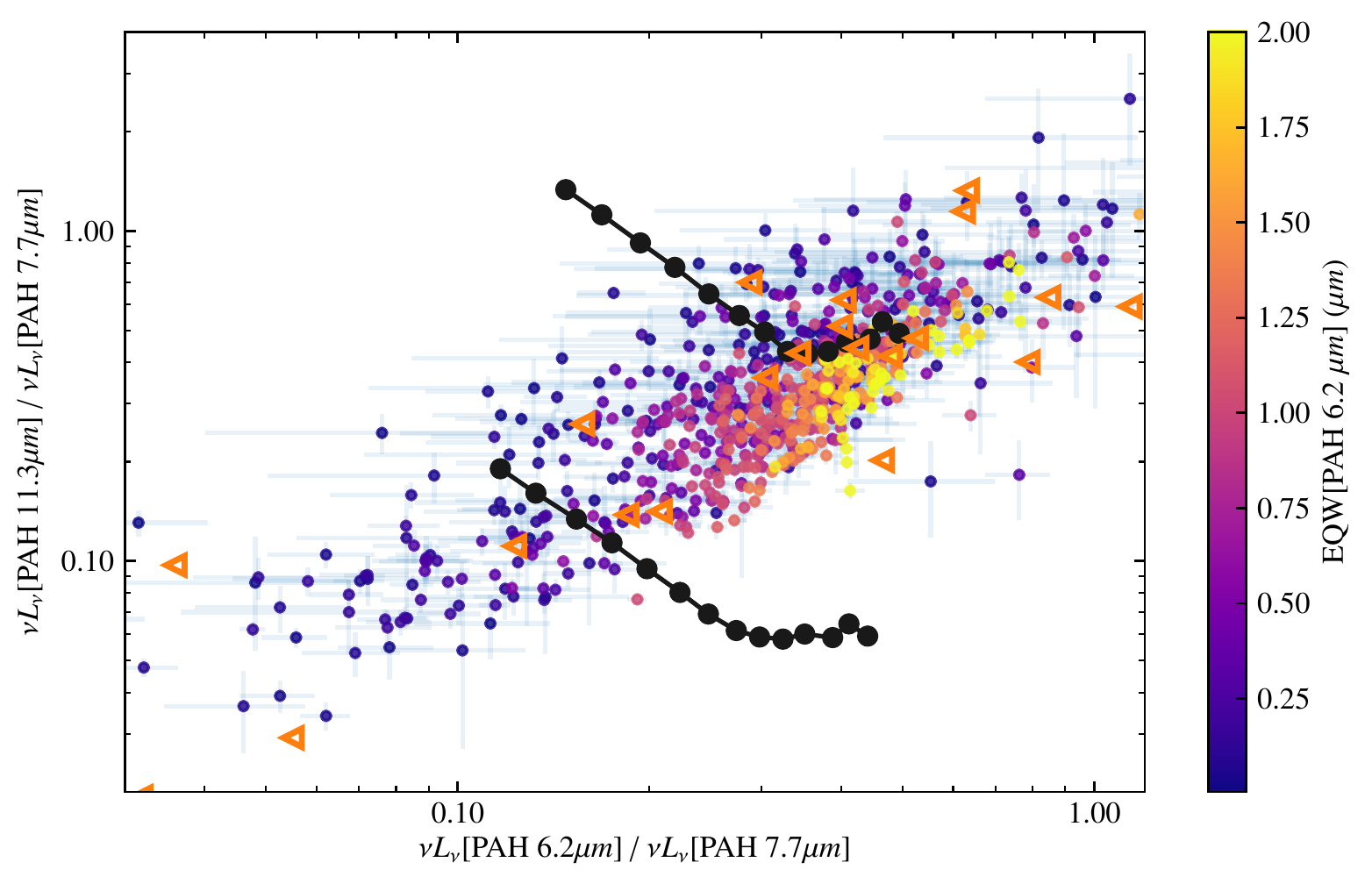}
	\caption{
PAH band ratios: The black lines correspond to the expected ratios for fully neutral (top) or fully ionized (bottom) PAH molecules of a given number of carbon atoms using \citet{draine07} models. The orange triangles are \eqwpah\ upper limits for objects with $<2\sigma$ detections of \molh S(3), PAH 7.7~\micronm\ and PAH 11.3~\micronm.
The colours of the points are the same as in previous figures, with blue  denoting AGN-dominated objects and yellow denoting SF-dominated objects, defined by having small and large values of \eqwpah\ respectively.}
\label{fig:relpah}    
\end{figure*}

Monochromatic continuum luminosity at 24~\micronm\ is commonly used to trace star-formation due to the warm dust associated with high-mass star-forming regions \citep{calzetti}. \citet{desai} and others find a linear trend between \eqwpah\ and 24~\micronm\ luminosity for the most luminous $z < 1.0$ ULIRGs, suggesting that at these redshifts, only galaxies with AGN contain large amounts of warm dust.  % They interpret their results to mean that large amounts of warm dust at these redshifts require the presence of an AGN. .. {\bf{are you sure? just say they conclude instead of "they interpret their results to mean..", also this is not super important for us because we don't talk about evolution with z.. so maybe let's take this out}}
In our sample, we find that although the majority of objects with large 24~\micronm\ luminosities have small \eqwpah, objects with small \eqwpah\ have diverse 24~\micronm\ luminosities. The 24~\micronm\ luminosities for these objects are indistinguishable from objects with larger values of \eqwpah. 
% In \autoref{fig:desai}, we have over-plotted with bold red crosses the objects from the trend described in \citet{desai}.
\autoref{fig:desai} shows that in our sample, we cannot identify the contribution of the AGN to the total MIR emission using only the 24~\micronm\ luminosities. As noted in \citet{desai}, despite their anti-correlation between \eqwpah\ and 24~\micronm\ luminosity for local ULIRGs, sub-millimetre galaxies can have high \eqwpah\ and high 24~\micronm\ luminosities. Furthermore, \citet{petric} find no correlation amongst LIRGs ($L_\mathrm{IR}<10^{11}\,\mathrm L_\odot$). As noted by \citet{desai, petric}, the tight correlation observed in ULIRGs between 24~\micronm\ luminosities and \eqwpah\ can be explained by the compact IR emission in ULIRGs and the relative high fraction of AGN dominated ULIRGs (40--60 per cent). ULIRGs also tend to be in the final stages of merging, while LIRGs span all stages of gravitational interactions. While a census of the merging stages in our sample is beyond the scope of this paper, we speculate that the galaxies in our sample have a wide range of morphologies and merger stages. Thus, it is not surprising that we do not find a relationship in our sample of mixed infrared luminosities and galaxy sub-classes. 

%Because the surroundings of AGN have warmer dust than those of star-forming galaxies, the flux-ratio of the MIR continuum at 30 and 15 \micronm\ is an estimate of the relative contributions of star-formation and AGN to the MIR emission \citep{brandl,veilleux}. We compare the \eqwpah\ with the continuum flux ratio. 80 per cent of our objects with low \eqwpah\ appear AGN dominated from their MIR continuum slope i.e $f_{nu}$(30 \micronm)/$f_{nu}$(15 \micronm) < 3.0. 15 per cent of our objects have \eqwpah\ $< 0.27$ but an MIR slope $> 3.0$. This may be due to objects with low metallicities \citep{ohalloran}. 

%\begin{figure}
%    \includegraphics[width=\columnwidth]{p62_f30f15_new.pdf}
%\caption{MIR Flux Ratio versus \eqwpah:  The solid black line is the empirical value of AGN dominance ($f_\nu [30\ \micron]$/ $f_\nu [15\ \micron] > 3.0$. The dashed line is the empirical \eqwpah\ AGN dominance threshold (\eqwpah\ $ < 0.27$).
%The orange triangles are \eqwpah\ upper limits for objects with $2\sigma$ \molh\ S(3), PAH 7.7 \micronm\ and PAH 11.3 \micronm.
%The colours of the points are the same as in previous figures, with blue  denoting AGN-dominated objects, yellow denoting SF-dominated objects, defined by having small and large values of \eqwpah\ respectively.}
%\label{fig:f30f15}
%\end{figure}

\begin{figure*}
\begin{center}
\includegraphics[width=\linewidth]{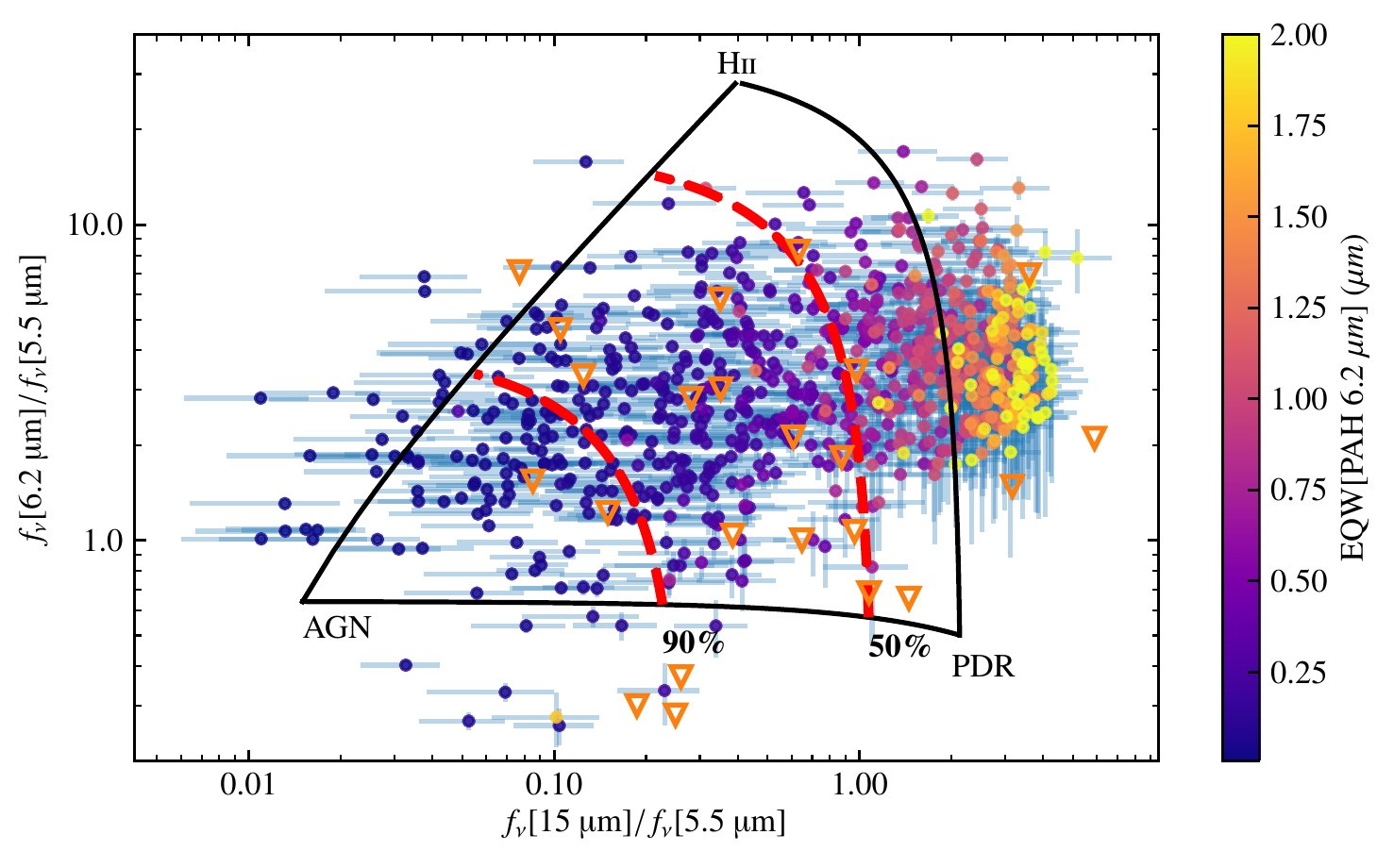}
	\caption{AGN selection comparison: 
The three vertices of the triangle are published values of an independently classified AGN (3C 273, \citealt{weedman05}), PDR (M17, \citealt{peeters})  and H \textsc{ii} region (NGC 7023, \citealt{peeters}). 
The red-dashed lines represent (left) 90 per cent and (right) 50 per cent AGN contribution to the MIR. The diagram compares the integrated continuum flux from 14--16~\micronm\ denoted $f_\nu(15\ \micron)$ to the integrated continuum flux from 5.3--5.8~\micronm\ denoted $f_\nu (5.5\ \micron)$. The $f_\nu(6.2\ \micron)$ values were derived as described in \autoref{sec:drude}. The orange triangles are \eqwpah\ upper limits for objects with $<2\sigma$ detections of \molh S(3), PAH 7.7~\micronm, and PAH 11.3~\micronm.
The colours of the points are the same as in previous figures, with blue  denoting AGN-dominated objects and yellow denoting SF-dominated objects, defined by having small and large values of \eqwpah\ respectively.}
\label{fig:laurent}   
\end{center}
\end{figure*}

\citet{laurent} combine both continuum emission and PAH EQW to estimate AGN contribution to the total IR. In \autoref{fig:laurent}, we use the revised version of the \citet{laurent} selection method presented in \citet{armus} which uses the the relative flux of the 6.2~\micronm\ PAH complex and 15~\micronm\ continuum versus the 5.5~\micronm\ continuum. Our method agrees with \citet{laurent}'s: 98 per cent of the objects that the \citet{laurent} criterion select as having 50 per cent or more AGN contribution have \eqwpah\ $< 0.27\ \micron$. Although the main purpose of this figure is to compare the \eqwpah\ selection to a common MIR AGN selection method, we also test if the correlation found in \citet{laurent} and \citet{armus} is driven by the shared dependence of both variables, $f_\nu[6.2\ \micron]/f_\nu[5.5\ \micron]$ and $f_\nu[15\ \micron]/f_\nu[5.5\ \micron]$, on the 5.5~\micronm\ flux. 
We perform a partial correlation analysis parametrized as:
\begin{equation}
\label{eq:parrcor}
r_{12,3} = \frac{r_{12} - r_{13}r_{23}}{\sqrt[]{1-r_{13}^{2}}\sqrt[]{1-r_{23}^{2}}}
\end{equation}
where the indices 1,2,3 refer to $f_{\nu}[6.2\ \micron]/f_{\nu}[5.5\ \micron]$, $f_{\nu}[15.5\ \micron]/f_{\nu}[5.5\ \micron]$, and $f_{\nu}[5.5 \micron]$ respectively. The correlation coefficients are the Spearman Rank correlation coefficients. We find the correlation is not dominated by the shared $f_{\nu}[5.5 \micron]$ values.  

As discussed in previous papers \citep[e.g.][]{petric}, low resolution spectra cannot be used to deblend the [\ion{Cl}{ii}]--[\ion{Ne}{v}] 14.322 \micronm\ lines. Furthermore, some AGNs do not show coronal line emission \citep[e.g. Mrk 231:][]{armus07}. After comparing multiple MIR AGN dominance criteria on our sample of low-resolution spectra, we use the \eqwpah\ criterion to select MIR AGN dominated host galaxies. 

\subsection{PAH Emission Features} \label{sec:p62}

The ratios of PAH emission line fluxes are determined by several factors including the size distribution and the ionization state of the PAH line emitting dust particles \citep{li2001,draine07}. The emission of the 6.2 and 7.7~\micronm\ bands are attributed to the radiative relaxation of the carbon-carbon stretching mode, which is more common in ionized PAH molecules \citep{tielens05}. The 11.3~\micronm\ feature emission, from carbon--hydrogen modes, drops its intensity by an order of magnitude between completely neutral and completely ionized PAH clouds. The ratio between the 6.2 and 7.7~\micronm\ features should not vary significantly as the ionization fraction changes \citep{li2001,draine07}. The relative power between two PAH bands depends on the distribution of grain sizes \citep{li2001,draine07}. Previous studies with the \textit{Spitzer Space Telescope} found dissimilar results concerning trends between AGN activity and the relative strengths of the PAH emission features. Some find evidence for preferential destruction of smaller PAHs by the AGN \citep[e.g.][]{smith07,odowd,wu10}. Others find a larger dispersion of relative strengths for AGN dominated objects but no preferential relative strength values \citep[e.g.][]{shipley, stierwalt}.  

\begin{figure}
\includegraphics[width=\columnwidth]{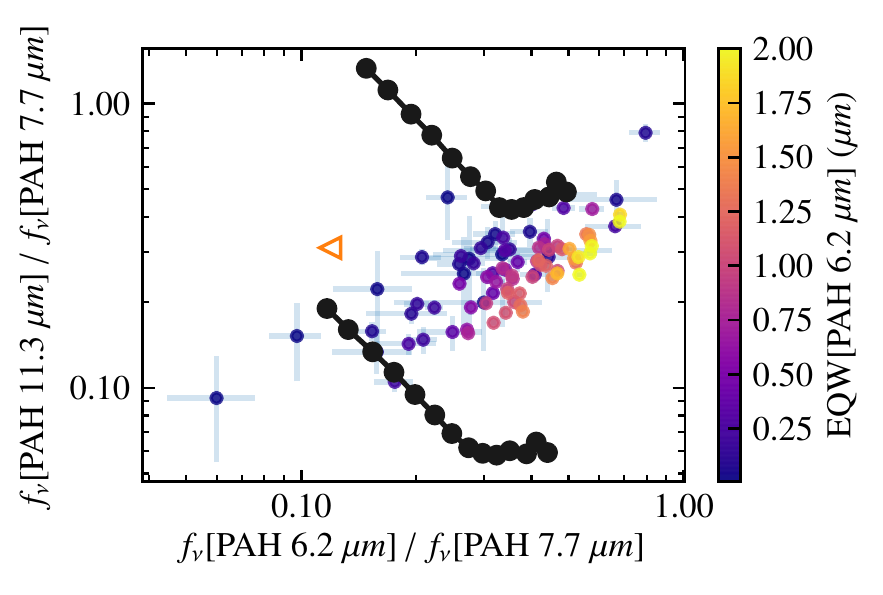}
	\caption{Stacked Spectra PAH band ratios: We show the normalized PAH relative strength ratios for our stacked spectra. Even in these high SNR spectra, there are AGN dominated sources below the theoretical line of complete ionization (bottom black line), and above the theoretical line of complete neutrality (top black line) \citep{draine07}. The open orange triangle is relative strength ratio calculated via a stack of the objects with only PAH 6.2~\micronm\ upper limits, but with $2\sigma$ detections of \molh S(3), PAH 11.3~\micronm, and PAH 7.7~\micronm.
The colours of the points are the same as in previous figures, with blue  denoting AGN-dominated objects and yellow denoting SF-dominated objects, defined by having small and large values of \eqwpah\ respectively.}
\label{fig:relpahstacks}    
\end{figure}

We compare measured ratios of $L[6.2\ \micron]/L[7.7\ \micron]$ and $L[11.3\ \micron]/L[7.7\ \micron]$ to the theoretical values for completely ionized and completely neutral dust grains from \citet{draine07} (\autoref{fig:relpah}). We find that that sources with \eqwpah\ $< 0.27$~\micronm, i.e. AGN dominated galaxies, have a wider range of relative strengths than the SF dominated objects and 20 per cent have ratios below the theoretical line of ionization. We calculate PAH ratios for our stacked spectra and find similar ranges of PAH relative strengths as compared to the unstacked spectra  (\autoref{fig:relpahstacks}). Several groups, e.g.~\citet{diamond,haan,odowd,shipley,stierwalt},  find that a small fraction of galaxies in their samples of nearby normal and IR luminous galaxies lie above or below the theoretical lines of pure neutrality or ionization. 

Our larger sample of objects with varying AGN MIR dominance significantly adds to this sample of outliers. \citet{draine07} models were calculated using a single Milky Way-based model. Our results highlight the potential need for more physical dust models to represent the diversity of extragalactic sources, as probed by their MIR emission. However our results are qualitatively consistent with \citet{odowd,shipley,stierwalt}: non-AGN form a tight locus but AGN dominated sources do not have a preferred location in the plot of "[6.2\micron]/L[7.7 \micron\ versus L[11.3 \micronm]/L[7.7 \micronm] (\autoref{fig:relpah},\autoref{fig:relpahstacks}). Nevertheless, we note that differences in morphologies, AGN sub-type, and metallicity may explain some of the scatter in the PAH properties of AGN hosts. Though this is beyond the scope of this paper, we provide the PAH luminosities of our sample in \autoref{tab:egpah} to assist future studies.

\subsection{Warm Molecular Gas and Dust Luminosity Relationships}

\begin{figure*}
\begin{center}
\includegraphics[]{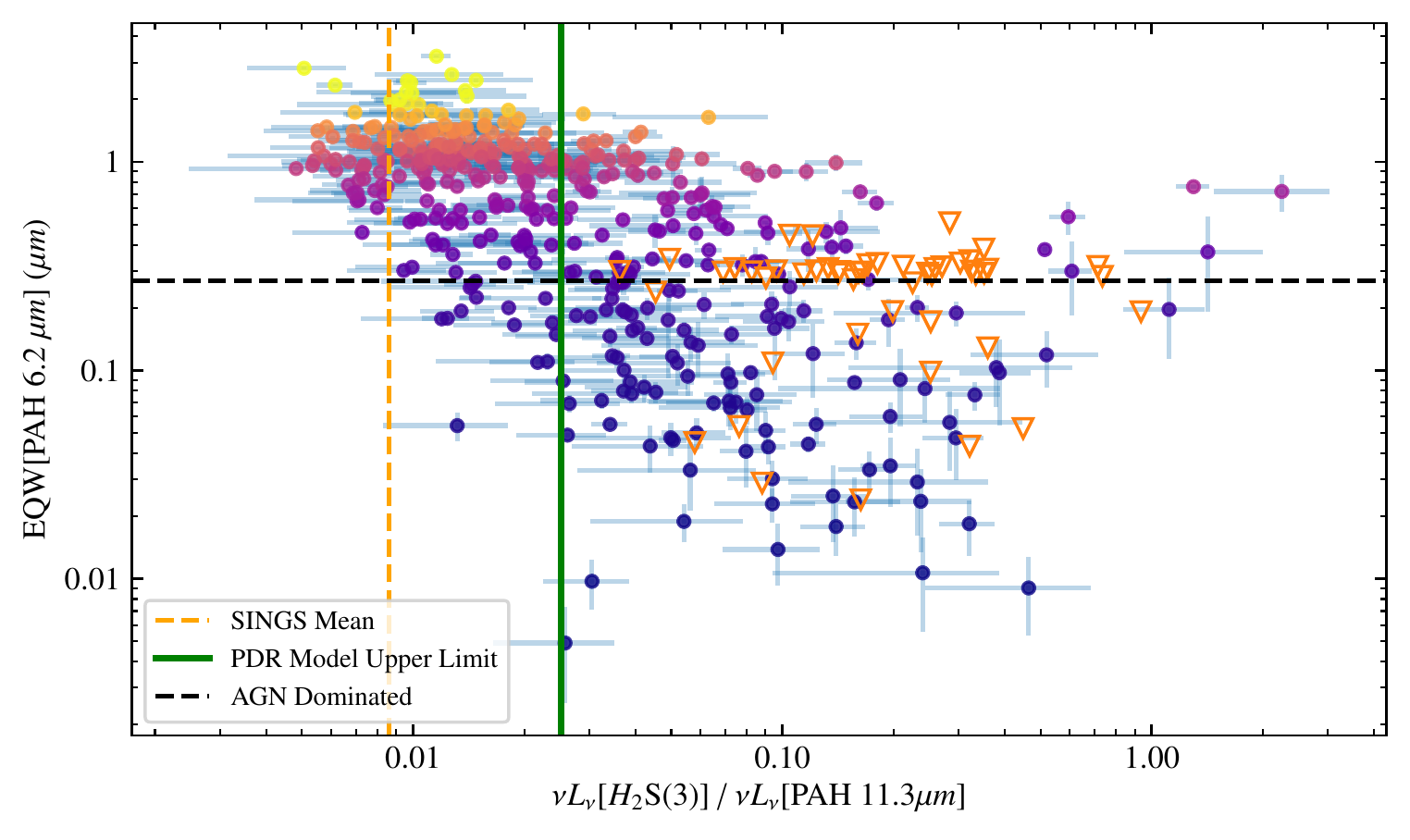}
	\caption{AGN activity versus molecular hydrogen excess emission: $2\sigma$ detections of the \eqwpah, $L$\big(\molh S(3)\big) and $L$(PAH[11.3~\micron]) emission features. The black dashed horizontal line represents the \eqwpah\ threshold ($<0.27$) to signify AGN dominance. The dashed orange vertical line is the mean \molh--PAH ratio from the SINGS normal star-forming galaxies sample \citep{roussel}. The vertical green line is the upper limit of the \molh\ to PAH ratio that is consistent with PDR emission as calculated via the Meudon PDR models \citep{petit} and presented in \citet{stierwalt}. The orange, open downward triangles are \eqwpah\ upper limits. The value of the correlation is $r_{s}=-0.6$ with $p_s\ll 0.001$. The orange triangles are \eqwpah\ upper limits for objects with $<2\sigma$ detections of \molh S(3), PAH 7.7~\micronm, and PAH 11.3~\micronm.
The colours of the points are the same as in previous figures, with blue  denoting AGN-dominated objects and yellow denoting SF-dominated objects, defined by having small and large values of \eqwpah\ respectively.}
\label{fig:excess}
\end{center}
\end{figure*}

In galaxies where star-formation processes dominate the IR emission, \molh\ and PAH emission are tightly correlated with an average value of $\mathrm{H_2/PAH} = 0.0065 \pm 0.001$ \citep{roussel}. This suggests that the bulk of \molh\ and PAH emission comes from gas and dust heated by similar sources. If star-forming regions emit a relatively constant amount of \molh\ relative to PAH emission, and if PAH EQW decreases in regions where the AGN contributes to the IR emission, then we expect higher ratios of \molh\ to PAH emission in sources with AGN. If the AGN heats the surrounding host material, then we may expect an additional warmer \molh\ component associated with the AGN. 

In \autoref{fig:excess}, we find that the ratio of molecular hydrogen to PAH emission is inversely proportional to the 6.2 PAH EQW, i.e. proportional to the AGN contribution to the total IR emission. We estimate the ratios of \molh\ to PAH emission for all sources in our sample with $2\sigma$ detections of \molh S(3), PAH 11.3~\micronm, and PAH 6.2~\micronm. The 11.3~\micronm\ feature is often used to estimate star-formation rates  \citep{peeters,calzetti,diamond}. \citet{zakamska2016} find that PAH emission may be suppressed in quasars. With a sample of lower-luminosity AGN, \citet{jensen} caution against using a simple relation between the 11.3~\micronm\ PAH flux and star-formation rates, though at large scales the method is reasonably reliable. We corroborate the PAH 11.3~\micronm\ flux invariance for lower-luminosity AGN on the large scales probed by the IRS spectrograph by finding a statistically significant weak correlation between \eqwpah\ and PAH 11.3~\micronm\ luminosity for a $\ge 2 \sigma$ PAH detected sub-sample comprised of 108 and 308 AGN and SF galaxies respectively ($r_{s}=0.15$, $p_s < 0.003$). To estimate what fraction of the observed \molh\ emission comes from gas in photo-dissociation regions, we divide the \molh S(3) 9.665~\micronm\ transition flux by the PAH 11.3\ \micronm\ flux.

In \autoref{fig:excess} we infer a large range of \molh\ to PAH ratios  (0.005--1.42). For \eqwpah\ $> 0.54$ \micronm, our values are consistent with the \molh\ to PAH ratios found in normal galaxies and SF dominated U/LIRGs \citep{rigopoulou,roussel,zakamska,stierwalt}. In \autoref{fig:excess}, we show the expected strong inverse correlation between SF (via increasing \eqwpah) and \molh\ to PAH ratio (via increasing $L[\mathrm{H_2S(3)}]/L[\mathrm{PAH\ 11.3\ \micron }]$). We plot the theoretically calculated upper limit presented in \citet{stierwalt} of the \molh\ to PAH ratio, assuming all the \molh\ is being fluorescently excited in PDRs \citep{petit}.

There is a statistically significant correlation between the \eqwpah\ and \molh\ to PAH ratio. Assuming the $L[$\molh S(3)$]$ normalized by $L[\mathrm{PAH\ 11.3\ \micron }]$ accounts for the \molh\ emission due to SF processes and \eqwpah\ traces the hot dust emission directly related to the power of the AGN, the anti-correlation between the \eqwpah\ and \molh\ to PAH ratio suggests that the luminosity of \molh\ scales with AGN activity ($r_{s}=-0.6$, $p_s \ll 0.001$). The median  $L[$\molh S(3)$]/L[\mathrm{PAH\ 11.3\ \micron }]$ is 0.17 for AGN-dominated objects and 0.06 for SF-dominated objects. We use a two-sample KS test to quantify the differences between the \molh\ to PAH ratio distributions of AGN and of star-formation dominated galaxies, and find that the distributions are different. We also find that the PAH 11.3 \micron\ emission is not correlated with the \eqwpah\ ($r_{s}=0.17$, $p_s \ll 0.001$), and thus our \eqwpah\ and \molh\ to PAH ratio is not due to the differences of the PAH 11.3 \micron\ emission between AGN and SF dominated galaxies. We perform a partial correlation analysis with the parametrization defined in \autoref{eq:parrcor}, and find the shared dependency on PAH fluxes is not driving the correlation.   
   
% Kirill replaced the following sentence with the previous sentence.
%We find the median $L[$\molh S(3)$]/L[\mathrm{PAH\ 11.3\ \micron }]$ for objects with \eqwpah\ $< 0.27$ \micronm\ to be 0.06 and 0.17 respectively, or an average factor of $>2$ \molh\ emission in SF dominated objects. 

We test whether our results are redshift dependent by splitting the $2\sigma$ \molh\ and PAH detections into equal bins of redshift space. We find the distribution of \molh\ to PAH does not change within each bin. We perform a two-sample KS test, and find that the distributions in each bin are statistically indistinguishable from one another. We check whether our AGN, SF dominated sub-samples are biased with respect to each other by quantifying whether the distributions of the $K$-band luminosities are consistent with being drawn from the same $K$-band luminosity distribution. They are: a two-sample KS test on the $K$ luminosities of AGN-dominated and SF-dominated sub-samples results in $D_\mathrm{KS} =0.09$ with $p_\mathrm{KS} = 0.6$.

For some of the most AGN MIR dominated sources, the reported \eqwpah\ is an upper limit; in these sources we only see continuum emission measure an upper limit for the PAH 6.2 \micron\ flux and \eqwpah\. As seen in \autoref{fig:excess}, all the objects with \eqwpah\ upper limits have \molh\ to PAH ratios larger than than most of the SF dominated systems. We estimate the effect the \eqwpah\ upper limits have on the \molh\ to PAH ratio relationship. The most conservative way to take the upper limits into account is to treat the limits as detections. Including the \eqwpah\ upper limits as detections, and calculating the Spearman correlation coefficient yields an even stronger anti-correlation between \eqwpah\ and the \molh\ to PAH ratio ($r_{s}=-0.65$, $p_s \ll 0.001$). If the actual values are lower, the anti-correlation is even stronger. If we assume that the \eqwpah\ could be any value between 0 and the upper limit, we can estimate the correlation strength between \eqwpah\ and \molh/PAH as follows:. for all of the objects with upper limits \eqwpah\, we draw a random value between 0 and the value of the upper limit from a uniform distribution; we then compute the Spearman $r$ coefficient using these randomly assigned values, and the actual detected values; we repeat this process 10,000 times, and measure the mean, median, minimum, and maximum of the distribution of Spearman $r$ coefficients as $-0.67$, $-0.67$, $-0.69$, and $-0.66$ respectively. While there is little physical basis behind choosing a uniform distribution to draw random values of \eqwpah\ upper limits, the fact that the Spearman $r$ coefficient is always less than the value excluding or assuming upper limits as detections shows that the reported relationship for detections is robust. 

In ULIRGs, there is no evidence for extinction affecting molecular hydrogen emission \citep{higdon,zakamska}. We test whether our sample is affected by extinction. We approximate the amount of extinction as proportional to the strength of the 9.7~\micronm\ silicate feature, a Si--O stretching resonance at 9.7~\micronm. % Kirill- don't need to explain that this is a stretching feature.
We measure the strength of the 9.7~\micronm\ silicate absorption (or emission) feature given by
\begin{equation}
    \tau_{9.7\ \textrm{\micron}} \equiv -\ln\left(\frac{f_\mathrm{\nu,obs}[9.7\ \micron]}{f_\mathrm{\nu,cont}[9.7\ \micron]}\right),
\end{equation}
where $f_\mathrm{\nu,obs}[9.7\ \micron]$ is the observed flux at 9.7~\micron\ and $f_\mathrm{\nu,cont}[9.7\ \micron]$ is the inferred continuum \citep{spoon,zakamska}. We provide the silicate strengths in \autoref{tab:egpah}. 

\begin{figure}
\begin{center}
\includegraphics[width=\columnwidth]{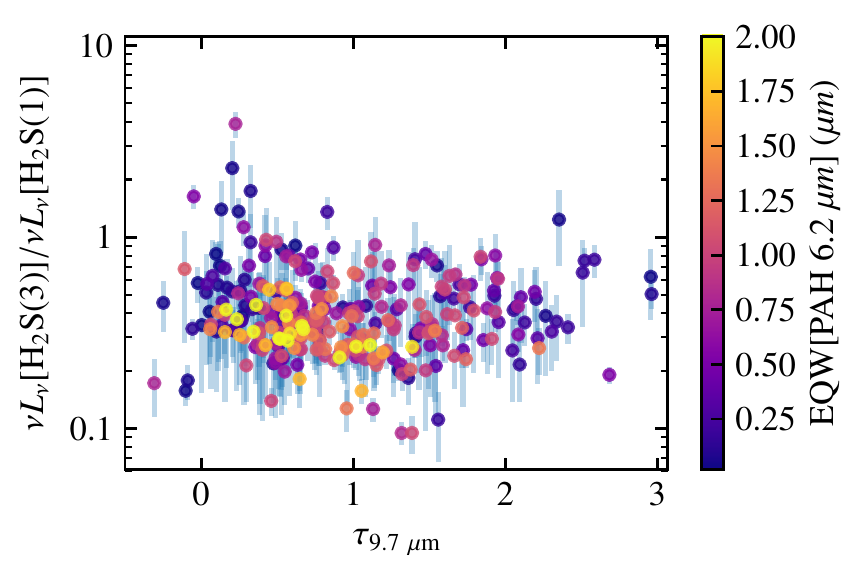}
	\caption{Molecular hydrogen emission versus silicate strength: \molh\ emission is not significantly affected by extinction. We calculate the apparent silicate strength and compare it to the ratio $\nu L_{\nu}[$\molh S(3)]/$\nu L_{\nu}[$\molh S(1)] ($r_{s}$, $p$-value is greater than 0.01, giving no evidence to discount the null hypothesis of no correlation). The colours of the points are the same as in previous figures, with blue  denoting AGN-dominated objects and yellow denoting SF-dominated objects, defined by having small and large values of \eqwpah\ respectively.}
\label{fig:silabs}
\end{center}
\end{figure}

\begin{figure*}
\includegraphics[]{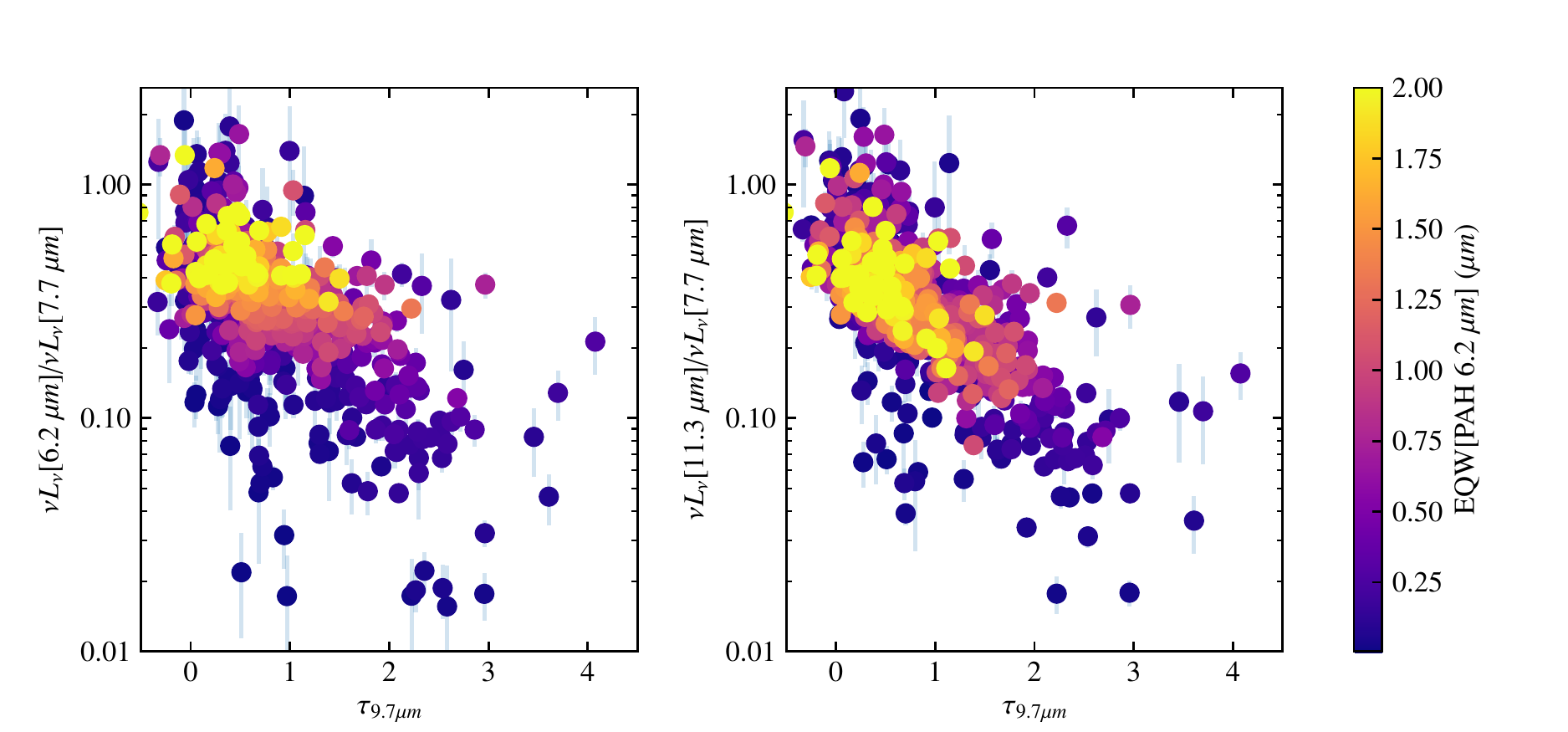}
	\caption{PAH band ratios versus silicate strength: We test how the diversity of PAH relative strengths relates to silicate obscuration. We use the 9.7~\micronm\ feature as tracer of obscuration.  The orange triangles are \eqwpah\ upper limits for objects with $<2\sigma$ detections of \molh S(3), PAH 7.7~\micronm\ and PAH 11.3~\micronm. The points are colour-coded by absorption strength, i.e. the red points have the greatest absorption, and the blue points the greatest emission. Light green points represent objects without significant silicate absorption or emission.}
%\textit{what does a typical reddening vector look like on this diagram? the correlation between position on diagram and silicate opacity seems pretty high\ldots -K}
\label{fig:pahtau_sep}    
\end{figure*}

\autoref{fig:silabs} shows that there is no statistically significant
trend between $\tau_{9.7\,\mathrm{\umu m}}$ and the ratio of the \molh S(3) and \molh S(1) transitions. Obscuration affects the measured PAH flux ratios. We plot each relative strength as a function of the silicate strengths in \autoref{fig:pahtau_sep}. As seen in \citet{zakamska}, the relationship found for $L($PAH[11.3 \micronm]$)/L($PAH[7.7\micronm]$)$ indicates similar effects for both SF and AGN dominated galaxies ($r_{s}=-0.76$, $p_{s} \ll 0.001$, $r_s=-0.72$, $p_{s} \ll 0.001$, for AGN, SF dominated objects respectively). For $L($PAH[6.2 \micronm]$)/L($PAH[7.7\micronm]$)$, we find the most SF dominated objects are located in a tight locus, and exhibit a much weaker correlation ($r_{s}=-0.19$, $p_{s} = 0.07$) than the rest of the sample. \citet{zakamska} explain the obscuration effects as evidence of PAHs existing behind the location of silicates and water ices in AGN dominated galaxies.                  

\subsection{Warm and Warmer Molecular Hydrogen Temperature Decomposition}

%The question we wish to answer is: are the distributions of \molh\ excitation temperatures the same or different in our AGN dominated and star-formation dominated samples? In this and the subsequent subsections, we estimate the typical temperatures of \molh\ in SF-dominated and AGN-dominated galaxies using two different overall approaches: (1) Two Temperature Distributions -- using all of the \molh\ lines simultaneously to separate two different \molh\ gas distributions within a given galaxy and (2) - Excitation Temperatures per Line Pair --  excitation temperatures of \molh\ transitions of equal parity without assuming a temperature distribution. Both (1) and (2) are the most standard ways of extricating the physical properties of the warm \molh\ gas in astrophysical sources. Within (1) and (2) we explore two different methods for each approach: (1A, 2A) represent the most common implementation in the literature and (1B, 2B) represent new algorithms we have developed for these approaches using Bayesian statistics. This variety of approaches and methods is necessary to test the integrity of our  conclusions  about the warm \molh\ gas. We summarize these definitions in \autoref{tab:var_names}.

We investigate if the distributions of \molh\ excitation temperatures we measure in AGN hosts differ from those of non-AGN galaxies. We estimate the typical temperatures of \molh\ in SF-dominated and AGN-dominated galaxies using two different approaches: (1) Two Temperature Distributions -- using all of the \molh\ lines simultaneously to separate two different \molh\ gas distributions within a given galaxy and (2) - Excitation Temperatures per Line Pair --  excitation temperatures of \molh\ transitions of equal parity without assuming multiple temperature distributions. Both (1) and (2) are the most standard ways of extricating the physical properties of the warm \molh\ gas in astrophysical sources. Within (1) and (2) we explore two different methods for each approach: (1A, 2B) represent the most common implementation in the literature and (1B, 2B) represent new algorithms we have developed for these approaches using Bayesian statistics. We use multiple methods to estimate \molh\ excitation temperatures to test if our conclusions about the warm \molh\ gas are robust. Methods (1A), (1B), and (2B) are performed on individual galaxies, while method (2A) is performed on both individual galaxies and the stacked spectra. We summarize the names and descriptions of our techniques in \autoref{tab:var_names}.

\begin{table}
	\caption{Summary of Warm \molh\ Temperature Analysis Algorithms}
    \begin{tabular}{| p{3cm} | p{2cm} | p{1.8cm} |}
    \hline
    Approach & Method A & Method B \\ \hline
    (1) - Two Temperature Decomposition & Least-Squares Line Fitting to the Excitation Diagrams & Marginalized Likelihood Analysis \\ \hline
    (2) - Excitation Temperatures & Means of the temperatures in a given transition & Hierarchical Bayesian Model \\ 
    \hline
    \end{tabular}
    \label{tab:var_names}
\end{table}

\textbf{\textsl{(1A) - Two Temperature Decomposition:}} For (1), the two-temperature decomposition, we aim to decompose the \molh\ excitation diagrams of the galaxies in our sample into two distributions: a warm and a warmer component. In both the unstacked and stacked spectra, the rotational transition ladders of the few galaxies in the dataset with high-significance detections of the \molh S(0) through \molh S(3) and \molh S(5) through \molh S(7) transitions cannot be described by a single excitation temperature; the higher-excitation \molh\ transitions tend to be at higher temperatures than the lower-excitation transitions. In some of these well-detected rotational transition ladders, one can see the saw-tooth pattern characteristic of a non-equilibrium ortho-to-para ratio \citep{neufeld,ogle}. This motivates the two-temperature decomposition approach for modelling the excitation diagram with one warm component at 100 K -- 300 K (denoted as $T_{1}$) and another warmer component at $> 300\ \mathrm K$ (denoted as $T_{2}$). Unfortunately, if we were to require $2\sigma$ detections of all lines at once, we would have fewer than 50 objects. Warm molecular hydrogen studies usually include upper limits for the non-significant detections in order to estimate the underlying temperature distribution. By analysing all of the \molh\ lines simultaneously, we are able to provide a mass estimate of the \molh\ in a given distribution. For (1A), we use the two-temperature decomposition algorithm as outlined in \citet{higdon}. This method and its variants are the most common techniques for extricating the warm ($T_{1}$) and warmer ($T_{2}$) components of the \molh\ gas \citep{roussel,ogle, petric18}.     

As \citet{roussel}, \citet{higdon}, and  \citet{petric18} find, the mass can be severely biased if the \molh S(0) flux is not detected. Despite the above issues, we test to see if there are systematic differences between the mass estimates of of warm \molh\ for the individual objects in our sample. We estimate the total \molh\ mass as
\begin{equation}
\label{eq:totmass}
M_\mathrm{tot} =\frac{4}{3}M_{o}, 
\end{equation}
where $M_{o}$ is the mass of the gas in the ortho state,
\begin{equation}
\label{eq:orthomass}
M_{o} = m_{\mathrm H_{2}}N_{T},
\end{equation}
with $m_{\mathrm H_{2}}$ being the mass of an \molh\ molecule and $N_{T}$ the total number of molecules. The total number of molecules in the $J^\mathrm{th}$ state is $N_{T} = N_{J}/f_{J}$, where $f_{J}$ is the partition function for the $J^\mathrm{th}$ state,
\begin{equation}
\label{eq:partitionfunction}
 f_{J} = \frac{g_{J} \exp[-E_{J}/kT_\mathrm{exc}]}{\Sigma_{J_{i},\mathrm{ortho}} g_{J_{i}} \exp[-E_{J_{i}}/kT_\mathrm{exc}]}
\end{equation}
where $i$ indexes the \molh\ transitions.

We fit \molh\ excitation diagrams ($E_{J}$ versus $\log(N_{i}/g_{i})$) to find the warm and warmer gas components, which uses a two component fit. Most of the pure-rotational \molh\ transitions are weak detections. Using only two components can be highly degenerate and difficult to constrain without \molh S(0) detections or stringent upper limits \citep{higdon,roussel,Hill,petric18}. Due to low detection rates of \molh S(0) in the majority of IRS low-resolution spectra, most two-temperature decomposition methods use upper limits of \molh S(0), so their mass estimates are rough approximations. We perform a two-temperature decomposition on the \molh\ excitation diagrams of our individual spectra. We only use spectra with at least two detected \molh\ transitions and include upper limits for non-detections. For objects where only the \molh S(1) and \molh S(3) are detected, we assume a single temperature distribution. We also test if an ortho-to-para ratio (OPR) of 3 is valid, and if not we calculate the OPR via
\begin{equation}
\mathrm{OPR} = \frac{\mathrm{OPR}_{\mathrm{high}\ T}}{3}
\frac{\sum_{o}(2I_{o}+1)(2J_{o}+1)\exp[-E_{o}/kT_\mathrm{exc}]}
{\sum_{p}(2I_{p}+1)(2J_{p}+1)\exp[-E_{p}/kT_\mathrm{exc}]}
\label{eq:OPR}
\end{equation}
where $o$, $p$ denote ortho and para respectively and $I_{p}$, $I_{o}$ are 0 and 1. $\mathrm{OPR}_{\mathrm{high}\ T}$ is equal to OPR in the high-temperature limit, i.e. $T > 200\ \mathrm K$, $\mathrm{OPR}= 3$. 

In the high-temperature OPR case we perform a Levenberg-Marquardt fitting algorithm \citep{mpfitfun} to determine the parameters of the $T_{1}$ and $T_{2}$ components ($T_{1}$ - lower temperature, $T_{2}$ - upper temperature). We calculate the mass and column density (as described in \autoref{eq:totmass}--\autoref{eq:partitionfunction}) of the warm and warmer component. In \autoref{tab:mean_temps_2temp} we provide the derived mean temperatures and total mass fractions of two gas distributions for AGN and SF dominated galaxies via two-temperature decomposition. We find the distribution parameters of the AGN, SF dominated galaxies to be statistically indistinguishable from one another. As mentioned earlier, a small minority of our sample has more than two 3$\sigma$ \molh\ detections. This severely affects the efficacy of the two-temperature decomposition method. As found in \citet{stierwalt}, when the \molh S(0) line is undetected, the temperature of the warm gas may be overestimated, and thus the warm mass component underestimated.  

For most sources, the masses and temperatures we derive are not well constrained by a fit, they are estimates of four unknown parameters (two masses and two temperatures) from four emission line fluxes. We are cognizant of the limitations of this approach, however this method together with the other methods of estimating masses and temperatures we present in this paper, allow us to consistently compare with other samples of galaxies analysed in a similar fashion. There are no obvious systematic errors in this method that would erroneously lead to trends between the warm molecular gas properties and the target's morphologies (mergers versus non-mergers) or AGN contribution to the IR emission from their host galaxy.

\begin{table}
    \caption{Method (1A) - Derived Mean Temperatures and Masses Fractions of Two Gas Distributions for Individual Galaxies via Two-Temperature Decomposition: columns 1, 2, and 3 indicate the $T_{1}$ component temperature, $T_{2}$ component temperature, and mass fraction of the warmer component to the total mass respectively. The rows indicate AGN, SF dominated as defined by \eqwpah\ $< 0.27$ \micronm, \eqwpah\ $> 0.54$ \micronm\ respectively.}
    \begin{tabular}{| p{1.65cm} | p{1.5cm} | p{1.8cm} | p{1.3cm} | }
    \hline
    Class & $T_\mathrm{1}$ & $T_\mathrm{2}$ & $\displaystyle{\frac{M_\mathrm{2}}{M_\mathrm{1} + M_\mathrm{2}}}$ \\
    & (Median, K) & (Median, K) & 
        \\ \hline
    AGN-Dominated & 198.3 $\pm$ 31.2 & 522.1 $\pm$ 169.4 & 0.13 $\pm$ 0.06  \\ \hline
    SF-Dominated & 192.9 $\pm$ 34.9 & 519.6 $\pm$ 276.0 & 0.11 $\pm$ 0.08 \\ 
    \hline
    \end{tabular}
    \label{tab:mean_temps_2temp}
\end{table}

\textbf{\textsl{(1B) - Two Temperature Decomposition Using a Marginalized Likelihood Analysis:}}
The standard two-temperature decomposition uses a minimum chi-squared to determine the optimal fit to the excitation diagram. Minimizing chi-squared in this case is equivalent to maximizing the likelihood. As noted earlier, the decomposition of the excitation diagram into two populations can be degenerate due to the covariances between the slope of the $T_{1}$ and $T_{2}$ \molh\ distribution. This motivates method (1B) - two temperature decomposition using a marginalized likelihood analysis, where we construct an algorithm which uses the entirety of the likelihood function. The (1B) algorithm infers the ratio of $T_{2}$ \molh\ to $T_{1}$ \molh\ by integrating over all possible values of the other parameters (e.g total mass). Unlike (1A), we treat the warm gas component and the total mass as a nuisance parameter. This allows us to fully examine the likelihood function function of the parameter we care most about: the warmer $T_{2}$ component. The $T_{1}$ component includes transitions that require more excitation energies than are typically found in PDRs. We hypothesize the greatest difference between AGN and SF dominated galaxies will be within these states. We first select targets where the signal-to-noise ratio of the PAH 6.2~\micronm\ line luminosity is at least 3, the \molh S(0), \molh S(1), \molh S(2), \molh S(3), and \molh S(5) lines all fall within the observed wavelength range, and the signal-to-noise ratio of the \molh\ line luminosities is at least 2. We convert the line luminosities and luminosity uncertainties to column densities and column density uncertainties. We do not replace marginal detections or non-detections with upper limits and instead keep the reported best-fit column densities and column density uncertainties.

We model each set of column densities as the superposition of a $T_{1}$ component and a $T_{2}$ component. We parametrize the relative amplitudes of the two components in terms of a ratio of column densities, $r(h)\equiv N_{\mathrm{2},J=2}/N_{\mathrm{1},J=2}$. We assign both components the same, possibly non-equilibrium, `local' (i.e. per-level, the quantity which is equal to 3/4 at ortho-to-para equilibrium regardless of the temperature) ortho-to-total fraction $f(o)$. We restrict the temperature of the $T_{1}$ component to be non-zero. We parametrize the temperature of the $T_{2}$ component as $T_\mathrm{2} = T_\mathrm{1} + \Delta T$, where we restrict $\Delta T$ to be non-zero. The likelihood function (and posterior probability distribution) of this model can take on a variety of shapes depending on which transitions we can detect at high signal-to-noise ratios.

To assess the uncertainties on the parameters, we generate samples from the posterior probability distribution using Markov chain Monte Carlo (MCMC). We have found that analytically marginalizing over the absolute amplitude dramatically improves convergence and mixing of MCMC, so we do not report any absolute column densities or masses.  Instead, we utilize the `local' ortho-to-total fraction $f(o)$; the ortho-to-para ratio $\mathrm{OPR}\equiv\sum N_{J_\mathrm{odd}}/\sum N_{J_\mathrm{even}}$; the ratio of $T_{1}$ to $T_{2}$ component column densities in the $J = 2$ level $r(h)$; the $T_{2}$ column density fraction relative to the total amount of emitting \molh\ ${f(\mathrm{2}) = N_\mathrm{2}/(N_\mathrm{1} + N_\mathrm{2})}$; the component temperatures $T_\mathrm{1}$ and $T_\mathrm{2}$; and the column density-weighted average temperature $T_\mathrm{avg}$.

\begin{figure}
   \includegraphics[width=.5\textwidth]{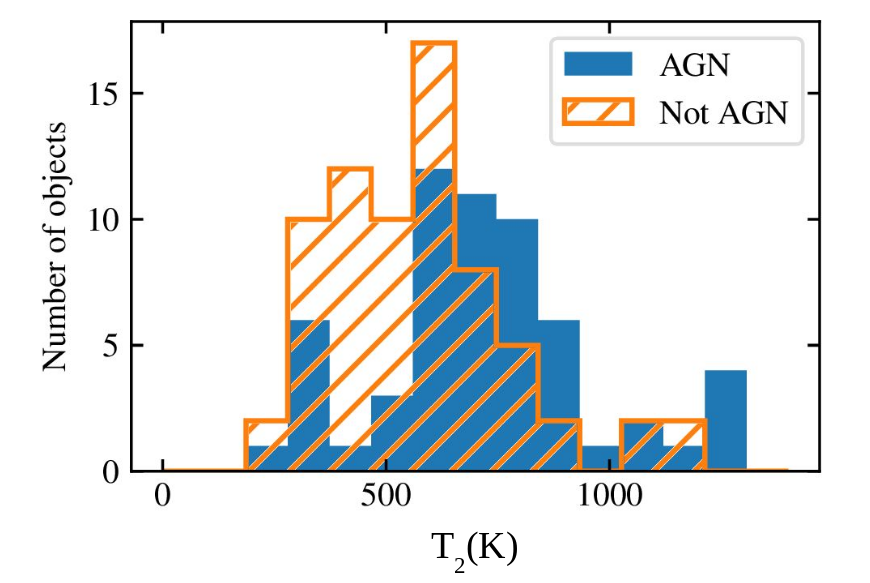}
   \label{fig:hothist} 
\caption{Method (1B) - Two-Temperature Decomposition Likelihood Analysis: warmer component temperature distribution histogram. The blue bins and striped orange bins are the AGN-dominated and star-formation dominated sub-samples of galaxies respectively.}

\end{figure}

We find that AGN-dominated galaxies typically have higher $T_\mathrm{2}$
than SF-dominated galaxies (\autoref{fig:hothist}). The difference between the two distributions is apparent by eye and is significant according to a two-sample KS test. The distributions of all other parameters reported from this analysis are consistent with being the same in the AGN-dominated and SF-dominated sub-samples, once again according to a two-sample KS test. 

Galaxies are complex systems, and in spatially unresolved mid-infrared spectroscopy, a given warm \molh\ transition represents the sum of different populations of \molh\ gas at different locations within a galaxy. In methods (1A) and (1B), we separate two gas components. This helps provide a more physical \molh\ gas parameter estimation, but this method suffers from a serious drawback; it requires well measured \molh\ transitions to accurately sample a wide range of excitation temperatures. The flux-limited nature of \molh\ detections makes secure temperature component estimates difficult, thus it is unsurprising we do not find a difference between AGN, not-AGN dominated sub-samples in (1A). Method (1B) attempts to overcome some of the technical problems of (1A), i.e. line-fitting noisy or under-sampled data by using the entirety of the likelihood function and marginalizing over parameters that we are less interested in. Method (1B) produces a more robust result. While (1A) produces bias on $T_{2}$ due to the large uncertainties of the higher \molh\ transitions; (1B) places this bias into the uncertainty of the $T_{2}$ by marginalizing over all the other allowed ways in which the \molh\ (1B) SED can vary. In (1B), we do find a statistical temperature difference in the warmer gas component: AGN have higher temperatures in their warmer component versus not-AGN dominated host galaxies. In the next section, we test if properties within a given line transition is statistically separable between the AGN and not-AGN dominated sub-samples.

\subsection{Warm Molecular Hydrogen Excitation Temperatures}

\textbf{\textsl{(2A) - Excitation temperatures per line pair:}}
For the unstacked spectra, we first utilize method (2A) which is the simplest approach of calculating the excitation temperatures via the following pairs of lines: (\molh S(0), \molh S(2)), (\molh S(3), \molh S(1)), (\molh S(5), \molh S(3)), and (\molh S(7), \molh S(5)). Using only transitions with $> 2\sigma$ significance, we calculate the excitation temperatures ($T_\mathrm{exc}$) of the gas in a given transition via pairs of lines. We compare the distributions of the temperatures between a sample of AGN-dominated and SF-dominated galaxies. As before, we define an AGN-dominated (SF-dominated) galaxy as one with \eqwpah\ $< 0.27\ \micron$ (\eqwpah\ $> 0.54$ \micronm). We do not include the 90 objects that have comparable AGN and SF contribution. Because the majority of the spectra do not have enough detections to confidently measure the ortho-to-para ratio, we choose to only measure excitation temperatures between states of the same parity. The column density, $N_{J+2}$, in the upper level of each transition assuming the gas is in local thermal equilibrium defined as
\begin{equation}
N_{J+2} = \frac{4 \pi D^{2}_{L} F_{J}}{A_{J+2 \to J}(E_{J+2} - E_{J})}
\end{equation}
where $D_{L}$ is the luminosity distance, $F_{J}$ is the line flux, $(E_{J+2} - E_{J})$ is the energy of the transition, and $A_J$ and $A_{J+2}$ are the Einstein coefficients \citep{einsteincoeffs}. The energy levels are
\begin{equation}
E_{J} = 85.35\,\mathrm K \cdot k_\mathrm{B}J(J+1) - 0.068\,\mathrm K \cdot k_\mathrm{B}J^{2}(J+1)^{2},
\end{equation}
where $k_\mathrm{B}$ is the Boltzmann constant. 
$T_\mathrm{exc}$ is then estimated via the relationship between $N_j$, $g_J$, $E_J$, and $T_\mathrm{exc}$,
%We then estimate the excitation temperature from the slope of the relationship between $\ln(N_{J}/g_{J})$ and $E_J$,
\begin{equation}
\frac{N_{J}}{g_{J}} = \exp \left( -\frac{E_{J}}{k_\mathrm{B}T_\mathrm{exc}} \right),
\end{equation}
where $g_{J} = 2J+1$ for even $J$ and $g_{J} = 3(2J+1)$ for odd $J$ assuming an equilibrium ortho-to-para ratio. The excitation temperature from transition pairs of the same parity is ${T_{u,l} = (E_{u} - E_{l})/\mathrm{ln}(N_{l}/N_{u} \times g_{u}/g_{l})}$ where $u$ and $l$ correspond to the upper and lower transition respectively. For example, the excitation temperature via the pair of transitions \molh S(3), \molh S(1) is represented as $T_{3,1}$.   

\begin{figure}
   \includegraphics[width=.5\textwidth]{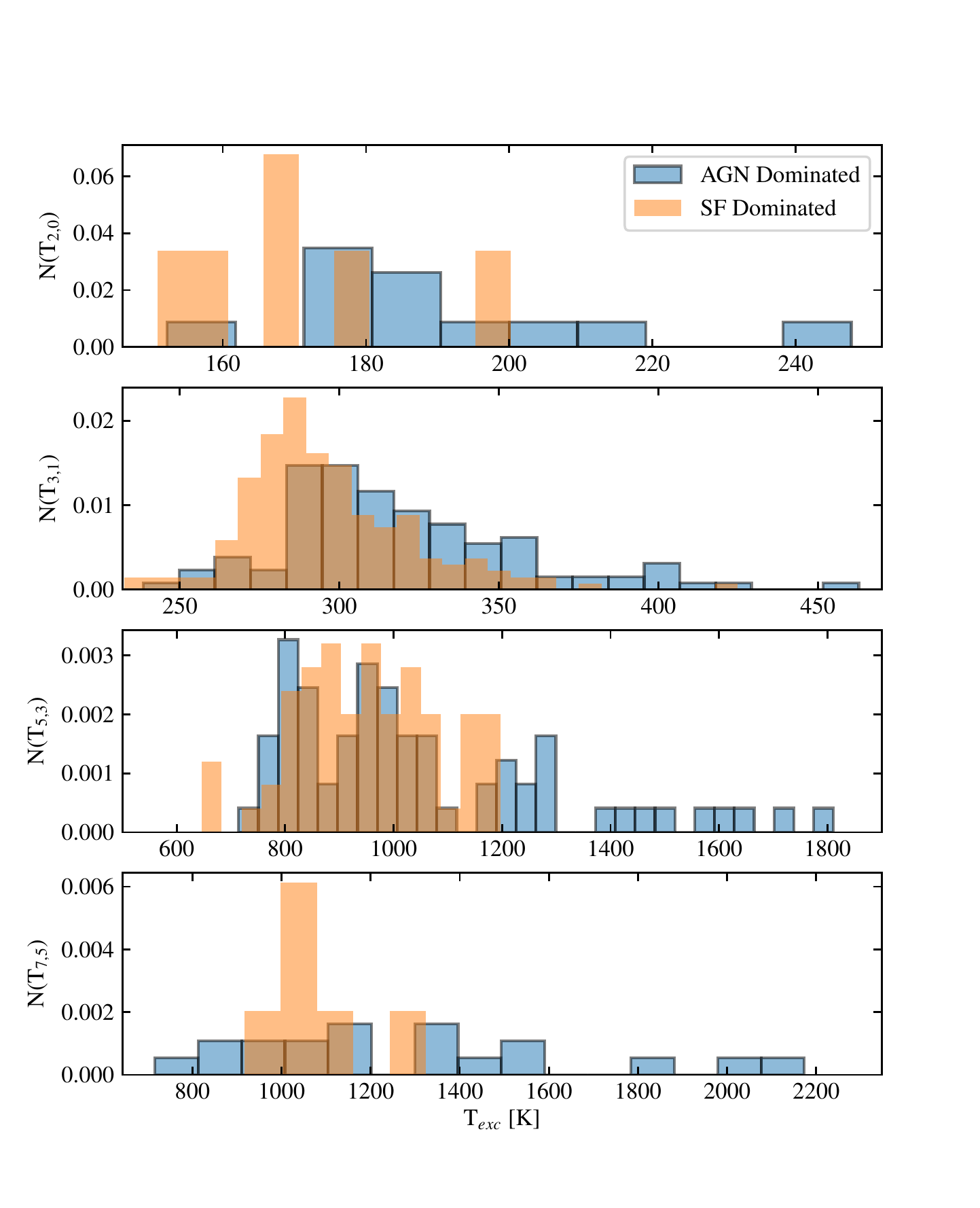}
\caption{Method (2A) - Excitation Temperature Analysis to Find Mean Temperatures of Individual $2\sigma$ transition temperature detections of the AGN, not-AGN sub-samples. The blue bins and the orange bins are the AGN-dominated and star-formation dominated sub-samples of galaxies respectively. The $y$-axis is the frequency per unit excitation temperature $T_{u,l}$.}
\label{fig:hist2A}
\end{figure}

As shown in \autoref{fig:hist2A}, we find the mean $T_\mathrm{exc}$ of the AGN dominated sub-sample marginally higher than the SF-dominated sub-sample in the highest transitions (i.e T$_{5,3}$,T$_{7,5}$. While this method is straightforward and simple, it is does not take full advantage of the dataset.  The majority of the spectra do not have strong individual detections of multiple \molh\ lines, and the exclusive selection criteria required for each pair of transitions reduces the sample size so drastically that we cannot make robust statistical inferences. For example, $T_{2,0}$ requires $2\sigma$ detections of both $f_{\nu}$[\molh S(0)] and $f_{\nu}$[\molh S(2)], and $T_{3,1}$ requires $2\sigma$ detections of both $f_{\nu}$[\molh S(1)] and $f_{\nu}$[\molh S(3)], but there are only 20 objects that satisfy both the $T_{3,1}$ and $T_{2,0}$ selection criteria. As seen in \autoref{tab:mean_temps_single}, the means between the distribution are mainly within a standard deviation of each other, but each excitation temperature has a tail of AGN dominated objects with significantly higher temperatures.       

\begin{table}
    \caption{Method (2A) - Excitation Temperature Analysis to Find Mean Temperatures of Individual $2\sigma$ transition temperature detections of the AGN, not-AGN sub-samples: column 1 indicates the excitation temperature $T_{u,l}$, column 2 and column 3 are the AGN sub-sample (\eqwpah\ < 0.27~\micronm) temperatures and the not-AGN dominated sub-sample (\eqwpah\ > 0.54~\micronm) temperatures respectively, column 4 is the number of objects with $2\sigma$ detections in each sub-sample, and column 5 is the $D_\mathrm{KS}$ statistic and p-value.}
    \begin{tabular}{| l | p{1.8cm} | p{1.8cm} | p{1.2cm}| p{1cm}|}
    \hline
    $T_{u,l}$ & not-AGN & AGN & Number & $D_\mathrm{KS}$, $p_\mathrm{KS}$\\
    & (Mean, K) & (Mean, K) & (not-AGN, AGN)
        \\ \hline
    $T_{2,0}$ & $171.1 \pm 16.8$ & $190.7 \pm 24.0$ & 6, 12 & 0.6,  0.2 \\ \hline
    $T_{3,1}$ & $298.6 \pm 38.7$ & $319.4 \pm 38.3$ & 115, 191 & 0.3, $\ll0.001$ \\ \hline
    $T_{5,3}$ &$949.4 \pm 133.9$ & $1051.8 \pm 257.9$ & 71, 86 & 0.1, 0.7 \\ \hline
	$T_{7,5}$ & $1084.9 \pm 133.9$ & $1294.5 \pm 395.2$ &  6, 19 & 0.5, 0.2\\
    \hline
    \end{tabular}
    \label{tab:mean_temps_single}
\end{table}

\begin{figure}
\begin{center}
\includegraphics[width=\columnwidth]{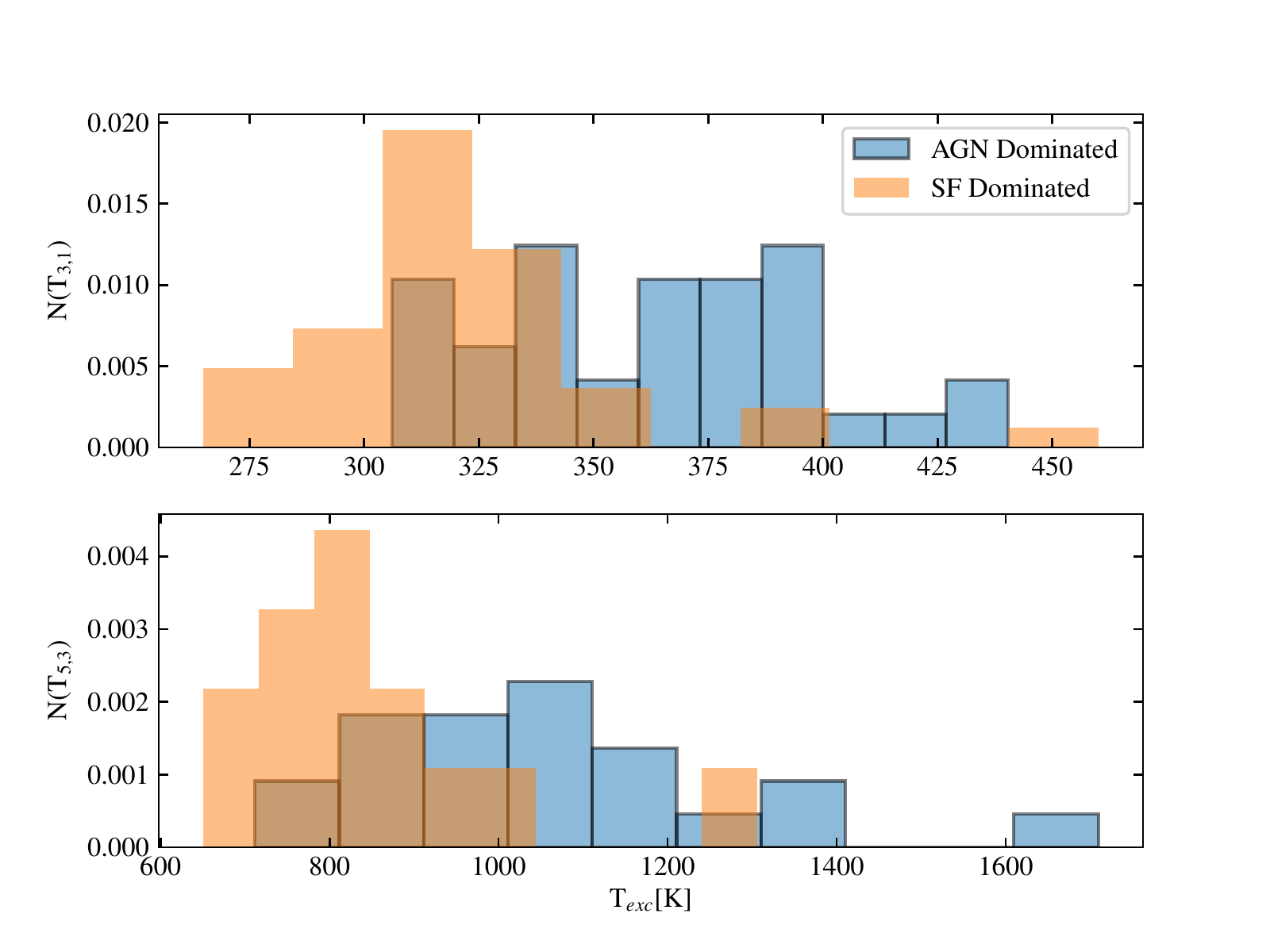}
\caption{Method (2A) - Excitation Temperature Analysis on Stacked Spectra to Find Mean Temperatures of Individual $2\sigma$ transition temperature detections of the AGN, not-AGN sub-samples. The blue bins and the orange bins are the AGN-dominated and star-formation dominated sub-samples of galaxies respectively. The $y$-axis is the frequency per unit excitation temperature $T_{u,l}$.}
\label{fig:stack_hist2a}
\end{center}
\end{figure}

We then employ method (2A) on the stacked spectra. We calculate excitation temperatures for the stacked spectra in which the \molh S(1), \molh S(3), \molh S(5) are at least $2\sigma$ detections. We exclude the \molh S(0) and \molh S(7) transitions from the stacked spectral excitation temperature analysis due to only a few stacks having detections in these transitions. We calculate the following temperatures using the following pairs of transitions that have the same parity: $T_{3,1}$ and $T_{5,3}$. In \autoref{fig:stack_hist2a}, we show the normalized density distributions of the excitation temperatures, and in \autoref{tab:mean_temps_stacks} we list the mean and standard deviation of the excitation temperature distributions. We find that in both the unstacked, and stacked space that the AGN dominated galaxies have a much wider range of excitation temperature distributions

Due to the normalization of the stacks, we cannot calculate the \molh\ mass. The stacks also rely wholly on the fundamental assumption that the \eqwpah\ is the sole separator between galaxy types. This assumption is useful for comparing AGN selection criteria, but the potential nuances between galaxy types and \molh\ emission within galaxies of similar \eqwpah\ would be lost. 
% Kirill - That's the result that actually gets reported. It's true that the AGN distribution is also wider, but that's not reported above.
%namely that the higher \molh\ transitions vary the greatest as a function of AGN dominance.  

\begin{table}
    \caption{Method (2A) - Excitation Temperature Analysis on Stacked Spectra to Find Mean Temperatures of Individual transition temperature detections of the AGN, not-AGN sub-samples: column 1 indicates the excitation temperature $T_{u,l}$, column 2 and column 3 are the AGN sub-sample (\eqwpah\ < 0.27~\micronm) temperatures and the not-AGN dominated sub-sample (\eqwpah\ > 0.54~\micronm) temperatures respectively, column 4 is the number of objects  in each sub-sample, and column 5 is the $D_\mathrm{KS}$ statistic and p-value.}
    \begin{tabular}{| l| p{1.8cm} | p{1.9cm} | p{1.2cm}| p{1cm}|}
    \hline
    $T_{u,l}$ & not-AGN & AGN & Number & $D_\mathrm{KS}$, $p_\mathrm{KS}$ \\
    & (Mean, K) & (Mean, K) & (not-AGN, AGN)
        \\ \hline
    $T_{3,1}$ & $321.37 \pm 34.13$ & $363.12 \pm 34.55$ & 36, 42 & 0.6, $\ll 0.001$ \\ \hline
    $T_{5,3}$ & $848.41 \pm 161.10$ & $1055.62 \pm 218.11$ & 22, 28 & 0.65, 0.0006\\ \hline
   \hline
    \end{tabular}
    \label{tab:mean_temps_stacks}
\end{table}

\textbf{\textsl{(2B) - Hierarchical modelling of the excitation temperature distribution:}} Methods (1A), (1B), and (2A) rely on measuring accurate excitation temperatures for galaxies individually. However, method (2B), hierarchical modelling of the excitation temperature distribution within a given sub-sample, can infer the distribution of excitation temperatures within the SF-dominated and AGN-dominated sub-samples without needing to measure excitation temperatures for any individual galaxy. A hierarchical model is one in which inference is done simultaneously over the parameters describing the population and the parameters describing the members of the population (see \citealt{gelman} for an in depth introduction to hierarchical modelling and \citealt{hogg} for a short but carefully explained astronomical example). 

Hierarchical modelling is more appropriate than doing an excitation analysis on a stacked spectrum for determining the mean excitation temperature of a population. This is the case because excitation temperature is non-linearly related to the observable, flux. As a result of this non-linearity, the excitation temperature of the mean (or median) of a collection of spectra will not, in general, be equal to the mean of the excitation temperatures of the individual spectra even when no noise is present. Our hierarchical model computes the mean of a collection of excitation temperatures derived from noisy flux measurements in a way that correctly accounts for the non-Gaussianity of their uncertainties.

A non-hierarchical modelling approach to characterizing the distribution of excitation temperatures in a population could be first calculating temperatures for each individual galaxy, then averaging those individual temperatures together. Then, the parameters describing the individual galaxies are fixed to some value which is then used to compute a population-level quantity. In hierarchical modelling, the parameters of individual galaxies are not held fixed. Parameters that vary in our hierarchical model include both the excitation temperature of each galaxy and the parameters of the distribution of excitation temperatures in the population. By integrating over all possible values of the individual galaxy parameters, we get a more robust estimate of the population-level parameters. 

\begin{figure*}
\begin{center}
\includegraphics[width=1\textwidth]{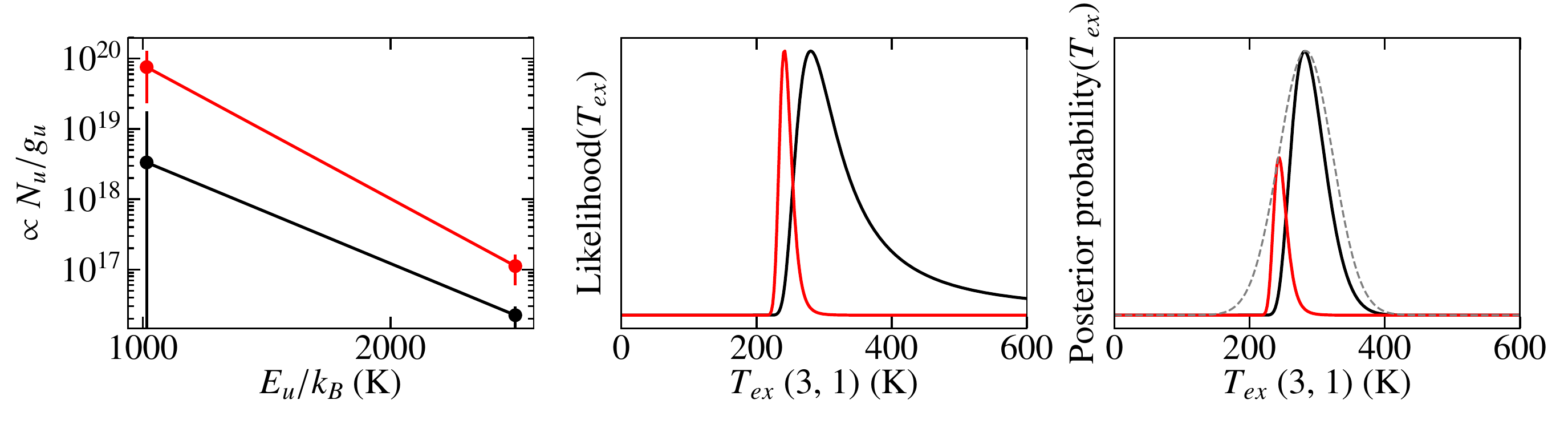}
	\caption{Method (2B) -- Example distribution mapping: The first panel shows the excitation diagram using \molh S(1) and \molh S(3) for two different random objects in our sample. The red line corresponds to a $3\sigma$ detection in both the S(1) and S(3) transitions. The black line corresponds to an object with a well constrained S(3) value, but a $1\sigma$ \molh S(1) detection. In the second panel, we show the likelihoods of the two objects. After finding the likelihoods for every detection for a given same parity pair, we then use the averages of the likelihoods to re-run the model, providing a more robust estimate of the mean temperature of the transition. In this example both objects are in our sample, and the grey-dashed line represents the estimated temperature distribution for a given transition for this sample. }
\label{fig:bayes_eg}
\end{center}
\end{figure*}

If we do not know the parameters of the prior distribution ahead of time, we can attempt to infer the prior parameters and the individual galaxy parameters at the same time. This approach is particularly useful when one has a large sample of galaxies, most of which have poorly constrained parameters. The black curve in the middle panel of \autoref{fig:bayes_eg} is an example of a poorly constrained excitation temperature likelihood function. By combining information from the black curve with information from the better-constrained red curve and many others, we can infer a prior over excitation temperatures. This prior is shown as a dashed grey curve in the third panel of the same figure.

The hierarchical Bayesian modelling method requires that we assume a functional form for the sample-level distribution. We assume the distribution of $T_\mathrm{exc}$ within each sample is a Gaussian with mean and standard deviation $T_\mathrm{exc,mean}$ and $T_\mathrm{exc,\sigma}$. If the $T_\mathrm{exc}$ of each galaxy in a sample were known to infinite precision, the probability of a ($T_\mathrm{exc, mean}$, $T_\mathrm{exc, \sigma}$) pair would be the product of a normal distribution with mean $T_\mathrm{exc, mean}$ and standard deviation $T_\mathrm{exc, \sigma}$. Instead, for each galaxy in our sample we have a likelihood function $\mathcal L(T_\mathrm{exc})$ over all possible values of $T_\mathrm{exc}$. The probability of a ($T_\mathrm{exc, mean}$, $T_\mathrm{exc, \sigma}$) pair as determined from the spectrum of a single galaxy is now given by an integral over the product of that galaxy's $T_\mathrm{exc}$ likelihood function and the (normal) distribution of $T_\mathrm{exc}$ values in our sample:
\begin{equation}
\int_{0}^{\infty} \frac{1}{\sqrt{2 \pi {T}_{\mathrm{exc}, \sigma}^2}} 
\exp \left[ {- \frac{ \left({T}_\mathrm{exc, mean} - {T}_\mathrm{exc} \right)^2}{2 {T}_\mathrm{exc, \sigma}^2}} \right]
\mathcal L({T}_\mathrm{exc}) \, \mathrm{d} {T}_\mathrm{exc}.
\end{equation}
The probability of a specific ($T_\mathrm{exc, mean}$, $T_\mathrm{exc, \sigma}$) determined from all the galaxies in our sample is the product of that integral evaluated for each galaxy. 
Our inference consists of mapping out the probability of $T_\mathrm{exc, mean}$ and $T_\mathrm{exc, \sigma}$ given the spectra in each sample.

\begin{figure}
\begin{center}
\includegraphics[width=\columnwidth]{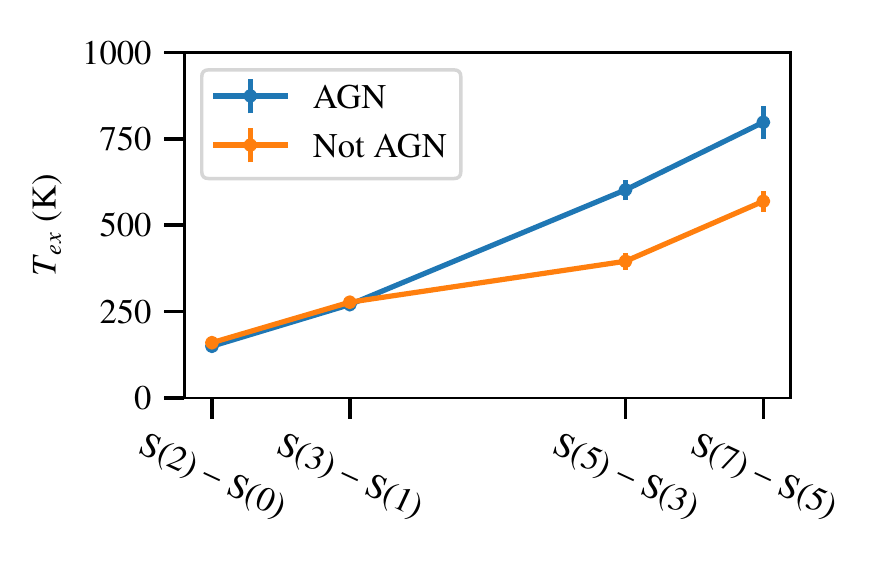}
	\caption{Method (2B) - Excitation temperature differences between AGN an non-AGN: We use a hierarchical Bayesian model to compute the excitation temperature distribution for the pairs of transitions listed in the figure within the AGN, not-AGN dominated subsamples. We find that the mean excitation temperatures of the higher \molh\ transitions distributions are higher in sources with an AGN than in sources without an AGN. The blue line represents the subsample with \eqwpah\ $< 0.27$ \micronm\ and the orange line shows all other sources. The points represent the expectation values for each distribution with one standard-deviation error-bars.}
\label{fig:bayes_results}
\end{center}
\end{figure}

We use MCMC with the \textsc{emcee} implementation of the affine invariant ensemble sampler \citep{emcee} to estimate the expectation value and standard deviation of $T_\mathrm{exc, mean}$ for each pair of transitions (\autoref{fig:bayes_results}). The excitation temperatures between the \molh S(2), \molh S(0) and \molh S(3), \molh S(1) energy levels are the same in both samples while the excitation temperatures between the higher-energy \molh S(5), \molh S(3) and \molh S(7), \molh S(5) energy levels are significantly higher in AGN dominated galaxies than in star-formation dominated galaxies.

Comparing the T$_{ex}$ values derived in (2A) and (2B), we find the T$_{2,0}$ and T$_{3,1}$ are similar between the two approaches. However the values of T$_{5,3}$ and T$_{7,5}$ from (2A) are lower than those derived from the (2B) analysis. This is potentially due to the fact that (2B) is less sensitive to noisy data than (2A) in two ways. First, the (2A) means are unweighted averages of noisy quantities. If the noise results in a tendency for positive bias in individual sources, the mean will be greater than is necessary. In (2B), the structure of the hierarchical model gives each object an effective "weight" that is proportional to how uncertain that object's T$_{ex}$ is. This weight is an entire distribution, because it includes information about the shape of p(T$_{ex}$) --- if there's asymmetry, extended tails, etc. Second, because the model is hierarchical, there is an informative prior for each individual object's T$_{ex}$. The information in this prior comes from all of the objects which increases the precision of the T$_{ex}$ model. Thus, the values derived from (2B) are more statistically reliable.

\section{Discussion} \label{sec:discussion}

We find an excess of \molh\ emission and a statistically significant temperature difference in the warmer gas component in AGN dominated galaxies. We find AGN dominated host galaxies on average have at least 5 times greater \molh\ to PAH ratios than SF-dominated host galaxies. In order to understand the nature of this excess \molh\ emission, we calculate the temperature of the warm \molh\ gas using two different approaches: (1) using all of the \molh\ lines simultaneously to determine a warm and warmer temperature component and (2) calculating excitation temperatures of \molh\ line pairs of equal parity. The two-temperature decomposition via likelihood analysis method shows no statistical difference between the AGN and not-AGN dominated sub-samples for the warm component, but the warmer component shows a $1\sigma$ 200 K median temperature difference in the AGN dominated and not-AGN dominated sub-samples. For (2A), the unstacked, stacked spectra show a roughly $2\sigma$ $T_\mathrm{exc, mean}$ difference for $(u,l) = (5,3)$ of 175.0~K, 210.0~K respectively. The unstacked spectra also show a roughly $2\sigma$ 276.0~K difference of $T_\mathrm{exc, mean}$ for $(u,l) = (7,5)$ between the AGN, not-AGN dominated sub-samples. Method (2B), the hierarchical Bayesian model, shows a roughly $2\sigma$ $T_\mathrm{exc, mean}$ difference for $(u,l) = 5,3$, $(u,l) = 7,5$  of 120.0~K, 200.0~K between the AGN dominated and not-AGN dominated sub-samples, respectively.

The warm gas component (100 K -- 300 K), is dominated by SF processes in both local IR AGN and not-AGN dominated host galaxies \citep{rigopoulou,higdon,roussel}. This leads to a strong correlation between the amount of \molh\ and PAH emission and a nearly constant ratio between the two, as both are by-products of star-formation. \citet{petric18} use high-resolution IRS spectra to find a population of LIRGs with enhanced \molh\ and broader, spectrally resolved \molh\ lines. They speculate that the broader profiles are due to bulk flows associated with AGN or high-mass star-formation. They find that AGN appear to have warmer gas and dust than non-AGN. However, few of their spectra have detections higher than \molh S(3), so they are unable to conduct the same thorough analysis we perform here. In our sample of objects that have $2\sigma$ or greater detections of  \molh S(3), PAH[11.3~\micronm], and PAH[6.2~\micronm], we find the most star-formation dominated objects ({\eqwpah\ $> 1.0$ \micronm}, 134 objects) have \molh\ to PAH ratios of $0.016 \pm 0.001$. For objects that are still considered star-formation dominated (\eqwpah\ $> 0.54$~\micronm\ \& \eqwpah\ $< 1.0$~\micronm, 104 objects), we find \molh\ to PAH ratios of $0.050 \pm 0.002$. These values are consistent with \citet{roussel} and \citet{stierwalt} results for star-forming dominated galaxies.

In ULIRGs, different authors draw different conclusions about the origins of the excess \molh\ emission. Although we are not splitting our sample in bins of IR, up to 58 per cent of ULIRGs contain an AGN \citep{yuan10}. \citet{higdon} find that the masses of warm \molh\ in ULIRGs are not correlated with the AGN contribution to the MIR emission, so they suggest that in ULIRGs the warm \molh\ emission comes from PDRs. However, using the \molh\ to PAH ratio as an indicator for warm \molh\ excess, \citet{zakamska} and \citet{Hill} do find more \molh\ than is expected from star-formation alone. Observations of \molh\ in AGN host galaxies with radio jets suggest that kinetic energy dissipation by shocks or cosmic rays can produce a factor of 300 or larger \molh\ to PAH values than normal star-forming galaxies \citep{ogle}. \citet{stierwalt} find a trend of increasing \molh\ excess with decreasing PAH equivalent width, but they note the dispersion is large for the objects with the lowest \eqwpah\ where they don't have many sources. \citet{stierwalt} also find that the galaxies with the most extreme \molh\ excess are mid- to late- stage mergers. They suggest that this excess may associated with powerful starbursts, and that the \molh\ may be excited by turbulence and shocks present in star-forming systems. Here, with a larger sample, we are able to show that the anti-correlation between \eqwpah\ and \molh\/PAH is valid down to the lowest \eqwpah\ values in our sample and hence may be due to the AGN. Of the most AGN dominated objects (\eqwpah\ $< 0.1$ \micronm, 51 objects) we find \molh\ to PAH ratios of $0.130 \pm 0.007$. For objects that are still considered AGN dominated with \eqwpah\ $> 0.1$  \micronm\ \& \eqwpah\ $< 0.27$ \micronm\ (85 objects), we find \molh\ to PAH ratios of $0.090 \pm 0.008$. These values are consistent with \citet{Hill} results for AGN dominated galaxies.  

In the literature, there is a lack of association between the temperature of the warm \molh\ and AGN activity. The greatest potential observable effect on the \molh\ gas would be seen in the higher temperature transitions, since these transitions are more difficult to excite from SF processes. These transitions are also difficult to observe, and thus, methods that rely on high signal to noise fluxes will be less effective. Methods that find excitation temperatures without separating the different temperatures have the problem of different components contributing to the flux of a given transition. For transitions that are easily excited by multiple physical processes, a single temperature would be inaccurate, but it has been found that the warmer gas component contributes on average only a few percent compared to the warm component in galaxies where SF dominates \citep{higdon,roussel}. The \molh S(3), \molh S(5), and \molh S(7) transitions all constrain the warmer temperature component of the gas, and in particular \molh S(5), \molh S(7) transitions have relatively little contribution from the warm gas component. This motivates methods (2A) and (2B), where a single temperature is assumed, and we focus only on the difference between the higher temperature transitions. The results of (2B) show significant differences in $T_\mathrm{exc, mean}$ for $(u,l) = 5,3$, $(u,l) = 7,5$ between the AGN, not-AGN sub-samples. These higher transitions require higher excitation temperatures, and have higher critical densities \citep{neufeld}. Although we cannot completely rule out density effects, our results show an average 200~K temperature difference in the transitions the AGN are likely to affect most. In the remaining text, we postulate the origin of the excess emission.  

Observations of AGN-dominated ULIRGs show evidence of fast outflows of molecular gas that are spatially extended on kiloparsec scales \citep{feruglio10,spoon13,rupke2013a,cicone14, aalto15}. Outflows can deplete galaxies of their gas and quench star-formation on timescales of 10$^6$--10$^8$ years \citep{sturm11,liu2013,brusa2015,rupke2017}. The origin of these molecular outflows is unclear, but one possibility is that radiative winds launched from regions close to the AGN create the outflow. The winds can provide an efficient way to couple the AGN's energy to the ISM \citep{tombesi}. Large fractions of kinetic energy are deposited in the ISM and can accelerate existing molecular gas. The propagating outflow interacts with the ISM via heating, entraining, and/or shocking gas clouds in its path \citep{cecil02, rupke2017}. A wide range of physical conditions exists in the outflow. One of the phases of the outflow is partially ionized regions where [\ion{O}{i}]$\lambda 6300$~\AA\ and [\ion{Fe}{ii}]  emission lines are produced. 

In ULIRGs, \citet{Hill} find a strong correlation between warm \molh\ and the ionized gas, suggesting that the excess warm \molh\ is excited in the same regions of outflow driven shocks. An alternative possibility is that the molecules form in the material swept up by the wind. \citet{richings} simulated an AGN wind interacting with a uniform medium and explored the possibility of in-situ molecule formation. Using a radiative transfer model, they computed the amount of warm \molh\ emission. The \molh S(0), \molh S(1), \molh S(2), and \molh S(3) level populations derived from the in-situ molecule formation AGN wind model described in \citet{richings,richingsB} indicate excitation temperatures of 400--547~K. This agrees with our result that AGN dominated galaxies have a distinct population of \molh. AGN driven jets, shocks, and winds are not the only processes than can lead to the excitation of \molh\ in the warm phase. Shocks produced by young stars, supernovae, and galaxy collisions can also excite these transitions. Any shock-producing mechanism related to solely SF processes will also occur in our SF dominated sub-sample. Thus, we postulate that the warm excess \molh\ emission and temperature difference is a direct consequence of AGN activity.

\section{Summary and Conclusion} \label{sec:conclusion}

We use MIR spectroscopy to evaluate the relationship between AGN and the ISM of their host galaxies. 
We analyse 2,015 objects low-resolution spectra \citep{cassis} with published spectroscopic redshifts \citep{ideos}. 
We correct mismatches between the different spectral orders and check the flux calibration of the spectra using \WISE\ photometry. We measure rotational \molh\ transitions, PAH emission in the 6.2, 7.7, and 11.3~\micronm\ bands, and summarize our results as follows:  

\begin{enumerate}
\item  We use the \eqwpah\ to separate our sample galaxies where the AGN 
contributes more than 50 per cent of the MIR luminosity and galaxies where star-formation contributes more than 50 per cent of the MIR luminosity.  
\item We find that the PAHs in AGN-dominated galaxies have a wider range of ionizations and sizes, and the effect of silicate absorption on grain size tracers is different for AGN dominated galaxies vs SF dominated galaxies. This may imply that the ISM in AGN hosts is more complex than the ISM of SF-dominated galaxies; without analysing the host morphologies we cannot separate the impact of the AGN on the ISM from that of any gravitational interactions. 
\item In AGN-dominated systems, we find an excess of molecular \molh\ emission relative to what we would measure if the molecular \molh\ originated solely from PDRs.
\item We assess the properties of the warm molecular gas applying Bayesian interpretations of the most commonly implemented techniques, and find statistically different temperature distributions between MIR SF dominated (\eqwpah\ > 0.54 \micronm) and MIR AGN dominated (\eqwpah\ < 0.27 \micronm) galaxies.  
\item We construct a hierarchical Bayesian model, and we find a 120~K temperature difference in $T_{5,3}$ between AGN-dominated galaxies and SF-dominated targets with greater than 3$\sigma$ significance. The difference in $T_{7,5}$ between these targets increases to 200 K with greater than 3$\sigma$ significance. This may suggest that the AGN heats the molecular gas in the inner $\sim5$ kpc probed by the IRS observations.

\end{enumerate}

\section*{Acknowledgements}
We thank D.A. Neufeld, T. Heckman, N. Flagey, K.R. Hall, J. Hamer, A. de la Vega, and R.C. Simons for useful discussions and insight. ELL is supported by the Maryland Space Grant Consortium. NLZ acknowledges support by Johns Hopkins University through the Catalyst Award and by the Institute for Advanced Study through the Deborah Lunder and Alan Ezekowitz Founders' Circle Membership. This research has made use of the NASA/IPAC Infrared Science Archive, which is operated by the Jet Propulsion Laboratory, California Institute of Technology, under contract with the National Aeronautics and Space Administration. This publication makes use of data products from the \textit{Wide-field Infrared Survey Explorer}, which is a joint project of the University of California, Los Angeles, and the Jet Propulsion Laboratory/California Institute of Technology, funded by the National Aeronautics and Space Administration. This publication makes use of data products from the Two Micron All Sky Survey, which is a joint project of the University of Massachusetts and the Infrared Processing and Analysis Center/California Institute of Technology, funded by the National Aeronautics and Space Administration and the National Science Foundation. We acknowledge the extensive use of the following Python packages: \textsc{pandas, scipy, ipython, matplotlib, sci-kit learn, emcee} \citep[respectively]{pandas, scipy,ipython,matplotlib,sci-kit,emcee}. This research made use of \textsc{astropy}, a community-developed core Python package for Astronomy \citep{astropy}.

\appendix

\bibliographystyle{mnras}
\bibliography{draft}

\begin{thebibliography}{}
\makeatletter
\relax
\def\mn@urlcharsother{\let\do\@makeother \do\$\do\&\do\#\do\^\do\_\do\%\do\~}
\def\mn@doi{\begingroup\mn@urlcharsother \@ifnextchar [ {\mn@doi@}
  {\mn@doi@[]}}
\def\mn@doi@[#1]#2{\def\@tempa{#1}\ifx\@tempa\@empty \href
  {http://dx.doi.org/#2} {doi:#2}\else \href {http://dx.doi.org/#2} {#1}\fi
  \endgroup}
\def\mn@eprint#1#2{\mn@eprint@#1:#2::\@nil}
\def\mn@eprint@arXiv#1{\href {http://arxiv.org/abs/#1} {{\tt arXiv:#1}}}
\def\mn@eprint@dblp#1{\href {http://dblp.uni-trier.de/rec/bibtex/#1.xml}
  {dblp:#1}}
\def\mn@eprint@#1:#2:#3:#4\@nil{\def\@tempa {#1}\def\@tempb {#2}\def\@tempc
  {#3}\ifx \@tempc \@empty \let \@tempc \@tempb \let \@tempb \@tempa \fi \ifx
  \@tempb \@empty \def\@tempb {arXiv}\fi \@ifundefined
  {mn@eprint@\@tempb}{\@tempb:\@tempc}{\expandafter \expandafter \csname
  mn@eprint@\@tempb\endcsname \expandafter{\@tempc}}}

\bibitem[\protect\citeauthoryear{{Aalto} et~al.,}{{Aalto}
  et~al.}{2015}]{aalto15}
{Aalto} S.,  et~al., 2015, \mn@doi [\aap] {10.1051/0004-6361/201526410}, \href
  {http://adsabs.harvard.edu/abs/2015A%26A...584A..42A} {584, A42}

\bibitem[\protect\citeauthoryear{{Allamandola}, {Tielens}  \&
  {Barker}}{{Allamandola} et~al.}{1989}]{allamandola}
{Allamandola} L.~J.,  {Tielens} A.~G.~G.~M.,   {Barker} J.~R.,  1989, \mn@doi
  [\apjs] {10.1086/191396}, \href
  {http://adsabs.harvard.edu/abs/1989ApJS...71..733A} {71, 733}

\bibitem[\protect\citeauthoryear{{Appleton} et~al.,}{{Appleton}
  et~al.}{2006}]{appleton}
{Appleton} P.~N.,  et~al., 2006, \mn@doi [\apjl] {10.1086/502646}, \href
  {http://adsabs.harvard.edu/abs/2006ApJ...639L..51A} {639, L51}

\bibitem[\protect\citeauthoryear{{Armus} et~al.,}{{Armus}
  et~al.}{2006}]{armus06}
{Armus} L.,  et~al., 2006, \mn@doi [\apj] {10.1086/500040}, \href
  {http://adsabs.harvard.edu/abs/2006ApJ...640..204A} {640, 204}

\bibitem[\protect\citeauthoryear{{Armus} et~al.,}{{Armus}
  et~al.}{2007a}]{armus07}
{Armus} L.,  et~al., 2007a, \mn@doi [\apj] {10.1086/510107}, \href
  {http://adsabs.harvard.edu/abs/2007ApJ...656..148A} {656, 148}

\bibitem[\protect\citeauthoryear{{Armus} et~al.,}{{Armus}
  et~al.}{2007b}]{armus}
{Armus} L.,  et~al., 2007b, \mn@doi [\apj] {10.1086/510107}, \href
  {http://adsabs.harvard.edu/abs/2007ApJ...656..148A} {656, 148}

\bibitem[\protect\citeauthoryear{{Assef} et~al.,}{{Assef} et~al.}{2013}]{assef}
{Assef} R.~J.,  et~al., 2013, \mn@doi [\apj] {10.1088/0004-637X/772/1/26},
  \href {http://adsabs.harvard.edu/abs/2013ApJ...772...26A} {772, 26}

\bibitem[\protect\citeauthoryear{{Assef}, {Stern}, {Noirot}, {Jun}, {Cutri}  \&
  {Eisenhardt}}{{Assef} et~al.}{2018}]{assef2017}
{Assef} R.~J.,  {Stern} D.,  {Noirot} G.,  {Jun} H.~D.,  {Cutri} R.~M.,
  {Eisenhardt} P.~R.~M.,  2018, \mn@doi [\apjs] {10.3847/1538-4365/aaa00a},
  \href {http://adsabs.harvard.edu/abs/2018ApJS..234...23A} {234, 23}

\bibitem[\protect\citeauthoryear{{Astropy Collaboration} et~al.,}{{Astropy
  Collaboration} et~al.}{2013}]{astropy}
{Astropy Collaboration} et~al., 2013, \mn@doi [\aap]
  {10.1051/0004-6361/201322068}, \href
  {http://adsabs.harvard.edu/abs/2013A\%26A...558A..33A} {558, A33}

\bibitem[\protect\citeauthoryear{{Baldwin}, {Phillips}  \&
  {Terlevich}}{{Baldwin} et~al.}{1981}]{baldwin}
{Baldwin} J.~A.,  {Phillips} M.~M.,   {Terlevich} R.,  1981, \mn@doi [\pasp]
  {10.1086/130766}, \href {http://adsabs.harvard.edu/abs/1981PASP...93....5B}
  {93, 5}

\bibitem[\protect\citeauthoryear{{Blecha}, {Snyder}, {Satyapal}  \&
  {Ellison}}{{Blecha} et~al.}{2018}]{blecha}
{Blecha} L.,  {Snyder} G.~F.,  {Satyapal} S.,   {Ellison} S.~L.,  2018, \mn@doi
  [\mnras] {10.1093/mnras/sty1274}, \href
  {http://adsabs.harvard.edu/abs/2018MNRAS.478.3056B} {478, 3056}

\bibitem[\protect\citeauthoryear{{Brandl} et~al.,}{{Brandl}
  et~al.}{2006}]{brandl}
{Brandl} B.~R.,  et~al., 2006, \mn@doi [\apj] {10.1086/508849}, \href
  {http://adsabs.harvard.edu/abs/2006ApJ...653.1129B} {653, 1129}

\bibitem[\protect\citeauthoryear{{Brusa} et~al.,}{{Brusa}
  et~al.}{2015}]{brusa2015}
{Brusa} M.,  et~al., 2015, \mn@doi [\mnras] {10.1093/mnras/stu2117}, \href
  {http://adsabs.harvard.edu/abs/2015MNRAS.446.2394B} {446, 2394}

\bibitem[\protect\citeauthoryear{{Buat} \& {Deharveng}}{{Buat} \&
  {Deharveng}}{1988}]{buat}
{Buat} V.,  {Deharveng} J.~M.,  1988, \aap, \href
  {http://adsabs.harvard.edu/abs/1988A%26A...195...60B} {195, 60}

\bibitem[\protect\citeauthoryear{{Burton}, {Hollenbach}  \& {Tielens}}{{Burton}
  et~al.}{1992}]{burton}
{Burton} M.~G.,  {Hollenbach} D.~J.,   {Tielens} A.~G.~G.,  1992, \mn@doi
  [\apj] {10.1086/171947}, \href
  {http://adsabs.harvard.edu/abs/1992ApJ...399..563B} {399, 563}

\bibitem[\protect\citeauthoryear{{Calzetti} et~al.,}{{Calzetti}
  et~al.}{2007}]{calzetti}
{Calzetti} D.,  et~al., 2007, \mn@doi [\apj] {10.1086/520082}, \href
  {http://adsabs.harvard.edu/abs/2007ApJ...666..870C} {666, 870}

\bibitem[\protect\citeauthoryear{{Cecil}, {Bland-Hawthorn}  \&
  {Veilleux}}{{Cecil} et~al.}{2002}]{cecil02}
{Cecil} G.,  {Bland-Hawthorn} J.,   {Veilleux} S.,  2002, \mn@doi [\apj]
  {10.1086/341861}, \href {http://adsabs.harvard.edu/abs/2002ApJ...576..745C}
  {576, 745}

\bibitem[\protect\citeauthoryear{{Chilingarian}, {Melchior}  \&
  {Zolotukhin}}{{Chilingarian} et~al.}{2010}]{kcor}
{Chilingarian} I.~V.,  {Melchior} A.-L.,   {Zolotukhin} I.~Y.,  2010, \mn@doi
  [\mnras] {10.1111/j.1365-2966.2010.16506.x}, \href
  {http://adsabs.harvard.edu/abs/2010MNRAS.405.1409C} {405, 1409}

\bibitem[\protect\citeauthoryear{{Cicone}, {Feruglio}, {Maiolino}, {Fiore},
  {Piconcelli}, {Menci}, {Aussel}  \& {Sturm}}{{Cicone}
  et~al.}{2012}]{cicone12}
{Cicone} C.,  {Feruglio} C.,  {Maiolino} R.,  {Fiore} F.,  {Piconcelli} E.,
  {Menci} N.,  {Aussel} H.,   {Sturm} E.,  2012, \mn@doi [\aap]
  {10.1051/0004-6361/201218793}, \href
  {https://ui.adsabs.harvard.edu/#abs/2012A&A...543A..99C} {543, A99}

\bibitem[\protect\citeauthoryear{{Cicone} et~al.,}{{Cicone}
  et~al.}{2014}]{cicone14}
{Cicone} C.,  et~al., 2014, \mn@doi [\aap] {10.1051/0004-6361/201322464}, \href
  {http://adsabs.harvard.edu/abs/2014A%26A...562A..21C} {562, A21}

\bibitem[\protect\citeauthoryear{{Cluver} et~al.,}{{Cluver}
  et~al.}{2010}]{cluver}
{Cluver} M.~E.,  et~al., 2010, \mn@doi [\apj] {10.1088/0004-637X/710/1/248},
  \href {http://adsabs.harvard.edu/abs/2010ApJ...710..248C} {710, 248}

\bibitem[\protect\citeauthoryear{{Cresci} et~al.,}{{Cresci}
  et~al.}{2015}]{cresci}
{Cresci} G.,  et~al., 2015, \mn@doi [\aap] {10.1051/0004-6361/201526581}, \href
  {http://adsabs.harvard.edu/abs/2015A\%26A...582A..63C} {582, A63}

\bibitem[\protect\citeauthoryear{{Desai} et~al.,}{{Desai} et~al.}{2007}]{desai}
{Desai} V.,  et~al., 2007, \mn@doi [\apj] {10.1086/522104}, \href
  {http://adsabs.harvard.edu/abs/2007ApJ...669..810D} {669, 810}

\bibitem[\protect\citeauthoryear{{Diamond-Stanic} \& {Rieke}}{{Diamond-Stanic}
  \& {Rieke}}{2010}]{diamond}
{Diamond-Stanic} A.~M.,  {Rieke} G.~H.,  2010, \mn@doi [\apj]
  {10.1088/0004-637X/724/1/140}, \href
  {http://adsabs.harvard.edu/abs/2010ApJ...724..140D} {724, 140}

\bibitem[\protect\citeauthoryear{{Donley} et~al.,}{{Donley}
  et~al.}{2012}]{donley12}
{Donley} J.~L.,  et~al., 2012, \mn@doi [\apj] {10.1088/0004-637X/748/2/142},
  \href {http://adsabs.harvard.edu/abs/2012ApJ...748..142D} {748, 142}

\bibitem[\protect\citeauthoryear{{Draine}}{{Draine}}{2003}]{draine03}
{Draine} B.~T.,  2003, \mn@doi [\apj] {10.1086/379118}, \href
  {http://adsabs.harvard.edu/abs/2003ApJ...598.1017D} {598, 1017}

\bibitem[\protect\citeauthoryear{{Draine} \& {Li}}{{Draine} \&
  {Li}}{2007}]{draine07}
{Draine} B.~T.,  {Li} A.,  2007, \mn@doi [\apj] {10.1086/511055}, \href
  {http://adsabs.harvard.edu/abs/2007ApJ...657..810D} {657, 810}

\bibitem[\protect\citeauthoryear{{Eisenhardt} et~al.,}{{Eisenhardt}
  et~al.}{2012}]{eisenhardt}
{Eisenhardt} P.~R.~M.,  et~al., 2012, \mn@doi [\apj]
  {10.1088/0004-637X/755/2/173}, \href
  {http://adsabs.harvard.edu/abs/2012ApJ...755..173E} {755, 173}

\bibitem[\protect\citeauthoryear{{Elitzur}}{{Elitzur}}{2012}]{elitzur12}
{Elitzur} M.,  2012, \mn@doi [\apjl] {10.1088/2041-8205/747/2/L33}, \href
  {http://adsabs.harvard.edu/abs/2012ApJ...747L..33E} {747, L33}

\bibitem[\protect\citeauthoryear{{Elvis} et~al.,}{{Elvis} et~al.}{1994}]{elvis}
{Elvis} M.,  et~al., 1994, \mn@doi [\apjs] {10.1086/192093}, \href
  {http://adsabs.harvard.edu/abs/1994ApJS...95....1E} {95, 1}

\bibitem[\protect\citeauthoryear{{Fabian}}{{Fabian}}{1999}]{fabian99}
{Fabian} A.~C.,  1999, \mn@doi [\mnras] {10.1046/j.1365-8711.1999.03017.x},
  \href {http://adsabs.harvard.edu/abs/1999MNRAS.308L..39F} {308, L39}

\bibitem[\protect\citeauthoryear{{Fabian}}{{Fabian}}{2012}]{fabian12}
{Fabian} A.~C.,  2012, \mn@doi [\araa] {10.1146/annurev-astro-081811-125521},
  \href {http://adsabs.harvard.edu/abs/2012ARA\%26A..50..455F} {50, 455}

\bibitem[\protect\citeauthoryear{{Feruglio}, {Maiolino}, {Piconcelli}, {Menci},
  {Aussel}, {Lamastra}  \& {Fiore}}{{Feruglio} et~al.}{2010}]{feruglio10}
{Feruglio} C.,  {Maiolino} R.,  {Piconcelli} E.,  {Menci} N.,  {Aussel} H.,
  {Lamastra} A.,   {Fiore} F.,  2010, \mn@doi [\aap]
  {10.1051/0004-6361/201015164}, \href
  {http://adsabs.harvard.edu/abs/2010A%26A...518L.155F} {518, L155}

\bibitem[\protect\citeauthoryear{{Foreman-Mackey}, {Hogg}, {Lang}  \&
  {Goodman}}{{Foreman-Mackey} et~al.}{2013}]{emcee}
{Foreman-Mackey} D.,  {Hogg} D.~W.,  {Lang} D.,   {Goodman} J.,  2013, \mn@doi
  [\pasp] {10.1086/670067}, \href
  {http://adsabs.harvard.edu/abs/2013PASP..125..306F} {125, 306}

\bibitem[\protect\citeauthoryear{Gelman, Carlin, Stern, Dunson, Vehtari  \&
  Rubin}{Gelman et~al.}{2013}]{gelman}
Gelman A.,  Carlin J.~B.,  Stern H.~S.,  Dunson D.~B.,  Vehtari A.,   Rubin
  D.~B.,  2013, Bayesian Data Analysis, Third Edition (Chapman \& Hall/CRC
  Texts in Statistical Science).
Chapman and Hall/CRC, \url {http://www.stat.columbia.edu/~gelman/book/}

\bibitem[\protect\citeauthoryear{{Genzel} et~al.,}{{Genzel}
  et~al.}{1998}]{genzel}
{Genzel} R.,  et~al., 1998, \mn@doi [\apj] {10.1086/305576}, \href
  {http://adsabs.harvard.edu/abs/1998ApJ...498..579G} {498, 579}

\bibitem[\protect\citeauthoryear{{Greene}, {Zakamska}, {Ho}  \&
  {Barth}}{{Greene} et~al.}{2011}]{greene2011}
{Greene} J.~E.,  {Zakamska} N.~L.,  {Ho} L.~C.,   {Barth} A.~J.,  2011, \mn@doi
  [\apj] {10.1088/0004-637X/732/1/9}, \href
  {http://adsabs.harvard.edu/abs/2011ApJ...732....9G} {732, 9}

\bibitem[\protect\citeauthoryear{{Greene}, {Zakamska}  \& {Smith}}{{Greene}
  et~al.}{2012}]{greene2012}
{Greene} J.~E.,  {Zakamska} N.~L.,   {Smith} P.~S.,  2012, \mn@doi [\apj]
  {10.1088/0004-637X/746/1/86}, \href
  {http://adsabs.harvard.edu/abs/2012ApJ...746...86G} {746, 86}

\bibitem[\protect\citeauthoryear{{Guillard} et~al.,}{{Guillard}
  et~al.}{2012}]{guillard}
{Guillard} P.,  et~al., 2012, \mn@doi [\apj] {10.1088/0004-637X/749/2/158},
  \href {http://adsabs.harvard.edu/abs/2012ApJ...749..158G} {749, 158}

\bibitem[\protect\citeauthoryear{{G{\"u}rkan}, {Hardcastle}  \&
  {Jarvis}}{{G{\"u}rkan} et~al.}{2014}]{gurkan}
{G{\"u}rkan} G.,  {Hardcastle} M.~J.,   {Jarvis} M.~J.,  2014, \mn@doi [\mnras]
  {10.1093/mnras/stt2264}, \href
  {http://adsabs.harvard.edu/abs/2014MNRAS.438.1149G} {438, 1149}

\bibitem[\protect\citeauthoryear{{Haan} et~al.,}{{Haan} et~al.}{2011}]{haan}
{Haan} S.,  et~al., 2011, \mn@doi [\apjs] {10.1088/0067-0049/197/2/27}, \href
  {http://adsabs.harvard.edu/abs/2011ApJS..197...27H} {197, 27}

\bibitem[\protect\citeauthoryear{{Heckman} \& {Best}}{{Heckman} \&
  {Best}}{2014}]{heckmanaa}
{Heckman} T.~M.,  {Best} P.~N.,  2014, \mn@doi [\araa]
  {10.1146/annurev-astro-081913-035722}, \href
  {http://adsabs.harvard.edu/abs/2014ARA\%26A..52..589H} {52, 589}

\bibitem[\protect\citeauthoryear{{Hern{\'a}n-Caballero}, {Spoon},
  {Lebouteiller}, {Rupke}  \& {Barry}}{{Hern{\'a}n-Caballero}
  et~al.}{2016}]{ideos}
{Hern{\'a}n-Caballero} A.,  {Spoon} H.~W.~W.,  {Lebouteiller} V.,  {Rupke}
  D.~S.~N.,   {Barry} D.~P.,  2016, \mn@doi [\mnras] {10.1093/mnras/stv2464},
  \href {http://adsabs.harvard.edu/abs/2016MNRAS.455.1796H} {455, 1796}

\bibitem[\protect\citeauthoryear{{Higdon}, {Armus}, {Higdon}, {Soifer}  \&
  {Spoon}}{{Higdon} et~al.}{2006}]{higdon}
{Higdon} S.~J.~U.,  {Armus} L.,  {Higdon} J.~L.,  {Soifer} B.~T.,   {Spoon}
  H.~W.~W.,  2006, \mn@doi [\apj] {10.1086/505701}, \href
  {http://adsabs.harvard.edu/abs/2006ApJ...648..323H} {648, 323}

\bibitem[\protect\citeauthoryear{{Hill} \& {Zakamska}}{{Hill} \&
  {Zakamska}}{2014}]{Hill}
{Hill} M.~J.,  {Zakamska} N.~L.,  2014, \mn@doi [\mnras]
  {10.1093/mnras/stu123}, \href
  {http://adsabs.harvard.edu/abs/2014MNRAS.439.2701H} {439, 2701}

\bibitem[\protect\citeauthoryear{{Hogg}, {Myers}  \& {Bovy}}{{Hogg}
  et~al.}{2010}]{hogg}
{Hogg} D.~W.,  {Myers} A.~D.,   {Bovy} J.,  2010, \mn@doi [\apj]
  {10.1088/0004-637X/725/2/2166}, \href
  {http://adsabs.harvard.edu/abs/2010ApJ...725.2166H} {725, 2166}

\bibitem[\protect\citeauthoryear{{Hollenbach} \& {Tielens}}{{Hollenbach} \&
  {Tielens}}{1999}]{hollenbach}
{Hollenbach} D.~J.,  {Tielens} A.~G.~G.~M.,  1999, \mn@doi [Reviews of Modern
  Physics] {10.1103/RevModPhys.71.173}, \href
  {http://adsabs.harvard.edu/abs/1999RvMP...71..173H} {71, 173}

\bibitem[\protect\citeauthoryear{{Hopkins}, {Hernquist}, {Cox}, {Di Matteo},
  {Robertson}  \& {Springel}}{{Hopkins} et~al.}{2006}]{hopkins06}
{Hopkins} P.~F.,  {Hernquist} L.,  {Cox} T.~J.,  {Di Matteo} T.,  {Robertson}
  B.,   {Springel} V.,  2006, \mn@doi [\apjs] {10.1086/499298}, \href
  {http://adsabs.harvard.edu/abs/2006ApJS..163....1H} {163, 1}

\bibitem[\protect\citeauthoryear{{Houck} et~al.,}{{Houck} et~al.}{2004}]{irs}
{Houck} J.~R.,  et~al., 2004, \mn@doi [\apjs] {10.1086/423134}, \href
  {http://adsabs.harvard.edu/abs/2004ApJS..154...18H} {154, 18}

\bibitem[\protect\citeauthoryear{Hunter}{Hunter}{2007}]{matplotlib}
Hunter J.~D.,  2007, \mn@doi [Computing In Science \& Engineering]
  {10.1109/MCSE.2007.55}, 9, 90

\bibitem[\protect\citeauthoryear{{Jarrett} et~al.,}{{Jarrett}
  et~al.}{2011}]{jarrett}
{Jarrett} T.~H.,  et~al., 2011, \mn@doi [\apj] {10.1088/0004-637X/735/2/112},
  \href {http://adsabs.harvard.edu/abs/2011ApJ...735..112J} {735, 112}

\bibitem[\protect\citeauthoryear{{Jensen} et~al.,}{{Jensen}
  et~al.}{2017}]{jensen}
{Jensen} J.~J.,  et~al., 2017, \mn@doi [\mnras] {10.1093/mnras/stx1447}, \href
  {https://ui.adsabs.harvard.edu/#abs/2017MNRAS.470.3071J} {470, 3071}

\bibitem[\protect\citeauthoryear{Jones, Oliphant, Peterson  et~al.}{Jones
  et~al.}{01  }]{scipy}
Jones E.,  Oliphant T.,  Peterson P.,   et~al., 2001--, {SciPy}: Open source
  scientific tools for {Python}, \url {http://www.scipy.org/}

\bibitem[\protect\citeauthoryear{{Karouzos}, {Woo}  \& {Bae}}{{Karouzos}
  et~al.}{2016}]{karouzos}
{Karouzos} M.,  {Woo} J.-H.,   {Bae} H.-J.,  2016, \mn@doi [\apj]
  {10.3847/1538-4357/833/2/171}, \href
  {http://adsabs.harvard.edu/abs/2016ApJ...833..171K} {833, 171}

\bibitem[\protect\citeauthoryear{{Kauffmann} et~al.,}{{Kauffmann}
  et~al.}{2003}]{kauffmann}
{Kauffmann} G.,  et~al., 2003, \mn@doi [\mnras]
  {10.1111/j.1365-2966.2003.07154.x}, \href
  {http://adsabs.harvard.edu/abs/2003MNRAS.346.1055K} {346, 1055}

\bibitem[\protect\citeauthoryear{{King}}{{King}}{2003}]{king03}
{King} A.,  2003, \mn@doi [\apjl] {10.1086/379143}, \href
  {http://adsabs.harvard.edu/abs/2003ApJ...596L..27K} {596, L27}

\bibitem[\protect\citeauthoryear{{Kormendy} \& {Ho}}{{Kormendy} \&
  {Ho}}{2013}]{kormendyaa}
{Kormendy} J.,  {Ho} L.~C.,  2013, \mn@doi [\araa]
  {10.1146/annurev-astro-082708-101811}, \href
  {http://adsabs.harvard.edu/abs/2013ARA\%26A..51..511K} {51, 511}

\bibitem[\protect\citeauthoryear{{Lacy} et~al.,}{{Lacy} et~al.}{2004}]{lacy04}
{Lacy} M.,  et~al., 2004, \mn@doi [\apjs] {10.1086/422816}, \href
  {http://adsabs.harvard.edu/abs/2004ApJS..154..166L} {154, 166}

\bibitem[\protect\citeauthoryear{{Lacy}, {Petric}, {Sajina}, {Canalizo},
  {Storrie-Lombardi}, {Armus}, {Fadda}  \& {Marleau}}{{Lacy}
  et~al.}{2007}]{lacy07}
{Lacy} M.,  {Petric} A.~O.,  {Sajina} A.,  {Canalizo} G.,  {Storrie-Lombardi}
  L.~J.,  {Armus} L.,  {Fadda} D.,   {Marleau} F.~R.,  2007, \mn@doi [\aj]
  {10.1086/509617}, \href {http://adsabs.harvard.edu/abs/2007AJ....133..186L}
  {133, 186}

\bibitem[\protect\citeauthoryear{{Lacy}, {Ridgway}, {Sajina}, {Petric},
  {Gates}, {Urrutia}  \& {Storrie-Lombardi}}{{Lacy} et~al.}{2015}]{lacy15}
{Lacy} M.,  {Ridgway} S.~E.,  {Sajina} A.,  {Petric} A.~O.,  {Gates} E.~L.,
  {Urrutia} T.,   {Storrie-Lombardi} L.~J.,  2015, \mn@doi [\apj]
  {10.1088/0004-637X/802/2/102}, \href
  {http://adsabs.harvard.edu/abs/2015ApJ...802..102L} {802, 102}

\bibitem[\protect\citeauthoryear{{Laurent}, {Mirabel}, {Charmandaris},
  {Gallais}, {Madden}, {Sauvage}, {Vigroux}  \& {Cesarsky}}{{Laurent}
  et~al.}{2000}]{laurent}
{Laurent} O.,  {Mirabel} I.~F.,  {Charmandaris} V.,  {Gallais} P.,  {Madden}
  S.~C.,  {Sauvage} M.,  {Vigroux} L.,   {Cesarsky} C.,  2000, \aap, \href
  {http://adsabs.harvard.edu/abs/2000A\%26A...359..887L} {359, 887}

\bibitem[\protect\citeauthoryear{{Le Petit}, {Nehm{\'e}}, {Le Bourlot}  \&
  {Roueff}}{{Le Petit} et~al.}{2006}]{petit}
{Le Petit} F.,  {Nehm{\'e}} C.,  {Le Bourlot} J.,   {Roueff} E.,  2006, \mn@doi
  [\apjs] {10.1086/503252}, \href
  {http://adsabs.harvard.edu/abs/2006ApJS..164..506L} {164, 506}

\bibitem[\protect\citeauthoryear{{Lebouteiller}, {Barry}, {Spoon},
  {Bernard-Salas}, {Sloan}, {Houck}  \& {Weedman}}{{Lebouteiller}
  et~al.}{2011}]{cassis}
{Lebouteiller} V.,  {Barry} D.~J.,  {Spoon} H.~W.~W.,  {Bernard-Salas} J.,
  {Sloan} G.~C.,  {Houck} J.~R.,   {Weedman} D.~W.,  2011, \mn@doi [\apjs]
  {10.1088/0067-0049/196/1/8}, \href
  {http://adsabs.harvard.edu/abs/2011ApJS..196....8L} {196, 8}

\bibitem[\protect\citeauthoryear{{Leger}, {D'Hendecourt}  \&
  {Defourneau}}{{Leger} et~al.}{1989}]{leger}
{Leger} A.,  {D'Hendecourt} L.,   {Defourneau} D.,  1989, \aap, \href
  {http://adsabs.harvard.edu/abs/1989A\%26A...216..148L} {216, 148}

\bibitem[\protect\citeauthoryear{{Li} \& {Draine}}{{Li} \&
  {Draine}}{2001}]{li2001}
{Li} A.,  {Draine} B.~T.,  2001, \mn@doi [\apj] {10.1086/323147}, \href
  {http://adsabs.harvard.edu/abs/2001ApJ...554..778L} {554, 778}

\bibitem[\protect\citeauthoryear{{Liu}, {Zakamska}, {Greene}, {Nesvadba}  \&
  {Liu}}{{Liu} et~al.}{2013}]{liu2013}
{Liu} G.,  {Zakamska} N.~L.,  {Greene} J.~E.,  {Nesvadba} N.~P.~H.,   {Liu} X.,
   2013, \mn@doi [\mnras] {10.1093/mnras/stt1755}, \href
  {http://adsabs.harvard.edu/abs/2013MNRAS.436.2576L} {436, 2576}

\bibitem[\protect\citeauthoryear{{Madau} \& {Dickinson}}{{Madau} \&
  {Dickinson}}{2014}]{madau}
{Madau} P.,  {Dickinson} M.,  2014, \mn@doi [\araa]
  {10.1146/annurev-astro-081811-125615}, \href
  {http://adsabs.harvard.edu/abs/2014ARA\%26A..52..415M} {52, 415}

\bibitem[\protect\citeauthoryear{{Maddox}, {Hewett}, {Warren}  \&
  {Croom}}{{Maddox} et~al.}{2008}]{maddox}
{Maddox} N.,  {Hewett} P.~C.,  {Warren} S.~J.,   {Croom} S.~M.,  2008, \mn@doi
  [\mnras] {10.1111/j.1365-2966.2008.13138.x}, \href
  {http://adsabs.harvard.edu/abs/2008MNRAS.386.1605M} {386, 1605}

\bibitem[\protect\citeauthoryear{{Markwardt}}{{Markwardt}}{2009}]{mpfitfun}
{Markwardt} C.~B.,  2009, in {Bohlender} D.~A.,  {Durand} D.,   {Dowler} P.,
  eds,  Astronomical Society of the Pacific Conference Series Vol. 411,
  Astronomical Data Analysis Software and Systems XVIII. p.~251 (\mn@eprint
  {arXiv} {0902.2850})

\bibitem[\protect\citeauthoryear{{Marshall}, {Herter}, {Armus}, {Charmandaris},
  {Spoon}, {Bernard-Salas}  \& {Houck}}{{Marshall} et~al.}{2007}]{cafe}
{Marshall} J.~A.,  {Herter} T.~L.,  {Armus} L.,  {Charmandaris} V.,  {Spoon}
  H.~W.~W.,  {Bernard-Salas} J.,   {Houck} J.~R.,  2007, \mn@doi [\apj]
  {10.1086/521588}, \href {http://adsabs.harvard.edu/abs/2007ApJ...670..129M}
  {670, 129}

\bibitem[\protect\citeauthoryear{{Mart{\'{\i}}nez-Sansigre}, {Rawlings},
  {Lacy}, {Fadda}, {Marleau}, {Simpson}, {Willott}  \&
  {Jarvis}}{{Mart{\'{\i}}nez-Sansigre} et~al.}{2005}]{martinez05}
{Mart{\'{\i}}nez-Sansigre} A.,  {Rawlings} S.,  {Lacy} M.,  {Fadda} D.,
  {Marleau} F.~R.,  {Simpson} C.,  {Willott} C.~J.,   {Jarvis} M.~J.,  2005,
  \mn@doi [\nat] {10.1038/nature03829}, \href
  {http://adsabs.harvard.edu/abs/2005Natur.436..666M} {436, 666}

\bibitem[\protect\citeauthoryear{{Mateos} et~al.,}{{Mateos}
  et~al.}{2012}]{mateos12}
{Mateos} S.,  et~al., 2012, \mn@doi [\mnras]
  {10.1111/j.1365-2966.2012.21843.x}, \href
  {http://adsabs.harvard.edu/abs/2012MNRAS.426.3271M} {426, 3271}

\bibitem[\protect\citeauthoryear{McKinney}{McKinney}{2010}]{pandas}
McKinney W.,  2010, in van~der Walt S.,  Millman J.,  eds, Proceedings of the
  9th Python in Science Conference. pp 51 -- 56

\bibitem[\protect\citeauthoryear{{Nenkova}, {Sirocky}, {Nikutta}, {Ivezi{\'c}}
  \& {Elitzur}}{{Nenkova} et~al.}{2008}]{nenkova}
{Nenkova} M.,  {Sirocky} M.~M.,  {Nikutta} R.,  {Ivezi{\'c}} {\v Z}.,
  {Elitzur} M.,  2008, \mn@doi [\apj] {10.1086/590483}, \href
  {http://adsabs.harvard.edu/abs/2008ApJ...685..160N} {685, 160}

\bibitem[\protect\citeauthoryear{{Nesvadba}, {Polletta}, {Lehnert}, {Bergeron},
  {De Breuck}, {Lagache}  \& {Omont}}{{Nesvadba} et~al.}{2011}]{nesvadba}
{Nesvadba} N.~P.~H.,  {Polletta} M.,  {Lehnert} M.~D.,  {Bergeron} J.,  {De
  Breuck} C.,  {Lagache} G.,   {Omont} A.,  2011, \mn@doi [\mnras]
  {10.1111/j.1365-2966.2011.18862.x}, \href
  {http://adsabs.harvard.edu/abs/2011MNRAS.415.2359N} {415, 2359}

\bibitem[\protect\citeauthoryear{{Neufeld} et~al.,}{{Neufeld}
  et~al.}{2006}]{neufeld}
{Neufeld} D.~A.,  et~al., 2006, \mn@doi [\apj] {10.1086/506604}, \href
  {http://adsabs.harvard.edu/abs/2006ApJ...649..816N} {649, 816}

\bibitem[\protect\citeauthoryear{{O'Dowd} et~al.,}{{O'Dowd}
  et~al.}{2009}]{odowd}
{O'Dowd} M.~J.,  et~al., 2009, \mn@doi [\apj] {10.1088/0004-637X/705/1/885},
  \href {http://adsabs.harvard.edu/abs/2009ApJ...705..885O} {705, 885}

\bibitem[\protect\citeauthoryear{{Ogle}, {Boulanger}, {Guillard}, {Evans},
  {Antonucci}, {Appleton}, {Nesvadba}  \& {Leipski}}{{Ogle}
  et~al.}{2010}]{ogle}
{Ogle} P.,  {Boulanger} F.,  {Guillard} P.,  {Evans} D.~A.,  {Antonucci} R.,
  {Appleton} P.~N.,  {Nesvadba} N.,   {Leipski} C.,  2010, \mn@doi [\apj]
  {10.1088/0004-637X/724/2/1193}, \href
  {http://adsabs.harvard.edu/abs/2010ApJ...724.1193O} {724, 1193}

\bibitem[\protect\citeauthoryear{{Ogle}, {Davies}, {Appleton}, {Bertincourt},
  {Seymour}  \& {Helou}}{{Ogle} et~al.}{2012}]{ogle12}
{Ogle} P.,  {Davies} J.~E.,  {Appleton} P.~N.,  {Bertincourt} B.,  {Seymour}
  N.,   {Helou} G.,  2012, \mn@doi [\apj] {10.1088/0004-637X/751/1/13}, \href
  {http://adsabs.harvard.edu/abs/2012ApJ...751...13O} {751, 13}

\bibitem[\protect\citeauthoryear{Pedregosa et~al.,}{Pedregosa
  et~al.}{2012}]{sci-kit}
Pedregosa F.,  et~al., 2012, \mn@doi [J. Mach. Learn. Res.]
  {10.1007/s13398-014-0173-7.2}, 12, 2825

\bibitem[\protect\citeauthoryear{{Peeters}, {Spoon}  \& {Tielens}}{{Peeters}
  et~al.}{2004}]{peeters}
{Peeters} E.,  {Spoon} H.~W.~W.,   {Tielens} A.~G.~G.~M.,  2004, \mn@doi [\apj]
  {10.1086/423237}, \href {http://adsabs.harvard.edu/abs/2004ApJ...613..986P}
  {613, 986}

\bibitem[\protect\citeauthoryear{{Peeters}, {Bauschlicher}, {Allamandola},
  {Tielens}, {Ricca}  \& {Wolfire}}{{Peeters} et~al.}{2017a}]{peeters17}
{Peeters} E.,  {Bauschlicher} Jr. C.~W.,  {Allamandola} L.~J.,  {Tielens}
  A.~G.~G.~M.,  {Ricca} A.,   {Wolfire} M.~G.,  2017a, \mn@doi [\apj]
  {10.3847/1538-4357/836/2/198}, \href
  {http://adsabs.harvard.edu/abs/2017ApJ...836..198P} {836, 198}

\bibitem[\protect\citeauthoryear{{Peeters}, {Bauschlicher}, {Allamandola},
  {Tielens}, {Ricca}  \& {Wolfire}}{{Peeters} et~al.}{2017b}]{peeters2017}
{Peeters} E.,  {Bauschlicher} Jr. C.~W.,  {Allamandola} L.~J.,  {Tielens}
  A.~G.~G.~M.,  {Ricca} A.,   {Wolfire} M.~G.,  2017b, \mn@doi [\apj]
  {10.3847/1538-4357/836/2/198}, \href
  {http://adsabs.harvard.edu/abs/2017ApJ...836..198P} {836, 198}

\bibitem[\protect\citeauthoryear{P\'erez \& Granger}{P\'erez \&
  Granger}{2007}]{ipython}
P\'erez F.,  Granger B.~E.,  2007, \mn@doi [Computing in Science and
  Engineering] {10.1109/MCSE.2007.53}, 9, 21

\bibitem[\protect\citeauthoryear{{Petric} et~al.,}{{Petric}
  et~al.}{2011}]{petric}
{Petric} A.~O.,  et~al., 2011, \mn@doi [\apj] {10.1088/0004-637X/730/1/28},
  \href {http://adsabs.harvard.edu/abs/2011ApJ...730...28P} {730, 28}

\bibitem[\protect\citeauthoryear{{Petric} et~al.,}{{Petric}
  et~al.}{2018}]{petric18}
{Petric} A.~O.,  et~al., 2018, preprint, \href
  {https://ui.adsabs.harvard.edu/#abs/2018arXiv180509926P} {p.
  arXiv:1805.09926} (\mn@eprint {arXiv} {1805.09926})

\bibitem[\protect\citeauthoryear{{Reyes} et~al.,}{{Reyes} et~al.}{2008}]{reyes}
{Reyes} R.,  et~al., 2008, \mn@doi [\aj] {10.1088/0004-6256/136/6/2373}, \href
  {http://esoads.eso.org/abs/2008AJ....136.2373R} {136, 2373}

\bibitem[\protect\citeauthoryear{{Richings} \&
  {Faucher-Gigu{\`e}re}}{{Richings} \& {Faucher-Gigu{\`e}re}}{2018a}]{richings}
{Richings} A.~J.,  {Faucher-Gigu{\`e}re} C.-A.,  2018a, \mn@doi [\mnras]
  {10.1093/mnras/stx3014}, \href
  {https://ui.adsabs.harvard.edu/#abs/2018MNRAS.474.3673R} {474, 3673}

\bibitem[\protect\citeauthoryear{{Richings} \&
  {Faucher-Gigu{\`e}re}}{{Richings} \&
  {Faucher-Gigu{\`e}re}}{2018b}]{richingsB}
{Richings} A.~J.,  {Faucher-Gigu{\`e}re} C.-A.,  2018b, \mn@doi [\mnras]
  {10.1093/mnras/sty1285}, \href
  {https://ui.adsabs.harvard.edu/#abs/2018MNRAS.478.3100R} {478, 3100}

\bibitem[\protect\citeauthoryear{{Rigopoulou}, {Kunze}, {Lutz}, {Genzel}  \&
  {Moorwood}}{{Rigopoulou} et~al.}{2002}]{rigopoulou}
{Rigopoulou} D.,  {Kunze} D.,  {Lutz} D.,  {Genzel} R.,   {Moorwood} A.~F.~M.,
  2002, \mn@doi [\aap] {10.1051/0004-6361:20020607}, \href
  {http://adsabs.harvard.edu/abs/2002A\%26A...389..374R} {389, 374}

\bibitem[\protect\citeauthoryear{{Roussel} et~al.,}{{Roussel}
  et~al.}{2007}]{roussel}
{Roussel} H.,  et~al., 2007, \mn@doi [\apj] {10.1086/521667}, \href
  {http://adsabs.harvard.edu/abs/2007ApJ...669..959R} {669, 959}

\bibitem[\protect\citeauthoryear{{Rowan-Robinson} \&
  {Crawford}}{{Rowan-Robinson} \& {Crawford}}{1989}]{rowan}
{Rowan-Robinson} M.,  {Crawford} J.,  1989, \mn@doi [\mnras]
  {10.1093/mnras/238.2.523}, \href
  {http://adsabs.harvard.edu/abs/1989MNRAS.238..523R} {238, 523}

\bibitem[\protect\citeauthoryear{{Rupke} \& {Veilleux}}{{Rupke} \&
  {Veilleux}}{2013}]{rupke2013a}
{Rupke} D.~S.~N.,  {Veilleux} S.,  2013, \mn@doi [\apj]
  {10.1088/0004-637X/768/1/75}, \href
  {http://adsabs.harvard.edu/abs/2013ApJ...768...75R} {768, 75}

\bibitem[\protect\citeauthoryear{{Rupke}, {G{\"u}ltekin}  \&
  {Veilleux}}{{Rupke} et~al.}{2017}]{rupke2017}
{Rupke} D.~S.~N.,  {G{\"u}ltekin} K.,   {Veilleux} S.,  2017, \mn@doi [\apj]
  {10.3847/1538-4357/aa94d1}, \href
  {http://adsabs.harvard.edu/abs/2017ApJ...850...40R} {850, 40}

\bibitem[\protect\citeauthoryear{{Sadjadi}, {Zhang}  \& {Kwok}}{{Sadjadi}
  et~al.}{2015}]{sadjadi}
{Sadjadi} S.,  {Zhang} Y.,   {Kwok} S.,  2015, \mn@doi [\apj]
  {10.1088/0004-637X/807/1/95}, \href
  {http://adsabs.harvard.edu/abs/2015ApJ...807...95S} {807, 95}

\bibitem[\protect\citeauthoryear{{Sales}, {Pastoriza}  \& {Riffel}}{{Sales}
  et~al.}{2010}]{sales}
{Sales} D.~A.,  {Pastoriza} M.~G.,   {Riffel} R.,  2010, \mn@doi [\apj]
  {10.1088/0004-637X/725/1/605}, \href
  {http://adsabs.harvard.edu/abs/2010ApJ...725..605S} {725, 605}

\bibitem[\protect\citeauthoryear{{Sanders}, {Phinney}, {Neugebauer}, {Soifer}
  \& {Matthews}}{{Sanders} et~al.}{1989}]{sanders}
{Sanders} D.~B.,  {Phinney} E.~S.,  {Neugebauer} G.,  {Soifer} B.~T.,
  {Matthews} K.,  1989, \mn@doi [\apj] {10.1086/168094}, \href
  {http://adsabs.harvard.edu/abs/1989ApJ...347...29S} {347, 29}

\bibitem[\protect\citeauthoryear{{Sauvage} \& {Thuan}}{{Sauvage} \&
  {Thuan}}{1992}]{sauvage92}
{Sauvage} M.,  {Thuan} T.~X.,  1992, \mn@doi [\apjl] {10.1086/186519}, \href
  {http://adsabs.harvard.edu/abs/1992ApJ...396L..69S} {396, L69}

\bibitem[\protect\citeauthoryear{{Sauvage} \& {Thuan}}{{Sauvage} \&
  {Thuan}}{1994}]{sauvage94}
{Sauvage} M.,  {Thuan} T.~X.,  1994, \mn@doi [\apj] {10.1086/174308}, \href
  {http://adsabs.harvard.edu/abs/1994ApJ...429..153S} {429, 153}

\bibitem[\protect\citeauthoryear{{Shipley}, {Papovich}, {Rieke}, {Dey},
  {Jannuzi}, {Moustakas}  \& {Weiner}}{{Shipley} et~al.}{2013}]{shipley}
{Shipley} H.~V.,  {Papovich} C.,  {Rieke} G.~H.,  {Dey} A.,  {Jannuzi} B.~T.,
  {Moustakas} J.,   {Weiner} B.,  2013, \mn@doi [\apj]
  {10.1088/0004-637X/769/1/75}, \href
  {http://adsabs.harvard.edu/abs/2013ApJ...769...75S} {769, 75}

\bibitem[\protect\citeauthoryear{{Silk} \& {Mamon}}{{Silk} \&
  {Mamon}}{2012}]{silk12}
{Silk} J.,  {Mamon} G.~A.,  2012, \mn@doi [Research in Astronomy and
  Astrophysics] {10.1088/1674-4527/12/8/004}, \href
  {http://adsabs.harvard.edu/abs/2012RAA....12..917S} {12, 917}

\bibitem[\protect\citeauthoryear{{Silk} \& {Rees}}{{Silk} \&
  {Rees}}{1998}]{silk98}
{Silk} J.,  {Rees} M.~J.,  1998, \aap, \href
  {http://adsabs.harvard.edu/abs/1998A\%26A...331L...1S} {331, L1}

\bibitem[\protect\citeauthoryear{{Skrutskie} et~al.,}{{Skrutskie}
  et~al.}{2006}]{2MASS}
{Skrutskie} M.~F.,  et~al., 2006, \mn@doi [\aj] {10.1086/498708}, \href
  {http://adsabs.harvard.edu/abs/2006AJ....131.1163S} {131, 1163}

\bibitem[\protect\citeauthoryear{{Smith} \& {Draine}}{{Smith} \&
  {Draine}}{2012}]{pahfit}
{Smith} J.~D.,  {Draine} B.,  2012, {PAHFIT: Properties of PAH Emission},
  Astrophysics Source Code Library (\mn@eprint {ascl} {1210.009})

\bibitem[\protect\citeauthoryear{{Smith} et~al.,}{{Smith}
  et~al.}{2007}]{smith07}
{Smith} J.~D.~T.,  et~al., 2007, \mn@doi [\apj] {10.1086/510549}, \href
  {http://adsabs.harvard.edu/abs/2007ApJ...656..770S} {656, 770}

\bibitem[\protect\citeauthoryear{{Spoon}, {Marshall}, {Houck}, {Elitzur},
  {Hao}, {Armus}, {Brandl}  \& {Charmandaris}}{{Spoon} et~al.}{2007}]{spoon}
{Spoon} H.~W.~W.,  {Marshall} J.~A.,  {Houck} J.~R.,  {Elitzur} M.,  {Hao} L.,
  {Armus} L.,  {Brandl} B.~R.,   {Charmandaris} V.,  2007, \mn@doi [\apjl]
  {10.1086/511268}, \href {http://adsabs.harvard.edu/abs/2007ApJ...654L..49S}
  {654, L49}

\bibitem[\protect\citeauthoryear{{Spoon} et~al.,}{{Spoon}
  et~al.}{2013}]{spoon13}
{Spoon} H.~W.~W.,  et~al., 2013, \mn@doi [\apj] {10.1088/0004-637X/775/2/127},
  \href {http://adsabs.harvard.edu/abs/2013ApJ...775..127S} {775, 127}

\bibitem[\protect\citeauthoryear{{Stern} et~al.,}{{Stern}
  et~al.}{2005}]{stern05}
{Stern} D.,  et~al., 2005, \mn@doi [\apj] {10.1086/432523}, \href
  {http://adsabs.harvard.edu/abs/2005ApJ...631..163S} {631, 163}

\bibitem[\protect\citeauthoryear{{Stern} et~al.,}{{Stern} et~al.}{2012}]{stern}
{Stern} D.,  et~al., 2012, \mn@doi [\apj] {10.1088/0004-637X/753/1/30}, \href
  {http://adsabs.harvard.edu/abs/2012ApJ...753...30S} {753, 30}

\bibitem[\protect\citeauthoryear{{Stierwalt} et~al.,}{{Stierwalt}
  et~al.}{2014}]{stierwalt}
{Stierwalt} S.,  et~al., 2014, \mn@doi [\apj] {10.1088/0004-637X/790/2/124},
  \href {http://adsabs.harvard.edu/abs/2014ApJ...790..124S} {790, 124}

\bibitem[\protect\citeauthoryear{{Stock} \& {Peeters}}{{Stock} \&
  {Peeters}}{2017}]{stock17}
{Stock} D.~J.,  {Peeters} E.,  2017, \mn@doi [\apj] {10.3847/1538-4357/aa5f54},
  \href {http://adsabs.harvard.edu/abs/2017ApJ...837..129S} {837, 129}

\bibitem[\protect\citeauthoryear{{Stone}, {Veilleux}, {Mel{\'e}ndez}, {Sturm},
  {Graci{\'a}-Carpio}  \& {Gonz{\'a}lez-Alfonso}}{{Stone}
  et~al.}{2016}]{stone16}
{Stone} M.,  {Veilleux} S.,  {Mel{\'e}ndez} M.,  {Sturm} E.,
  {Graci{\'a}-Carpio} J.,   {Gonz{\'a}lez-Alfonso} E.,  2016, \mn@doi [\apj]
  {10.3847/0004-637X/826/2/111}, \href
  {https://ui.adsabs.harvard.edu/#abs/2016ApJ...826..111S} {826, 111}

\bibitem[\protect\citeauthoryear{{Sturm} et~al.,}{{Sturm}
  et~al.}{2011}]{sturm11}
{Sturm} E.,  et~al., 2011, \mn@doi [\apjl] {10.1088/2041-8205/733/1/L16}, \href
  {http://adsabs.harvard.edu/abs/2011ApJ...733L..16S} {733, L16}

\bibitem[\protect\citeauthoryear{{Tielens}}{{Tielens}}{2005}]{tielens05}
{Tielens} A.~G.~G.~M.,  2005, The Physics and Chemistry of the Interstellar
  Medium.
Cambridge University Press

\bibitem[\protect\citeauthoryear{{Tombesi}, {Mel{\'e}ndez}, {Veilleux},
  {Reeves}, {Gonz{\'a}lez-Alfonso}  \& {Reynolds}}{{Tombesi}
  et~al.}{2015}]{tombesi}
{Tombesi} F.,  {Mel{\'e}ndez} M.,  {Veilleux} S.,  {Reeves} J.~N.,
  {Gonz{\'a}lez-Alfonso} E.,   {Reynolds} C.~S.,  2015, \mn@doi [\nat]
  {10.1038/nature14261}, \href
  {http://adsabs.harvard.edu/abs/2015Natur.519..436T} {519, 436}

\bibitem[\protect\citeauthoryear{{Trump} et~al.,}{{Trump}
  et~al.}{2015}]{trump15}
{Trump} J.~R.,  et~al., 2015, \mn@doi [\apj] {10.1088/0004-637X/811/1/26},
  \href {http://adsabs.harvard.edu/abs/2015ApJ...811...26T} {811, 26}

\bibitem[\protect\citeauthoryear{{Turner}, {Kirby-Docken}  \&
  {Dalgarno}}{{Turner} et~al.}{1977}]{einsteincoeffs}
{Turner} J.,  {Kirby-Docken} K.,   {Dalgarno} A.,  1977, \mn@doi [\apjs]
  {10.1086/190481}, \href {http://adsabs.harvard.edu/abs/1977ApJS...35..281T}
  {35, 281}

\bibitem[\protect\citeauthoryear{{Villar-Mart{\'{\i}}n}, {Arribas}, {Emonts},
  {Humphrey}, {Tadhunter}, {Bessiere}, {Cabrera Lavers}  \& {Ramos
  Almeida}}{{Villar-Mart{\'{\i}}n} et~al.}{2016}]{villar}
{Villar-Mart{\'{\i}}n} M.,  {Arribas} S.,  {Emonts} B.,  {Humphrey} A.,
  {Tadhunter} C.,  {Bessiere} P.,  {Cabrera Lavers} A.,   {Ramos Almeida} C.,
  2016, \mn@doi [\mnras] {10.1093/mnras/stw901}, \href
  {http://adsabs.harvard.edu/abs/2016MNRAS.460..130V} {460, 130}

\bibitem[\protect\citeauthoryear{{Weedman}, {Feldman}, {Balzano}, {Ramsey},
  {Sramek}  \& {Wuu}}{{Weedman} et~al.}{1981}]{weedman}
{Weedman} D.~W.,  {Feldman} F.~R.,  {Balzano} V.~A.,  {Ramsey} L.~W.,  {Sramek}
  R.~A.,   {Wuu} C.-C.,  1981, \mn@doi [\apj] {10.1086/159133}, \href
  {http://adsabs.harvard.edu/abs/1981ApJ...248..105W} {248, 105}

\bibitem[\protect\citeauthoryear{{Weedman} et~al.,}{{Weedman}
  et~al.}{2005}]{weedman05}
{Weedman} D.~W.,  et~al., 2005, \mn@doi [\apj] {10.1086/466520}, \href
  {http://adsabs.harvard.edu/abs/2005ApJ...633..706W} {633, 706}

\bibitem[\protect\citeauthoryear{{Weinberger} et~al.,}{{Weinberger}
  et~al.}{2018}]{illustris2017}
{Weinberger} R.,  et~al., 2018, \mn@doi [\mnras] {10.1093/mnras/sty1733}, \href
  {https://ui.adsabs.harvard.edu/#abs/2018MNRAS.479.4056W} {479, 4056}

\bibitem[\protect\citeauthoryear{{Wright} et~al.,}{{Wright}
  et~al.}{2010}]{wright}
{Wright} E.~L.,  et~al., 2010, \mn@doi [\aj] {10.1088/0004-6256/140/6/1868},
  \href {http://adsabs.harvard.edu/abs/2010AJ....140.1868W} {140, 1868}

\bibitem[\protect\citeauthoryear{{Wu} et~al.,}{{Wu} et~al.}{2010}]{wu10}
{Wu} Y.,  et~al., 2010, \mn@doi [\apj] {10.1088/0004-637X/723/1/895}, \href
  {https://ui.adsabs.harvard.edu/#abs/2010ApJ...723..895W} {723, 895}

\bibitem[\protect\citeauthoryear{{Wylezalek} et~al.,}{{Wylezalek}
  et~al.}{2017}]{dominika}
{Wylezalek} D.,  et~al., 2017, \mn@doi [\mnras] {10.1093/mnras/stx246}, \href
  {https://ui.adsabs.harvard.edu/#abs/2017MNRAS.467.2612W} {467, 2612}

\bibitem[\protect\citeauthoryear{{Yuan}, {Kewley}  \& {Sanders}}{{Yuan}
  et~al.}{2010}]{yuan10}
{Yuan} T.~T.,  {Kewley} L.~J.,   {Sanders} D.~B.,  2010, \mn@doi [\apj]
  {10.1088/0004-637X/709/2/884}, \href
  {https://ui.adsabs.harvard.edu/#abs/2010ApJ...709..884Y} {709, 884}

\bibitem[\protect\citeauthoryear{{Yuan}, {Strauss}  \& {Zakamska}}{{Yuan}
  et~al.}{2016}]{yuan}
{Yuan} S.,  {Strauss} M.~A.,   {Zakamska} N.~L.,  2016, \mn@doi [\mnras]
  {10.1093/mnras/stw1747}, \href
  {http://adsabs.harvard.edu/abs/2016MNRAS.462.1603Y} {462, 1603}

\bibitem[\protect\citeauthoryear{{Zakamska}}{{Zakamska}}{2010}]{zakamska}
{Zakamska} N.~L.,  2010, \mn@doi [\nat] {10.1038/nature09037}, \href
  {http://adsabs.harvard.edu/abs/2010Natur.465...60Z} {465, 60}

\bibitem[\protect\citeauthoryear{{Zakamska}, {G{\'o}mez}, {Strauss}  \&
  {Krolik}}{{Zakamska} et~al.}{2008}]{zakamska08}
{Zakamska} N.~L.,  {G{\'o}mez} L.,  {Strauss} M.~A.,   {Krolik} J.~H.,  2008,
  \mn@doi [\aj] {10.1088/0004-6256/136/4/1607}, \href
  {http://esoads.eso.org/abs/2008AJ....136.1607Z} {136, 1607}

\bibitem[\protect\citeauthoryear{{Zakamska} et~al.,}{{Zakamska}
  et~al.}{2016}]{zakamska2016}
{Zakamska} N.~L.,  et~al., 2016, \mn@doi [\mnras] {10.1093/mnras/stv2571},
  \href {http://adsabs.harvard.edu/abs/2016MNRAS.455.4191Z} {455, 4191}

\bibitem[\protect\citeauthoryear{{Zhang} \& {Kwok}}{{Zhang} \&
  {Kwok}}{2015}]{zhang}
{Zhang} Y.,  {Kwok} S.,  2015, \mn@doi [\apj] {10.1088/0004-637X/798/1/37},
  \href {http://adsabs.harvard.edu/abs/2015ApJ...798...37Z} {798, 37}

\makeatother
\end{thebibliography}

\appendix

\bsp	% typesetting comment
\label{lastpage}
\end{document}